\documentclass[a4paper,11pt]{article}
\pdfoutput=1

\usepackage{jcappub}
\usepackage[T1]{fontenc}
\usepackage[dvipsnames]{xcolor}
\usepackage{makecell}
\usepackage{ulem}

\newcommand{\be}{\begin{equation}}
\newcommand{\ee}{\end{equation}}
\newcommand{\bea}{\begin{eqnarray}}
\newcommand{\eea}{\end{eqnarray}}

\newcommand{\kk}{{\langle K\rangle}}
\newcommand{\BB}{\mathcal{B}}
\newcommand{\FF}{\mathcal{F}}

\def\em#1{{\it #1}}

\newcommand{\CL}{{\tt ${\mathcal C}$osmo${\mathcal L}$attice}~}

\title{\boldmath Biased Domain Wall Networks and their Gravitational Waves}

\author[1,2]{Davide Barbini,}
\author[3]{Alessio Notari,}
\author[1]{Oriol Pujolàs,}
\author[2,1]{Fabrizio Rompineve,}
\author[4]{Francisco Torrentí}

\affiliation[1]{Institut de Física d'Altes Energies (IFAE), Campus UAB, Facultat Ciencies Nord, Bellatera, Barcelona, Spain}
\affiliation[2]{Departament de F\'isica, Universitat Aut\`onoma de Barcelona, 08193 Bellaterra, Barcelona, Spain}
\affiliation[3]{Dipartimento di Fisica, Sapienza University of Rome and INFN,
Piazzale Aldo Moro 2, I-00185, Italy}
\affiliation[4]{Department of Mathematics, Universidad Carlos III de Madrid, Avenida de la Universidad 30, 28911 Legan\'es, Madrid, Spain.}
\emailAdd{dbarbini@ifae.es, alessio.notari@uniroma1.it,   pujolas@ifae.es, frompineve@ifae.es, francisco.torrenti@uc3m.es}

\abstract{Cosmic Domain Wall networks are among the most interesting sources of a stochastic Gravitational Wave (GW) background from the early Universe. We present a thorough analysis of their annihilation, with a focus on scenarios where the collapse is induced by a \em{population bias}, whereby one of two degenerate vacua is initially preferred over the other. Our state-of-the-art $3+1$ lattice field theory simulations in the expanding Universe reveal that the network decays around the temperature $T_\text{ann}\sim T_s\,\mathcal{B}_s^{0.8}$, where $\mathcal{B}_s$ quantifies the preference for one vacuum over the other at the onset  of the scaling regime at the temperature $T_s$. Furthermore, we obtain the spectrum of GWs from such networks, and provide a detailed comparison with the alternative \em{potential bias} annihilation mechanism that relies on a small explicit symmetry breaking in the potential. \em{En passant}, we update results on the evolution of these networks and on their GWs, and clarify existing disagreements in the recent literature. 
Our results sharpen the phenomenological viability
of spontaneously broken discrete symmetries, and provide GW spectra that Pulsar Timing Arrays (PTAs) and ground-based interferometers (LIGO-Virgo-KAGRA) can readily use in their searches for a cosmological GW background.}

\begin{document}

\maketitle

\section{Introduction}

The rapid contemporary progress towards the detection of a stochastic gravitational wave (GW) background (SGWB) offers strong incentive to investigate well-motivated cosmological sources and to extract the features of their GW spectra. This is particularly true in light of the reported strong evidence for a SGWB at nHz frequencies by Pulsar Timing Array collaborations~\cite{NANOGrav:2023gor, EPTA:2023fyk, Reardon:2023gzh, Xu:2023wog, Miles:2024rjc}. The origin of the observed signal is yet to be established, and several cosmological sources provide as good of an interpretation of the data as the astrophysical origin from supermassive black hole binaries, if not a better one (at least if the astrophysical signal is modeled under simplifying assumptions~\cite{NANOGrav:2023hvm, EPTA:2023xxk}).

Cosmic Domain Walls (DWs)~\cite{Zeldovich:1974uw, Kibble:1976sj} (see also~\cite{Vilenkin:2000jqa} for an introduction to the subject) have long been pointed out as hypothetical strong sources of GWs in the early Universe~\cite{Vilenkin:1981zs}, and improving the description of their GW signal~\cite{Hiramatsu:2012sc, Hiramatsu:2013qaa} has been the aim of several recent works~\cite{Kitajima:2023cek, Ferreira:2024eru, Notari:2025kqq, Cyr:2025nzf, Babichev:2025stm, Blasi:2025tmn}. Their relevance for PTAs and LIGO-Virgo-KAGRA (LVK) searches has correspondingly been highlighted~\cite{Ferreira:2022zzo, NANOGrav:2023hvm, Blasi:2023sej, Gouttenoire:2023ftk, Jiang:2022svq, LIGOScientific:2025kry}. 

The crucial property of domain wall networks for cosmology is that, shortly after formation upon the spontaneous breaking of a  discrete symmetry (via e.g.~the Kibble mechanism), they achieve a scaling regime, in which a roughly fixed number of walls per Hubble volume persists in the Universe, independently of the (post-inflationary) background. This implies that DW networks tend to dominate over other components and spoil the original FRW spacetime~\cite{Zeldovich:1974uw}. However, this conclusion relies on the exactness of the underlying discrete symmetry. The presence of a small symmetry breaking term leads to a scaling-like evolution for a possibly long epoch, before inducing annihilation due to a so-called \em{potential bias} pressure~\cite{Sikivie:1982qv}. If this event occurs when the DWs make up a significant fraction of the energy budget of the Universe (above $\simeq 1\%$), their GWs are visible at current and/or near-future GW observatories. Only recently~\cite{Kitajima:2023cek, Ferreira:2024eru, Notari:2025kqq, Cyr:2025nzf, Babichev:2025stm}, it has been appreciated that the GW signal from viable DW networks is set by this late annihilation epoch, rather than by the preceding evolution in the scaling regime. Correspondingly, it has been shown that the amplitude of the signal is more than one order of magnitude larger than previously assumed~\cite{Ferreira:2024eru, Notari:2025kqq, Cyr:2025nzf} and that the spectral shape is altered~\cite{Notari:2025kqq, Cyr:2025nzf, Babichev:2025stm}. These improvements have been achieved by means of (classical) lattice field theory simulations \cite{Figueroa:2020rrl,Baeza-Ballesteros:2025tme}, which enable the computation of the non-linear evolution of DW networks, as well as the extraction of their GW spectra. Additionally, recent works \cite{Blanco-Pillado:2025gzs,Blanco-Pillado:2026nyh} have started to provide analytic insight on the mechanisms behind an enhanced GW emission at high frequencies compared to traditional estimates, in the thin wall limit.

From a modern perspective on global symmetries, a potential bias represents a well-motivated ingredient to obtain a cosmologically viable network. However, there are relevant scenarios in which this annihilation mechanism is not available or is not efficient enough to induce DW annihilation before domination (a graceful exit from a DW-dominated Universe is problematic~\cite{Vilenkin:2000jqa}). First, if the discrete symmetry is gauged (as in some scenarios that address the strong CP problem, imposing e.g. CP symmetry, à la Nelson-Barr~\cite{Nelson:1983zb, Barr:1984qx, Choi:1992xp}, see also~\cite{Asadi:2022vys}), then a potential bias would be forbidden by gauge symmetry. Second, even when the symmetry is global, the maximum size of a potential bias might be constrained by the microphysical model, thereby impeding annihilation before DW domination. A particularly relevant variant of this problem occurs in the post-inflationary QCD axion with $N_\text{dw}>1$~\cite{Peccei:1977ur, Weinberg:1977ma, Wilczek:1977pj, Sikivie:1982qv}, where the presence of a Peccei-Quinn breaking term is constrained by the strong CP problem, and annihilation may thus occur only when the axions radiated by DWs overclose the Universe, see e.g.~\cite{Ferrer:2018uiu}.

In this work, we thus investigate an alternative ingredient to induce DW annihilation that does not rely on an explicit breaking of the underlying discrete symmetry. Rather, we consider the possibility that a preference for one of the vacua is present at the epoch of the (spontaneous) symmetry breaking phase transition. This is realized if the initial configuration of the field that undergoes the phase transition is asymmetric, for instance if it is homogeneously displaced from the maximum of the potential, such that the probability of populating one vacuum rather than the other one during the phase transition deviates from $P=0.5$. While certainly much less studied than the potential bias mechanism, such a \em{population bias} scenario has been known to induce efficient domain wall annihilation for a long time~\cite{Lalak:1993ay, Coulson:1995nv, Larsson:1996sp, Hindmarsh:1996xv},
and has been numerically studied in more recent works~\cite{Correia:2014kqa,Correia:2018tty, Krajewski:2021jje, Gonzalez:2022mcx, Kitajima:2023kzu}, mostly by means of two-dimensional lattices, although a physical understanding of the annihilation process in this case is still somewhat lacking. Additionally, 2D simulations may not be well suited for inferring realistic properties of cosmological population-biased networks: indeed, the initial configuration of such a network depends crucially on the number of spatial dimensions $D$, since so does the probability threshold $P_c$ for percolation of both vacua. In particular, for $D=3$ the threshold is $P_c<0.5$, whereas for $D=2$ the threshold is closer to $P_c=0.5$  \cite{ZallenScher,Schandarin:1989sr}. This implies that, in the presence of a small population bias, a 3D DW network initially percolates, whereas this appears not to be the case in 2D.

\begin{figure*}
    \centering
    \includegraphics[width=\textwidth]{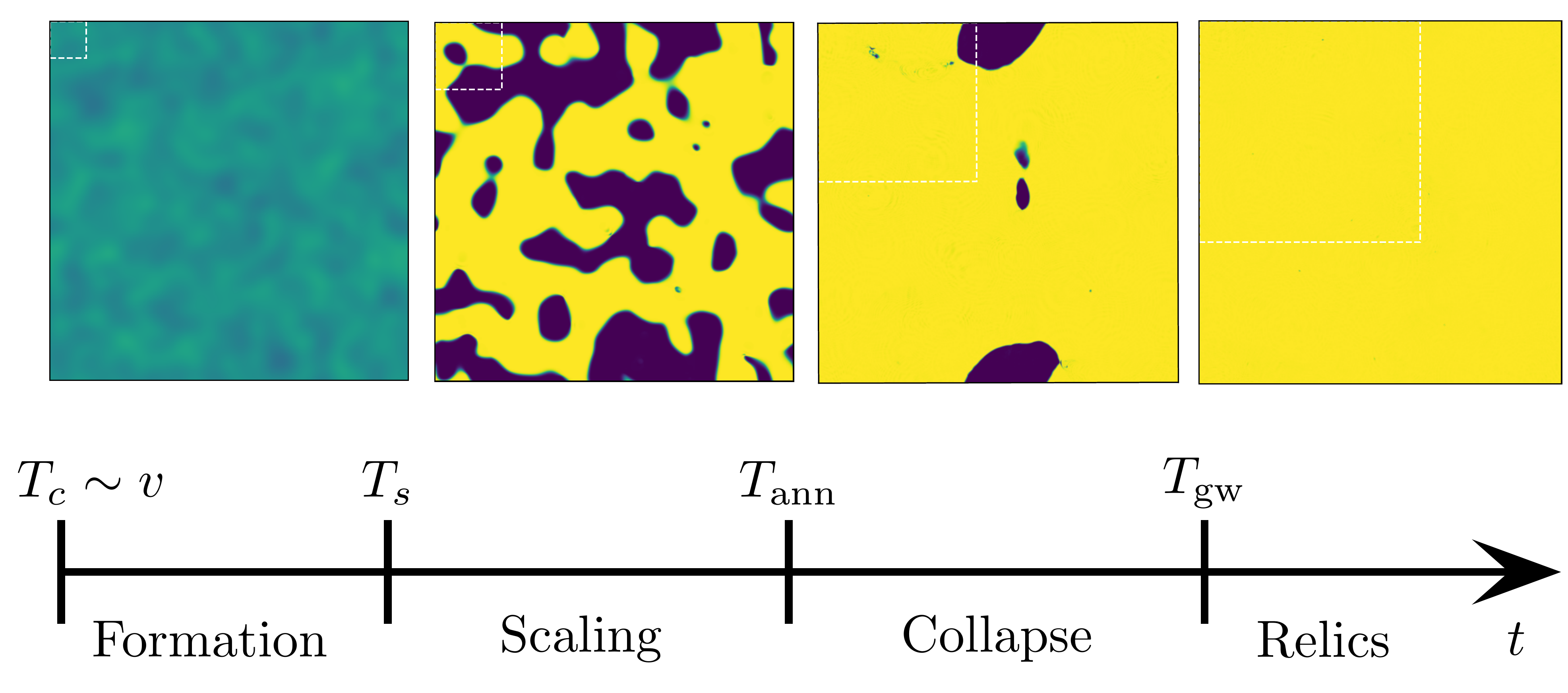}\\
    \caption{Sketch of the evolution of biased domain wall networks with cosmic time $t\propto T^{-2}$. The temperatures reported on the ticks indicate: the critical temperature of the symmetry-breaking phase transition $T_c$; the onset of the scaling regime $T_s$; the start of the annihilation phase $T_\text{ann}$; and the end of gravitational wave production $T_\text{gw}$. Representative 2D snapshots of the field values for each of the corresponding epochs, as obtained in our 3D numerical simulations with $N=4096$, $L=49$, are also shown (dark and yellow shading: less and more populated vacuum respectively). The squares in each panel have sides of length $1/H$, where $H$ is the Hubble rate.} \label{fig:evolution} 
\end{figure*}

Our main aims in this work are to improve the description of the evolution of population-biased networks and to provide the spectrum of gravitational waves radiated by the annihilation of the network. We focus on the simplest $\mathbb{Z}_2$ quartic potential, and use numerical strategies that are similar to those employed in~\cite{Ferreira:2024eru, Notari:2025kqq}, focusing on three-dimensional lattices. When the initial preference for one of the vacua is tiny, the network forms (percolates) and achieves the scaling regime as usual (a sketch of the evolution of biased networks is shown in Fig.~\ref{fig:evolution}). However, the fraction of the Universe in the initially less-populated minimum decreases sharply around a temperature $T_\text{ann}$, similarly to what happens in the presence of a potential bias. Nonetheless, we shall see that the decrease of the FV fraction in a population-biased network is more gradual. Owing to this similarity between the two scenarios, we shall often refer to the less-populated minimum as the ``false vacuum'', although we stress that the two vacua are actually degenerate. We then numerically extract the evolution of the false vacuum (FV) fraction in networks with initial population biases as small as $O(1-3)\%$.

We are able to follow the late annihilation of networks with such small biases thanks to, among other improvements, the use of a so-called \textit{fattening} technique~\cite{Press:1989yh}, by which one artificially keeps the comoving width of the domain walls constant,
thereby overcoming some limitations due to the loss of resolution in our grids. Furthermore, we aim to provide a relation between the annihilation temperature $T_\text{ann}$ and the population bias size, for a biased network. While we will not consider any specific origin for the population bias, we notice that inflationary diffusion provides a simple way to induce an initially asymmetric homogeneous value of the field, and scenarios in which population bias has been discussed in this context exist, see e.g.~\cite{Casini:2001ai, Hebecker:2016vbl, Kitajima:2023kzu}.

For gravitational waves, similarly to potential-biased networks, we find that the signal is dominated by the annihilation phase. While the cause and the dynamics of the annihilation are different in the two cases, we find overall similar GW spectra, with deviations in the detailed behaviour of the near-peak region and the UV tail of the spectrum, as well as a somewhat different enhancement of the amplitude compared to the scaling regime estimate.

A secondary aim of our work is to critically assess some claims in the recent literature on potential bias annihilation~\cite{Cyr:2025nzf, Babichev:2025stm}, by also presenting a new set of high-resolution simulations that broadly confirm the results of~\cite{Notari:2025kqq}.

This work is structured as follows. In Sec.~\ref{sec:PopBias} we introduce the model under investigation and provide a qualitative description of the growth of the bias in the presence of an initial preference for one of the vacua. In Sec.~\ref{sec:model}, we provide details about our numerical strategy. We then present results on the evolution of the network in the presence of an initial population bias in Sec.~\ref{sec:dynamics}, and a comparison with the potential bias scenario. In Sec.~\ref{sec:GWs}, we characterize the gravitational wave spectrum produced by the collapsing network and discuss its differences with respect to the potential bias scenario.
In Sec.~\ref{sec:comparison}, we comment on recent literature on the potential bias scenarios.
In Sec.~\ref{sec:summary}, we relate our results to GW observations, and offer our conclusions. The paper also includes App.~\ref{app:Technicalities} and App.~\ref{app:fattening}, where we discuss the technicalities of our simulations in more detail.

\section{Population bias} \label{sec:PopBias}

\subsection{Model} \label{sec:modmodel}

In this work, we focus on the simplest field theory that exhibits domain wall formation, namely, a real scalar field with potential:
\be V(\phi) = \frac{\lambda}{4} (\phi^2 - v^2)^2 \ , \label{eq:quartic-pot}\ee
featuring two degenerate minima at $\phi =v$ and $\phi = -v$. Several other well-motivated possibilities could also be considered.

Upon spontaneous symmetry breaking in the early Universe, which can occur, for instance, via the usual Kibble mechanism around the critical temperature $T_c\sim v$, a DW network forms and soon attains a scaling regime, with approximately one domain wall per Hubble patch. At large scales, this network is essentially characterized by two quantities: the (approximate) physical width of the walls $\delta_{w} \equiv m^{-1}$, where $m \equiv \sqrt{2 \lambda} v$ is the mass around the minima; and their tension (i.e.~the mass per unit area) $\sigma \equiv \int_{-v}^{+v}\sqrt{2V(\phi)}d\phi=  2 \sqrt{2 \lambda} v^3 / 3$. The energy density of the network is therefore
\be \rho_{\rm dw} = \frac{\sigma A a^2}{Va^3} = 2 \mathcal{A} \sigma H \ , \hspace{0.5cm} \mathcal{A} \equiv \frac{A}{V} \frac{1}{2 a H} \ ,  \label{eq:area-parameter} \ee
where  the \textit{area parameter} $\mathcal{A}$ is defined in terms of the total comoving area $A$ of the domain walls in a region of comoving volume $V$ and $H$ is the Hubble rate. In the scaling regime, the area parameter is approximately constant and of order one for the potential considered here (see also~\cite{Ferreira:2024eru, Notari:2025kqq}).

The volume fractions of any of the vacua are  of course central quantities in population-biased networks. In a $\mathbb{Z}_2$ model, we can focus on the fraction of one of the two vacua, which we take to be the disappearing vacuum for definiteness. In an abuse of language, we shall borrow the terminology for potential bias~\cite{Ferreira:2024eru} and refer to it as the \em{false vacuum} (FV) \em{fraction} 
\begin{equation}
    \mathcal{F}\equiv\frac{V_-}{V_-+V_+}~,
\end{equation} 
where $V_-$ and $V_+$ refer to the volumes occupied by each vacua, and understanding that it is now ``false'' only because we choose to bias the initial condition towards the positive minimum in our simulations.

We are interested in an initial condition with non-zero but very small bias, $\mathcal{F} \approx 0.5$, which causes the network to eventually annihilate, with the Universe (our simulation box) ending up in the positive vacuum, i.e.~$\mathcal{F} \rightarrow 0$ at sufficiently late times. The following quantity thus provides a natural definition of \em{population bias},
\begin{equation}\label{Bdefinition}
\mathcal{B} \equiv \frac12 - \mathcal{F} = \frac12 \frac{V_+-V_-}{V_++V_-} ~.
\end{equation}

In terms of this quantity, the summary of how annihilation proceeds is as follows: the network starts with a tiny value of $\BB$ and spends a long time in a regime that is virtually indistinguishable from the usual cosmological scaling behaviour. We refer to this as \textit{quasi-scaling} regime (although we will often drop the ``quasi'', in a slight abuse of language). In simulations, it is observed that the bias $\BB$ grows with time, with two distinct phases. As $\BB\ll1$, the bias grows as a power law, which using conformal time reads:
\begin{equation}\label{powerlaw}
    \BB(\eta)= \BB_s\,\left(\frac{\eta}{\eta_s}\right)^p~,
\end{equation}
being characterized by the \em{bias growth exponent} $p$ only. Here, $\eta_s$ and $\BB_s\ll1$ denote the conformal time at which  the network reaches the quasi-scaling epoch and the initial bias at that time respectively. Note that   \eqref{powerlaw}  already implies a power law behaviour for the annihilation time (setting $\BB$ to a fixed threshold below $0.5$ at $\eta_\text{ann}$)
\begin{equation}
\label{etaannB}    \eta_\text{ann}\sim\eta_s \, \BB_s^{\,-1/p}~.
\end{equation}
We shall see that this expectation is roughly confirmed by our simulations in Sec.~\ref{sec:dynamics}.

Once $\BB$ has grown beyond a certain threshold, the disappearance of the FV becomes much faster, and is better characterized as
\begin{equation}\label{exp}
    \FF\sim \frac12\,\exp\left[-\left(\frac{\eta}{\eta_{\text{ann}}}\right)^{\tilde p}\,\right] \ ,
\end{equation}
with $\tilde p$ not necessarily coincident with  the initial exponent $p$ in \eqref{powerlaw}.

This markedly different time dependence only manifests, of course, that the DW network is in extremely different configurations. Initially, in scaling or with a very small $\BB$, the DW network percolates the whole volume, with basically all of the network consisting of one single connected DW component. However, as $\BB$ grows, percolation is lost and the network must eventually evolve into a collection of isolated closed walls. A rough idea for when the transition between the two regimes of growth happens is naturally provided by the percolation threshold $\BB_c$. Results from percolation theory on continuous random Gaussian fields in 3D suggest that the percolation threshold is around $\FF_c\sim 0.1-0.2$ \cite{ZallenScher,Schandarin:1989sr}, or equivalently $\BB_c\sim 0.3-0.4$. 

Our numerical simulations, presented in Sec.~\ref{sec:dynamics} (in particular~\eqref{eq:FfvFunc2}, \eqref{eq:pp} and Fig.~\ref{fig:PopBiasFits}), certainly go below this rough threshold and provide evidence that the best fitting function to the numerically extracted FV fraction is of the form~\eqref{exp} with a running exponent $p$.

\subsection{Qualitative picture}\label{sec:analytics}

Let us provide here a qualitative discussion on the causes of bias growth, focusing on the initial stages when the network is still percolated, and $\mathcal{B}\ll 1$.

To start, let us comment on well-known analytic work in this direction~\cite{Hindmarsh:1996xv}, which posits $\mathcal{A}\sim \exp\left(- {\rm const }\,\mathcal{B}_s^2~\eta^D\right)$, strongly relying on an assumption of Gaussianity of the field. In 2D the relation above implies $\mathcal{A}\sim \exp\left(-\eta^2\right)$, however the numerical results of~\cite{Correia:2014kqa} show that $\mathcal{A}\sim \exp\left(-\eta\right)$ provides a much better fit to simulation data. Furthermore, extending the suppression of~\cite{Hindmarsh:1996xv} to the FV fraction and 
expanding for small $\eta$ leads to a power-law growth of $\BB$ (with a  prefactor that is quadratic in the bias) and in 3D  it predicts $p=3$. This is quite far from the behaviour observed in our simulations (see Sec.~\ref{sec:dynamics}), which, to our knowledge, are the first ones to obtain $p$ numerically in 3D. This motivates us to attempt a different strategy to gain some  insight. 

In the quasi-scaling epoch, that is while the network is percolated, the latter consists essentially of an infinite DW that spans the whole universe. As usual, the DW  is bent on the scale of the Hubble scale so that there is one large DW every Hubble patch and there is a negligible amount of DWs (DW area) in the form of sub-Hubble closed DWs. 

It is clear that introducing a population bias has a double effect: by definition, one of  the volume fraction decreases, say $V_-$ to follow our previous choice. At the same time though, the spatial geometry of the interface (the DW  network) must change accordingly. In particular, one expects that the extrinsic curvature (of the spatial geometry) $K$ of the DW network is also sensitive to $\BB$. 

The extrinsic curvature  plays an important role in the dynamics of population biased networks because, as per the Nambu-Goto (NG) equation (see e.g.~\cite{Vilenkin:2000jqa}), it is the one and only driving force that pushes the DWs when the pressure bias is absent. To see this, consider the NG equation for closed DWs of radius $R(\eta)$ for cylindrical and spherical geometries ($n=1, 2$ respectively), which follows from the NG Lagrangian $L =-\sigma \;  a^3(\eta) R^n(\eta)\sqrt{1-R'(\eta)^2}$, where $R$ is the comoving coordinate, $\eta$ the conformal time and $a$ is the FLRW scale factor:
\begin{equation}\label{NG}
    R''+ \left(\frac{n }{R} + \frac{3 a' }{a} R'\right)\left(1-R'^2\right)=0~.
\end{equation}
Aside from the Hubble friction term, one recognizes a force proportional to the extrinsic curvature $K\propto n/R$. Locally, the motion of each piece of the DW is driven by its value of $K$. 

This suggests that what controls the bias growth in a network is a global (average) value of the trace of the extrinsic curvature $\langle K\rangle$, which is a well defined quantity (for instance, it can be defined as a large volume limit of the ratio of area-integrals $\int K dA/\int dA$). Certainly, $\langle K\rangle$ seems in principle calculable, including whether/how  it depends on the bias $\BB$. It only requires a global direction of the normal vector to the surface, but this is certainly possible because at percolation the DW surface divides the whole volume in two equal parts, and we are free to point the normal from, say, the $V_-$ to the $V_+$ volume. For an unbiased network i.e. at $\BB=0$,
one expects $\langle K\rangle=0$  at large volume because the network is random and unbiased, and so there are equally concave and convex regions of both volumes overall (in spite of  $K\neq0$   generically everywhere).

There are two arguments indicating that $\langle K\rangle$ depends on $\BB$. The first one comes from the geometry of random surfaces, which has been much studied using level set techniques. The level sets of Gaussian random fields indeed share many qualitative similarities to the (spatial) geometries defined by DW  networks. It is a well known result that $\langle K\rangle$ is in fact exactly linear in the equivalent notion to $\BB$ -- the normalized level set \cite{Schmalzing:1997uc,Pranav:2018pnu,Matsubara:2003yt}. 

Second, the most important properties of the unbiased network are: $V_-=V_+$ and $\langle K\rangle=0$. We can replace the network by  simpler configurations with the same properties. The simplest consists in a pair of closed DWs, enclosing $+$ and $-$ vacua each, with the same radii $R_\pm$. Or similarly, with a pair of cylinders. Other shapes, more faithful to the actual topology of the network could be used. 
In either case, $\BB=\kk=0$ for equal radii, $R_\pm=R_0$. 

Increasing, say, $R_+$ introduces both nonzero $\BB$ and $\kk$.
For these proxy configurations, ($n=1,2$ for cylinder and sphere respectively), setting
\begin{equation}\label{Rpm}
R_\pm = R_0  \pm \delta  ~
\end{equation}
and identifying $V_\pm\propto R_\pm^3$, $K_\pm\propto 1/R_\pm$, leads to 
both a population bias and a net driving force at leading order in $\delta$,
\begin{equation}
    \BB  \sim    \frac32 \frac{\delta}{R_0}
\qquad
{\rm and }
\qquad
\frac{K_+-K_-}{K_++K_-} \sim   - \frac{\delta}{R_0}~,
\end{equation}
which is expected to  translate into a tendency of $\BB$ to grow. 

We can track the evolution of $\BB(\eta)$ for these two shapes from the NG equation~\eqref{NG}. This logic suggests how to extend it to more general shapes that are still characterized by a single scale $R$. 
The first step is that we can identify $n/R$ as (some measure of) the extrinsic curvature, which is the driver of the motion for that shape. With this in mind we can think of $n$ as a generic number. We may try to define what is the value of $n$ that characterizes the network by imagining that we cut it in cubes of size equal to the correlation length. In scaling, there is one large DW that cuts each cube in two statistically equal (in volume) sides.  
It is clear that the parameter $n/R$ corresponds to the  value for $\kk_{\text{r.m.s.}}$ for this region. 
Note that in scaling the network is percolated and in every Hubble patch there is a large DW. The realistic shapes in each patch are not closed, nevertheless an equal partition of $+$ and $-$ persists in each Hubble patch. Yet, the typical shape of the DWs is not flat, therefore the extrinsic curvature does not vanish locally, being of order Hubble scale. The typical shape of a DW crossing the Hubble  patch is actually quite flat so a good guess for $n$ (that plays a role of $\kk$ times the correlation length) is in the range $0 - 1$. 

Second, we linearize the NG equation in the bias with the prescription of \eqref{Rpm} with $R_0\equiv R_0(\eta), \delta\equiv \delta(\eta)$. 
Assuming that the scaling network is statistically homogeneous and isotropic, then a given shape appears with all possible orientations and local velocities. 
Therefore, marginalizing over velocities effectively cancels the terms that are odd in $R_0'$ while the even terms are replaced by the typical (r.m.s.) wall speed, $v$.
Thus we obtain
\begin{equation}\label{delta}
    \delta''(\eta) + 3\left(1-3\,v^2\right)\frac{\delta'(\eta)}{\eta}  - \frac{n }{R_0^2} \left(1-v^2\right) \delta(\eta)=0~,
\end{equation}
The length scale $R_0$ is meant to represent the typical size of the objects at any moment in the unbiased scaling network so it is identified as the (time dependent) correlation length, which is of order of the comoving Hubble length  $\eta$. Hence, it is natural to take
\begin{equation}
    R_0(\eta) = \varepsilon\; \eta
\end{equation}
where $\varepsilon$ is an $O(1)$ value that retains information on the sampling of sizes in the ensemble defined by the scaling network itself.

With this identification, Eq.~\eqref{delta} admits simple power law solutions. Expressing them directly in terms of the bias $\BB(\eta) = 3\delta/(2R_0)$, the solutions  are $\BB \sim \eta^p$, with a growing  mode and a decaying mode. For the growing mode,
\begin{equation}\label{p_analytic}
p = -2+\frac92\,v^2- \sqrt{\left(1-\frac{9}{2}v^2\right)^2 + \frac{1-v^2} {\varepsilon^2} \,n }\; .
\end{equation}

Interestingly, this relates the growth exponent to geometric and kinematic quantities ($n, v$ and  $\varepsilon$) that characterize the unbiased network. 
The computation of these quantities in realistic networks is a well defined task which we leave for future work. Let us notice that similar notions arise in the Velocity-One-Scale (VOS) model~\cite{Avelino:2005kn,Martins:2016ois}. It is however not clear that the VOS prescription provides the appropriate measure of (some of) these parameters. 

Overall, in this section we have provided a qualitative picture of bias growth in population-biased networks. We now provide a quantitative analysis based on lattice simulations.

\section{Lattice formulation} \label{sec:model}

Let us outline some basic features of the lattice field theory simulations whose results we present in the following sections (more details can be found in App.~\ref{app:Technicalities}). We solve the scalar field dynamics in 3+1-dimensional comoving lattices with $N$ points per dimension and side length $L$, using a modified version of \CL \cite{Figueroa:2021yhd}. We thus numerically solve the equation of motion for $\phi = \phi (\vec{x}, \eta)$:
\be \phi'' - \nabla^2 \phi + 2 \mathcal{H} \phi' =  - a^2 \partial_{\phi} V \ , \label{eq:EOMsc}
\ee
where $d\eta\equiv dt/a(t)$ is conformal time, $t$ is cosmic time, and $\mathcal{H} \equiv \frac{a'}{a} = a H$, with $H\equiv\frac{1}{a}\frac{da}{dt}$. We set initial conditions as:
\be \phi (\vec{x},\eta_i) = \bar{\phi} (\eta_i)+ \delta \phi (\vec{x}, \eta_i) \ , \label{eq:ICondHom}\ee
where $\bar{\phi}_{\rm i} \equiv \bar{\phi} (\eta_i) \ll v$ is the initial homogeneous component and $\delta \phi (\vec{x}, \eta_i)$ represents small fluctuations that vary randomly throughout space. We consider a fixed radiation-dominated background, so that $a(\eta) \propto \eta$ and the energy density of the DW network in the scaling regime behaves as $\rho_{\rm dw} \propto H \sim \eta^{-2}$. In order to mimic the initial conditions in a Kibble-like phase transition, we impose Gaussian fluctuations with  white-noise power spectrum:
\begin{equation}
\langle \delta \phi^2 \rangle_{\rm i}^{k=k_\text{cut}} \equiv  \int_0^{k_\text{cut}} d \, {\rm log} k \, \Delta_{ \phi} (k) \ , \quad   \Delta_{\phi} (k) \equiv \frac{k^3}{2 \pi^2} |\phi_k|^2= c \frac{k^3}{4 \pi^2 m}\ , \label{eq:PowSc} \end{equation}
where $c$ is a dimensionless constant that sets the amplitude of the power spectrum and $k_\text{cut}$ is an ultraviolet cutoff.  We choose $k_\text{cut} = 2 m$ and $c=10^{-10}$ as in~\cite{Ferreira:2024eru, Notari:2025kqq} (these choices simply accelerate the approach to the scaling regime, without affecting physical results). While the power spectrum~\eqref{eq:PowSc} indeed reproduces the expectation of causality-limited correlations that characterize a post-inflationary thermal phase transition, we also set $H(\eta_i=0) \equiv H_i = m$ at the beginning of the simulation for numerical convenience. These initial conditions do not correspond to those of a thermal phase transition, since in that case $H_i\sim T_c^2/M_p\ll m$, at least when $\lambda$ is not too small. Nonetheless, the impact of these differences on our results is negligible, since we are interested in networks that annihilate after reaching the scaling regime.

We quantify the presence of an initial population bias by the dimensionless \textit{bias parameter},
\be b_i \equiv \frac{ \bar{\phi}_i }{\sqrt{\langle \delta \phi^2 \rangle_i^{k_{\rm cut}} } } = 2\pi\sqrt{\frac{3m}{c \, k_{\rm cut}^3}}\bar{\phi}_i \ , \label{eq:biaspar} \ee
where $\langle \delta \phi^2 \rangle_i^{k=k_{\rm cut}} \equiv c k_{\rm cut}^3/(12 m \pi^2)$ is the variance integrated over the momentum range $k \in [0, k_{\rm cut}]$. For our particular choice of $k_{\rm cut}=2m$, this reduces to $b_i = \sqrt{\frac{3}{2c}} \frac{\pi}{m} \bar{\phi}_i$. This range includes the comoving momenta excited during the initial linear regime, i.e.~when the field starts decaying towards the potential minima before DW formation. We have checked that our results depend only on $b_i$, and not separately on $\bar{\phi}_i$ and $c$, see App.~\ref{app:Technicalities} for more details. Importantly, $b_i$ does not coincide with the bias at the onset of the scaling regime, as we will discuss in the next section. 

We use units $v=1$ and $m=\sqrt{2 \lambda} v = 1$, i.e. we set $\lambda = 1/2$, with time and space both dimensionless. In these units, the DW tension is $\sigma = 2/3$, $H_i=m=1$, so that $a(\eta)=1+\eta$ and $H(\eta)=(1+\eta)^{-2}$. These choices are consistent with~\cite{Ferreira:2024eru, Notari:2025kqq}.

\begin{figure*}
    \centering
    \includegraphics[width=0.85\textwidth]{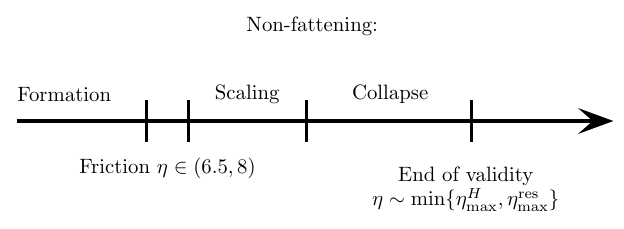}\\
    \includegraphics[width=0.85\textwidth]{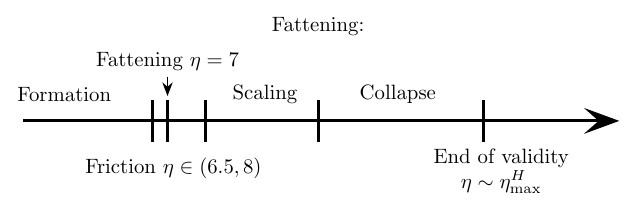}
    \caption{Sketch of the different phases in the numerical simulations of domain walls. \textit{Top}: simulation without fattening. \textit{Bottom}: simulation with fattening.} \label{fig:sketch_simulation} 
\end{figure*}

In our work, we use two numerical strategies to evolve the DW network and observe its collapse:
\begin{itemize}
    \item In one set of simulations, we solve a discretized version of Eq.~\eqref{eq:EOMsc} during the entire dynamical evolution. The time range of our simulations is then limited by two conditions. First, the need to resolve the width of the domain walls for as long as they are present in our simulations. Since the physical lattice spacing $\Delta x_p$ increases as $a(t)$, the condition $\delta_w\gtrsim \Delta x_p$ leads to\footnote{Our definition for $\eta_{\rm max}^{\rm res}$ \eqref{eq:tmax} is obtained by equating the comoving momentum associated with the domain wall width, $k_{\rm w} \equiv 2 \pi a \delta_w^{-1}$, to the maximum momentum captured by the lattice $k_{\rm max} \equiv \sqrt{3} \pi /\Delta x$, which corresponds to the diagonal of the Fourier lattice. However, this expression must be interpreted only as an estimate of the time at which loss-of-resolution effects may become relevant. } 
    \begin{equation}
        \eta \lesssim \eta_{\rm max}^{\rm res} \equiv \frac{\sqrt{3}}{2}\, \frac{N}{L} - 1  \ .
        \label{eq:tmax}
    \end{equation}
    Moreover, our results are only reliable if the lattice contains a sufficiently large number of Hubble patches $\mathcal{N}_H(\eta)$, i.e.~if
    \begin{equation}
        \mathcal{N}_H(\eta) \equiv \left( \frac{a(\eta)\, L}{H^{-1}(\eta)} \right)^3 \sim \frac{L^3}{\eta^{3}} \gg 1 \hspace{0.3cm} \Rightarrow \hspace{0.3cm} \eta \lesssim \eta_{\rm max}^{\rm H} = L \ .
        \label{eq:tmaxpatch}
    \end{equation}
    The maximum time for which domain walls can be reliably analyzed in this first set of simulations is then given by the minimum of~\eqref{eq:tmax} and~\eqref{eq:tmaxpatch}.
    \item In order to extend the range of biases explored, we also perform a second set of simulations using a so-called \textit{fattening} procedure~\cite{Press:1989yh} after the formation of the domain walls, so that the physical width of the DWs grows with time. This allows us to disregard condition~\eqref{eq:tmax}, leaving ~\eqref{eq:tmaxpatch} as the only limitation. In order to achieve this, we evolve the field according to the modified equation of motion,
    \be \phi'' + \alpha \mathcal{H} \phi' - \nabla^2 \phi = - a^{\beta} \partial_{\phi} V \label{eq:EOMfattening} \ee
    with $\alpha = 3 $ and $\beta = 0$, rather than~\eqref{eq:EOMsc}. The choice of $\beta$ leads to $\delta_w\sim a$, while the choice of $\alpha$ allows us to reproduce the rate at which the field settles into the minima of the physical case, i.e.~$\langle \phi - v \rangle_{\rm rms} \propto a^{-\alpha /2 - \beta/4} = a^{-3/2}$. While this new equation of motion is not physical, it has been shown to closely reproduce the physical dynamics of the network once the DW width is sufficiently smaller than the Hubble radius (thin-wall regime) \cite{Press:1989yh, Sousa:2010zza}. A similar strategy has previously been applied to both 3D and 2D domain wall networks, with biased and unbiased initial conditions \cite{Larsson:1996sp,Correia:2014kqa,Correia:2018tty,Coulson:1995nv,Oliveira:2004he}. However, in this work, rather than applying \eqref{eq:EOMfattening} from the beginning of the simulation, we initially evolve the field using the physical equation of motion \eqref{eq:EOMsc}, and apply fattening only after DW formation. Indeed, the presence of fattening in the initial stages of DW formation would induce a greater growth in the population bias than in simulations without fattening, see App.~\ref{app:fattening} for more details. Our novel strategy avoids this issue while still allowing us to follow DWs through a much longer dynamical range. Overall, our second set of simulations solves the equations:
    \begin{equation}
        \begin{cases}
            \phi'' - \nabla^2 \phi + 2 \mathcal{H} \phi' =  - a^2 \partial_{\phi} V & \eta \leq \eta_F \\
            \phi'' - \nabla^2 \phi + 3 \mathcal{H} \phi' =  - a(\eta_F)^2 \partial_{\phi} V & \eta>\eta_F
        \end{cases},
        \label{eq:EOMfatact}
    \end{equation}
    where $\eta_F$ is chosen to be around the formation time of the DW network. The factor $a(\eta_F)$ is necessary to match the DW widths at time $\eta_F$.
\end{itemize}

In addition, in order to accelerate the emergence of the scaling regime and to reduce the presence of homogeneous field oscillations even after DW formation, we incorporate an initial stage of \textit{friction} (or \textit{diffusion}) immediately after domain wall formation, as in~\cite{Notari:2025kqq}, in both sets of simulations. This is implemented through an arbitrary, unphysical increase of the friction term in the equation of motion, see App.~B.2 of \cite{Notari:2025kqq} for more details. In our set-up, domain walls form at $\eta \sim 6$, while the friction stage is active only at times $6.5 \lesssim \eta \lesssim 8$ approximately. A sketch of the timeline for both simulations, with and without fattening, can be found in Fig.~\ref{fig:sketch_simulation}.

Let us also mention that, when solving Eq.~\eqref{eq:EOMsc} (or equivalently, Eq.~\eqref{eq:EOMfattening}), one must replace the Laplacian with discretized operators that approximate the continuum expressions up to a certain order in the lattice spacing $\delta x \equiv L/N$. The lattice results presented in Sec.~\ref{sec:dynamics}, which study only the scalar field dynamics, have been obtained with the second-order-accurate spatial derivatives implemented in \CL by default. However, the lattice simulations presented in Sec.~\ref{sec:GWs}, which include the gravitational waves, instead use fourth-order-accurate derivatives, which improve the accuracy of the spectrum at mid-to-high frequencies, see App.~B.3 of \cite{Notari:2025kqq} for details. Finally, we mention that, while the simulations including only the scalar field have been carried out with double precision, simulations with GWs have been obtained with single precision to minimize memory consumption. We have verified that the results obtained with single and double precision are compatible.

Both of these aspects, among others, distinguish our work from other recent studies on biased and unbiased domain walls, such as~\cite{Cyr:2025nzf, Babichev:2025stm, Blasi:2025tmn}. We comment on their impact and compare with those works in Sec.~\ref{sec:comparison}.

\section{Evolution and Collapse of Biased Networks} \label{sec:dynamics}

We are now in a position to present results from our two sets of lattice field theory simulations. \em{Physical simulations} are performed on lattices of side length $L=80$ and $N=4096$ points per dimension (unless otherwise stated), which, according to Eq.~\eqref{eq:tmax}, allow to resolve the domain walls until time $\eta_{\rm max}^{\rm res} \approx 43$. We simulate four different realizations for several values of the initial bias, in the range $b_i \in [3.8 \cdot 10^{-3}, 3.8 \cdot 10^{-2}]$. Indeed, significantly larger initial biases prevent the formation and scaling of the domain walls, while smaller biases lead to annihilation times beyond the range of our simulations.
\textit{Simulations with fattening}, on the other hand, are performed on lattices with $N=3840$ and side length $L=240$. These can be evolved until $\eta \approx 240$ while still having $\mathcal{N}_H(\eta) > 1$, as they are only limited by~\eqref{eq:tmaxpatch}. The advantages of this second set of simulations over the previous set are clear: first, the much larger simulation range allows us to probe longer annihilation times, induced by smaller biases. Second, larger biases can be simulated in lattices containing many more Hubble patches. The former allows us to consider networks that have more reliably achieved the scaling regime before collapse becomes efficient, while the latter allows us to extract the same parameters as in simulations without fattening, but with smaller statistical uncertainties. For this set of simulations, we choose the initial bias parameters in the range $b_i \in [2.3 \cdot 10^{-3}, 7.7 \cdot 10^{-3}]$, and obtain three different realizations for each value of the bias.

\begin{figure*}
    \centering
    \includegraphics[width=0.47\textwidth]{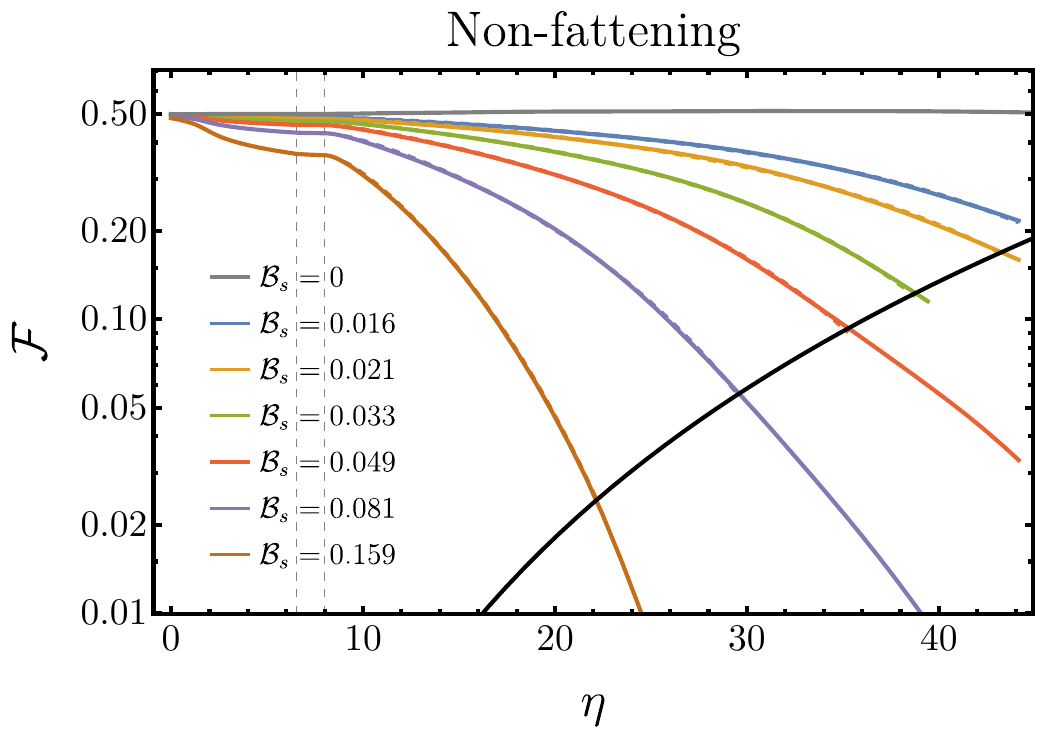} \hspace{0.3cm}
    \includegraphics[width=0.47\textwidth]{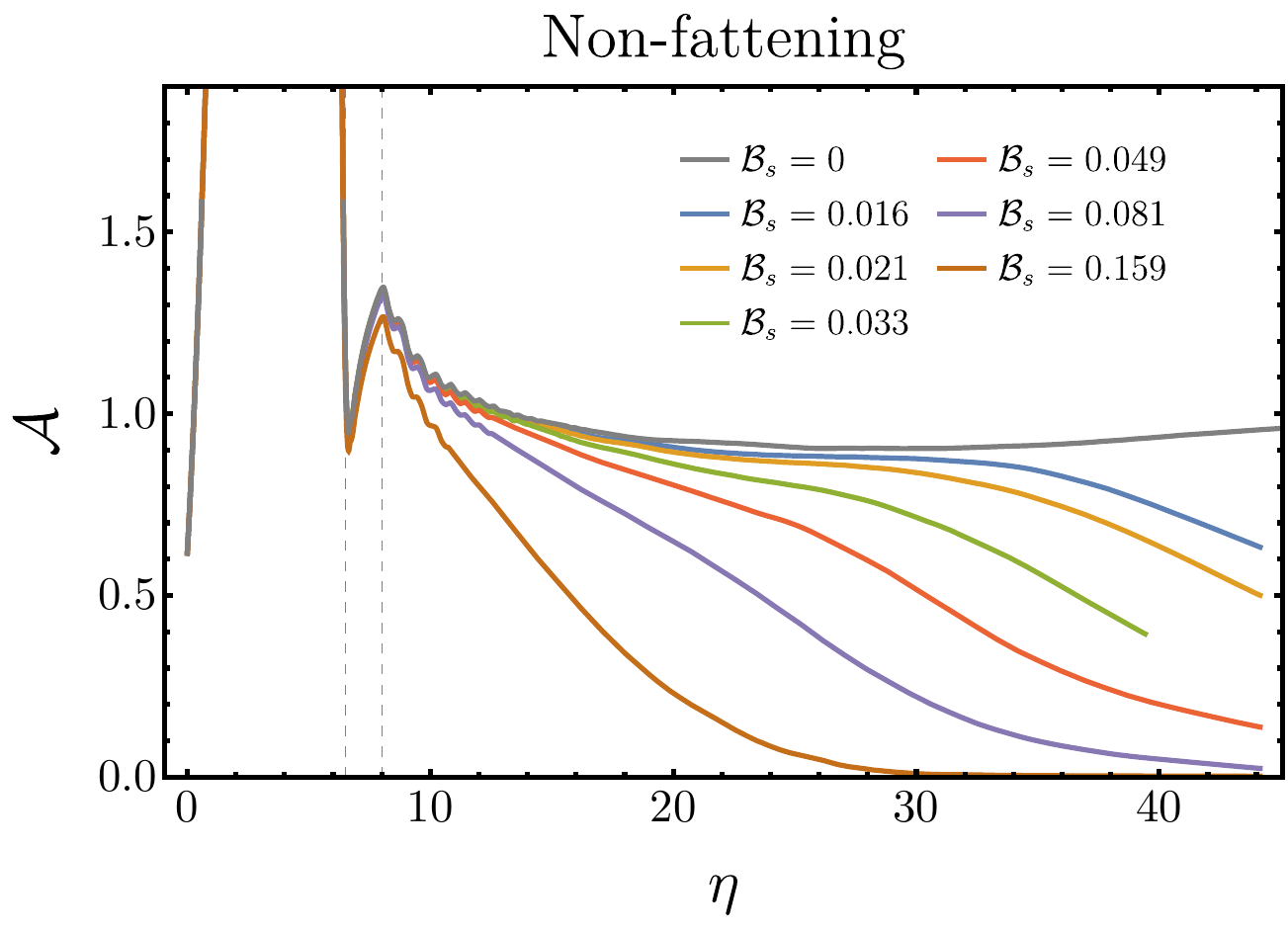} 
    \includegraphics[width=0.47\textwidth]{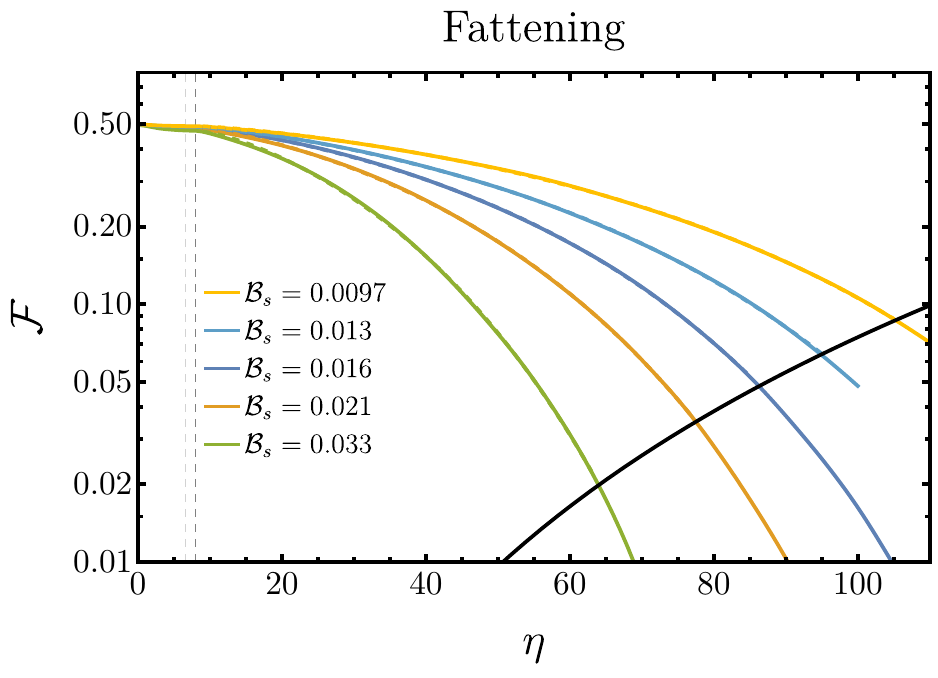} \hspace{0.3cm}
    \includegraphics[width=0.47\textwidth]{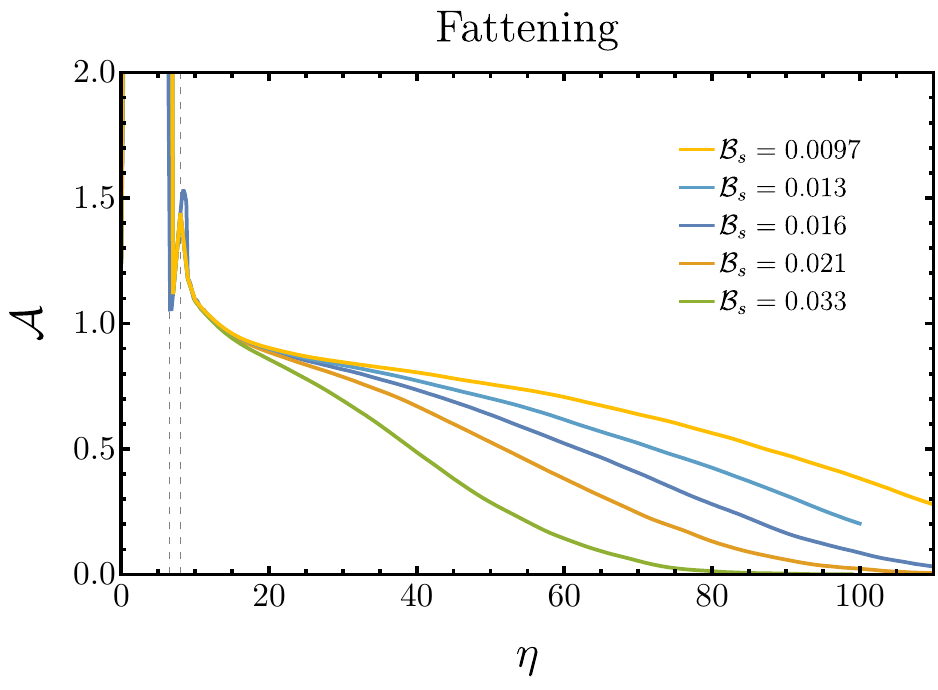}  
    \caption{\textit{Left:} Evolution of the FV fraction for different initial biases and one specific realization of initial conditions (solid lines). \textit{Right:} Evolution of the area parameter for the same initial biases. \textit{Top:} Results from simulations without fattening. \textit{Bottom:} Results from simulations with fattening. In the FV plots, dashed lines show the best fits to the function \eqref{eq:FfvFunc2}, see the main text. They are barely distinguishable because of the quality of the fits. The solid black lines in those plots show the inverse number of Hubble volumes~\eqref{eq:FfvHub}. The vertical dashed lines bound the range during which friction is active. The top panels have been obtained from lattice simulations with $N=4096$ and $L=80$ without fattening. The bottom panels have been obtained from lattice simulations with $N=3840$ and $L=240$.\textcolor{blue}{to fix}} \vspace{0.5cm} \label{fig:PopBiasFV} 
\end{figure*}

As already outlined in previous works~\cite{Ferreira:2024eru, Notari:2025kqq}, the DW network forms around $\eta\approx 6$ for the potential under consideration, when $H_i=m$. Nonetheless, an initial bias can grow even during this initial formation epoch, before domain walls are actually present in our simulations. We find it convenient to employ the measure of  bias in~\eqref{Bdefinition}, i.e. $\mathcal{B} \equiv 1/2 - \mathcal{F}$  with the  ``FV'' fraction $\mathcal{F}$ identified with the fraction of the lattice volume with $\phi<0$. 
Initially, $\mathcal{F} \approx 0.5$, with the bias~\eqref{eq:ICondHom} only inducing tiny deviations from this value. 
At the initial time, we find this quantity to be related to the initial bias parameter~\eqref{eq:biaspar} according to
\be \frac{\mathcal{B} (\eta_i)}{b_i} = 0.3 - 0.4 \ .\ee

The evolution of the false vacuum fraction $\mathcal{F}$ (left panels) and of the area parameter defined in~\eqref{eq:area-parameter} (right panels) is shown in Fig.~\ref{fig:PopBiasFV} for different bias values, for one specific realization of initial conditions (the same realization for all values of $b_i$\footnote{That is, the initial three-dimensional distribution of fluctuations is identical for all values of $b_i$, while the amplitude of the homogeneous mode changes according to the desired bias.}). Rather than the initial bias parameters $b_i$ or $\mathcal{B}(\eta_i)$, we present our results in terms of the population bias $\mathcal{B}_s=\mathcal{B}(\eta_s)$ at a time $\eta_s=9$ shortly after the end of the friction epoch (the duration of this epoch is indicated in Fig.~\ref{fig:PopBiasFV} by the two dashed vertical lines). The reason for this choice is that we observe an initial growth of the bias (well visible in the upper left panel in the figure), which we believe may, in principle, depend on the specific model, as well as on the precise initial spectrum of fluctuations, whereas the behaviour in the scaling regime is expected to be more independent of the model and initial conditions. Therefore, we consider $\mathcal{B} (\eta_s)$ to contain the physically relevant information about the bias size after the initial formation epoch, shortly before the onset of the scaling regime. Numerically, in our simulations we find\footnote{Notice that, as the initial conditions are Gaussian, computing $\mathcal{B}(\eta_i)$ from $b_i$ can be done analytically, while at time $\eta_s$ the field has evolved in a non-trivial way, and the value of $\mathcal{B}(\eta_s)$ is not straightforward to compute analytically.}
\be 
\label{eq:insbias}
\frac{\mathcal{B}_s}{b_i} = 4.1 - 4.3 \ , \ee
with the specific number depending on the choice of $b_i$ in the range that we have considered and on the particular realization.

After $\eta_s$, the proper evolution of the biased network starts, and the FV fraction decreases in a characteristic way. 
Ideally, we would like the simulated network to evolve in the scaling regime for some time before annihilating. Inspection of the area parameter in the top-right panel of Fig.~\ref{fig:PopBiasFV} shows that this is approximately the case only for the smallest biases that we have been able to simulate. In order to avoid possible contaminations, we thus discard the two largest bias values when inferring an analytic behaviour of the FV fraction. Nonetheless, it is worth stressing that simulations with even smaller bias sizes, which are currently beyond our numerical capabilities, may be desirable to obtain more reliable results.

We find the behaviour of the FV fraction after $\eta_s=9$  to be very well captured by the following function:
\begin{equation}
\mathcal{F} (\eta) = \mathcal{F}(\eta_{\rm *}) \exp \left[ - \left( \frac{\eta - \eta_{s}}{\Delta \eta_{\rm ann} }\right)^{p + \alpha\frac{\eta - \eta_{s}}{\Delta \eta_{\rm ann}}} \right] \ , \label{eq:FfvFunc2}    
\end{equation}
where $p$, $\alpha$ and $\Delta \eta_{\rm ann}$ are fitting parameters. 

Eq.~\eqref{eq:FfvFunc2} is very similar to the template proposed in \cite{Ferreira:2024eru} to describe the evolution of the FV fraction in the presence of a small potential bias term, with  two main differences:
\begin{itemize}
    \item An offset in conformal time has been added, $\eta \rightarrow \eta - \eta_{s}$, since we discard data from the initial epoch $\eta\leq\eta_s$. Indeed, as already mentioned above, the initial formation phase may not fully represent the physically relevant case of a Kibble-like transition, and the introduction of unphysical friction halts the evolution of the FV fraction. This is also the reason for the notation $\Delta\eta_{\text{ann}}= \eta_{\text{ann}}-\eta_s$, although in our fits we extract $\Delta\eta_{\text{ann}}$ directly. Of course, the impact of the shift is reduced in simulations with $\eta_{\text{ann}}\gg\eta_s$, which correspond to the physically more realistic case.
    \item A running parameter $\alpha$ in the exponent, which we have empirically found to improve the fit of simulation data significantly.
\end{itemize}

In addition, in obtaining the values of the fitting parameters we also exclude data that correspond to  
\be \mathcal{F}< \mathcal{F}^{(H)} \equiv  \frac{(1+ \eta)^3}{L^3} \, .\label{eq:FfvHub} \ee
As explained in~\cite{Ferreira:2024eru}, FV fractions that are smaller than $\mathcal{F}^{(H)}$ are expected to capture only FV regions of sub-Hubble size, because of the finite box size, since $\mathcal{F}^{(H)}$ is proportional to the inverse of the number of Hubble volumes at any given time. In a realistic cosmological setup, a certain (eventually very tiny) amount of super-Hubble false vacuum regions can instead persist until late times, and affect the decrease of the FV fraction. The curve corresponding to \eqref{eq:FfvHub} is plotted in black in the left panels of Fig.~\ref{fig:PopBiasFV}. Overall, for each fit, we only consider times after 
$\eta_s = 9$ and until the condition $\mathcal{F} = \mathcal{F}^{(H)}$ holds.

Eq.~\eqref{eq:FfvHub} reinforces the advantage of fattening in the specific case of annihilating domain wall networks, as with larger $L$ we can substantially decrease $\mathcal{F}^{(H)}$, widening the range of useful data. This is evident when comparing the upper and lower left panels in Fig. \ref{fig:PopBiasFV}.

\begin{figure*}
    \includegraphics[width=0.47\textwidth]{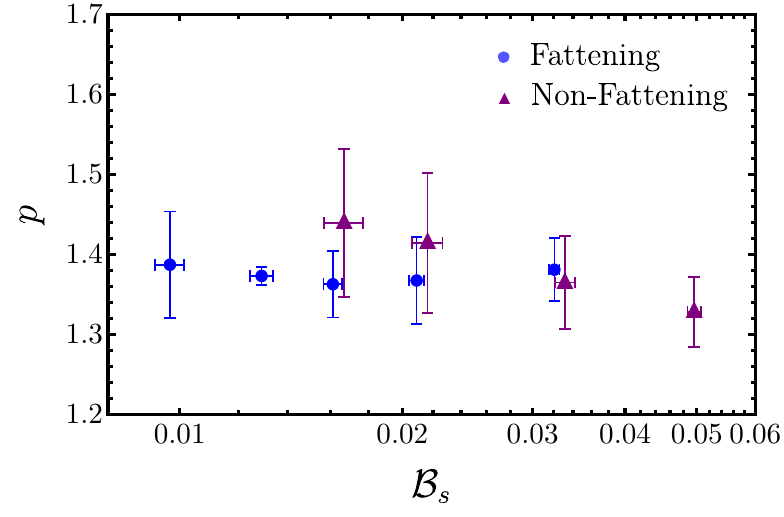} \hspace{0.3cm}
    \includegraphics[width=0.47\textwidth]{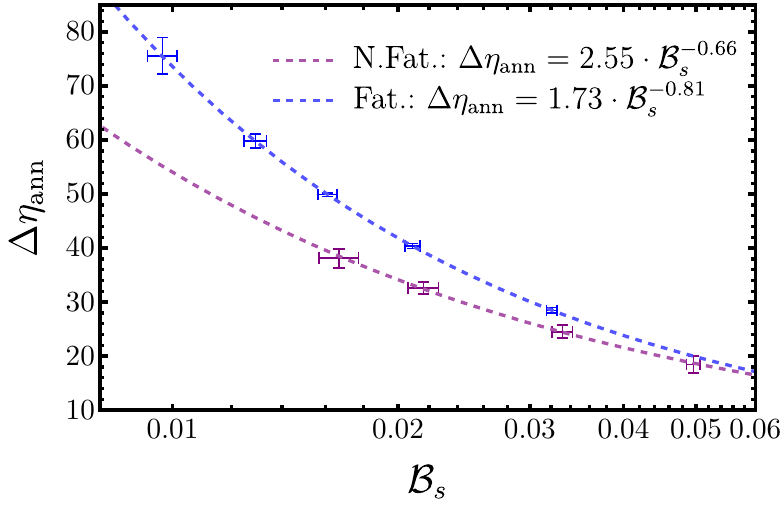}
    \includegraphics[width=0.47\textwidth]{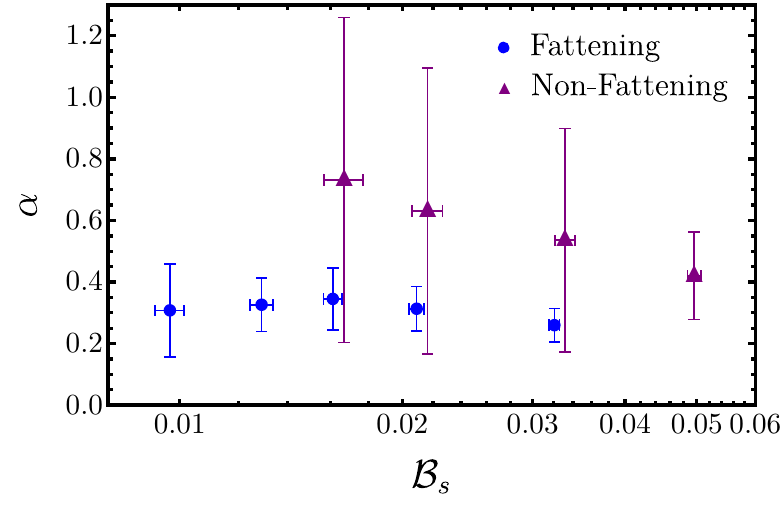}\centering
    \caption{Values of $p$ (top-left), $\Delta\eta_{\rm ann}$ (top-right), and $\alpha$ (bottom) obtained by fitting the FV fraction to the template \eqref{eq:FfvFunc2} for simulations with different biases $\mathcal{B}_s$. Purple/blue triangles/dots represent the values obtained from simulations without/with fattening. The triangles/dots show, for each value of $\mathcal{B}_s$, the average of four/three simulations with different random realizations, while the error bars correspond to one standard deviation. In the upper right plot, we also report the best fit curves for $\Delta \eta_{\rm ann}$, see Eq.~\eqref{eq:EtaAnn}.} \label{fig:PopBiasFits}
\end{figure*}

Results for $p$ (top-left), $\Delta\eta_{\rm ann}$ (top-right), and $\alpha$ (bottom) for different values of $\mathcal{B}_s = \mathcal{B} (\eta_s)$ are shown in Fig.~\ref{fig:PopBiasFits}. The error bars correspond to the one-standard-deviation uncertainty obtained by averaging results from different random realizations (as explained above, we actually fix $b_i$ rather than $\mathcal{B}_s$, thereby causing a small uncertainty in the latter parameter and not only in the fit parameters $p,\alpha$ and $\Delta\eta_{\text{ann}}$). 
Results for $p$ and $\alpha$ from fattening simulations are in reasonable agreement with those from physical simulations, and results from fattening simulations exhibit smaller uncertainties due to the much larger number of Hubble volumes in the corresponding simulations, especially for small bias sizes, where annihilation occurs at late times and is thus poorly captured in physical simulations.
On the other hand, we observe that physical simulations yield smaller values of $\Delta \eta_{\rm ann}$ compared to fattening simulations. We interpret this mismatch as a shortcoming of simulations without fattening, which struggle to fully capture the later stages of collapse due to their more limited validity in time. This is further reinforced by the observation that the deviation between simulations with and without fattening increases as $\Delta \eta_{\rm ann}$ increases. 
Due to these limitations, we present results mostly based on fattening, for the precise characterization of the FV evolution.

Averaging over all simulations with fattening, we obtain:
\be
\text{Population bias:}\quad p = 1.37 \pm 0.01 \ , \hspace{0.3cm} \alpha = 0.30 \pm 0.04 \ . \label{eq:pp}\ee

The dependence of $\Delta\eta_{\text{ann}}$ on $\mathcal{B}_s$ is particularly relevant for phenomenological applications (in particular, to estimate the amplitude of GWs from the collapsing network, see below). A power-law template provides an excellent fit to both sets of simulations. From simulations with fattening, we find
 \be \frac{\Delta \eta_{\rm ann}}{\eta_s} \simeq 0.2~\mathcal{B}_s^{-0.8} \  ,\label{eq:EtaAnn} \ee
with negligible statistical uncertainties. We can compare these results with the 
expectation $\eta_{\text{ann}}\propto \mathcal{B}_s^{-1/p}$, already discussed in Sec.~\ref{sec:analytics}, see~\eqref{etaannB}. We notice that~\eqref{eq:EtaAnn} agrees with this relation to within $\sim 10\%$. We consider this as satisfactory, since our simulations include an initial phase of friction. Additionally, let us emphasize that the errors in Eqs.~\eqref{eq:pp} and \eqref{eq:EtaAnn} arise solely from averaging results with different random initial conditions, while more relevant systematic errors may exist, related e.g.~to our inability to fully simulate the late stages of the collapse for small biases. 
As commented above, results for smaller initial biases are in principle more trustworthy because the network remains in an approximate scaling regime for a longer period of time. However, for our smallest value of  $b_i$, we still have $\mathcal{F} (\eta_{\rm max}^{\rm res}) \sim 0.2$ at the final time.\footnote{Potentially, a second caveat may be related to finite lattice size effects. Indeed, we have simulated various DW networks with no initial population bias (i.e.~fixing $b_i = 0$ exactly), and observed that an ``artificial bias'' develops at late times, due to the reduced number of Hubble volumes. However, the maximal size of such a bias is $|\mathcal{B}_{\text{art}} (\eta)| \lesssim 0.07$ in either vacuum direction at the end of our simulations. This is always much smaller than the values of our physical bias $\mathcal{B}$ in our simulations, therefore we do not expect it to affect our results.}

Our results can be compared with the recent work of~\cite{Kitajima:2023kzu}, which studied population bias using 2D simulations (see also previous works~\cite{Coulson:1995nv, Larsson:1996sp, Correia:2014kqa, Gonzalez:2022mcx}) and reported good fits with the template~\eqref{eq:FfvFunc2}, keeping $p=1$ and $\alpha = 0$ fixed. While these results, together with ours, may suggest the extrapolation $p \approx N_{\rm dim}/2$ where $N_{\rm dim}$ is the number of spatial dimensions, we note that: first, the dynamical range studied in~\cite{Kitajima:2023kzu} is more limited than the one investigated here; second, and most importantly, population bias in 2D may exhibit important physical differences from the realistic 3D case, since the threshold for percolation varies significantly with the number of dimensions, as mentioned in the introduction.

\subsection{Comparison with the potential bias scenario} \label{sec:PotBias}

\begin{figure*}
    \centering
    \includegraphics[width=0.47\textwidth]{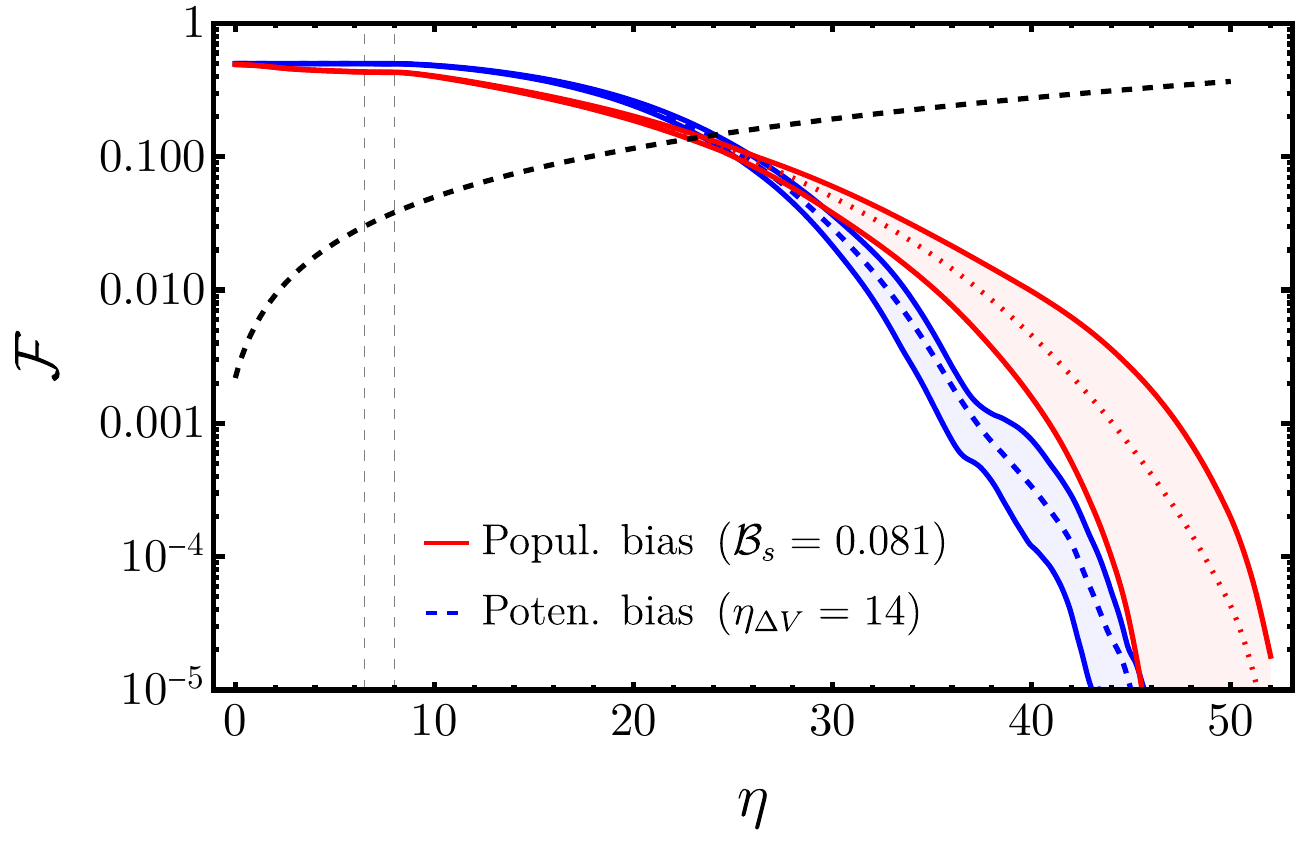} \hspace{0.43cm}
    \includegraphics[width=0.46\textwidth]{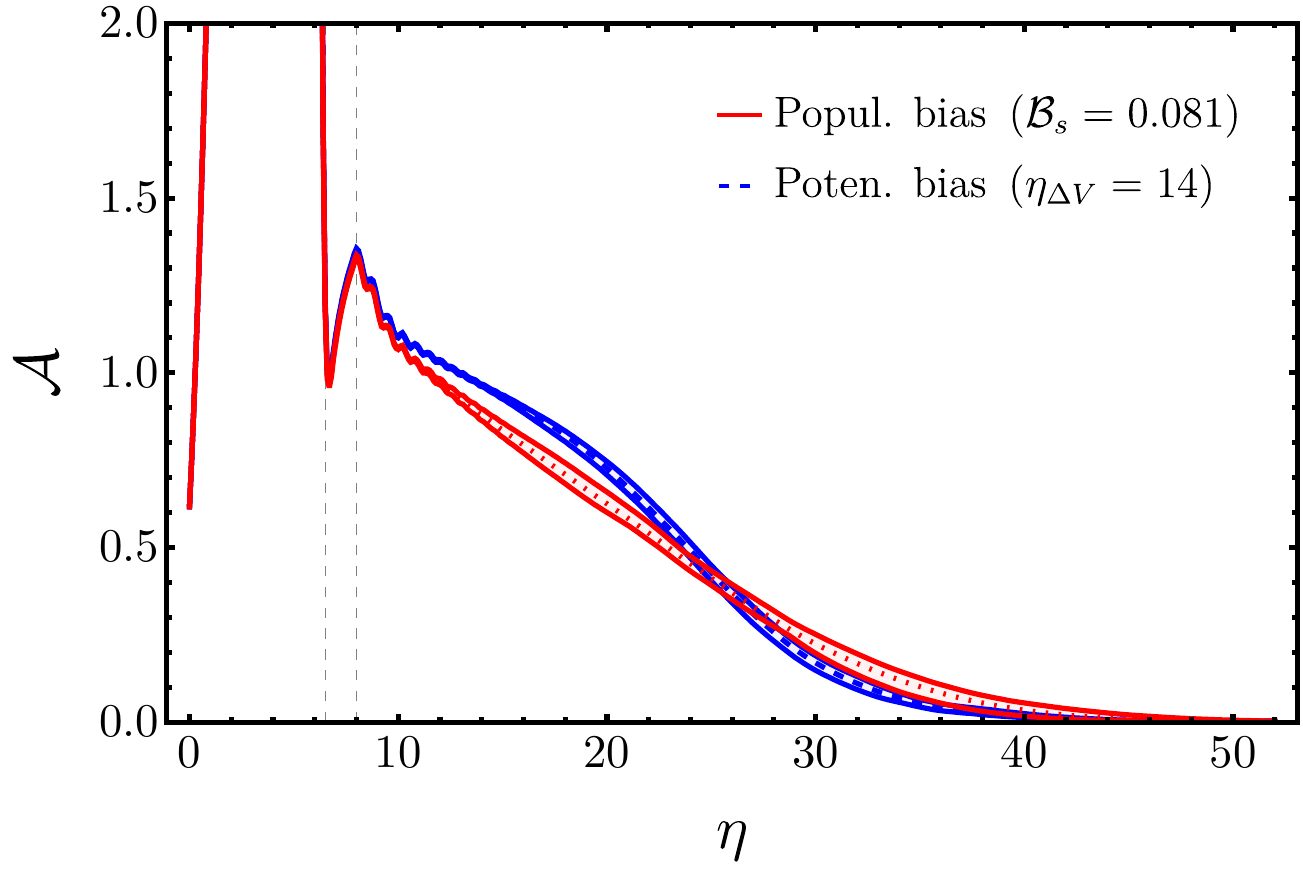}
    \caption{\textit{Left:} Evolution of the FV fraction for the population bias (red) and potential bias (blue) scenarios, obtained from physical lattice simulations with $N=4228$ and $L=110$. The dashed (blue) and dotted (red) lines in each case show the average of five random realizations, while the solid lines show the maximum and minimum values of each quantity. The black dashed line corresponds to the inverse number of Hubble volumes. \textit{Right:} Evolution of the area parameter in the same cases. } \label{fig:FVApPotPob} 
\end{figure*}

Phenomenologically viable domain wall networks can also annihilate due to a small bias term in the potential, which explicitly breaks the underlying discrete symmetry. In both scenarios, the network is expected to experience a possibly long epoch of scaling evolution, before undergoing collapse. We expect differences between the two scenarios to arise mostly in the collapse phase, specifically in the decrease of the false vacuum fraction and the area parameter. For the sake of comparing the two scenarios, let us consider here a potential-biased model in which a $\mathbb{Z}_2$-breaking term $V_\text{bias} = qv\phi^3$ with $q \ll 1$ is added to~\eqref{eq:quartic-pot}, thereby splitting the two minima by $\Delta V=2q(1+9q^2)^{3/2}v^4$. The network then starts collapsing around the time $\eta_{\Delta V}$ at which the condition $\sigma H(\eta_{\Delta V})=\Delta V$ is satisfied. 

This scenario has already been thoroughly studied by some of us in~\cite{Ferreira:2024eru, Notari:2025kqq}. Nonetheless, here we present results based on five new physical simulations with the specific choice $\eta_{\Delta V}=14$ and with lattice parameters $N=4228$ and $L=110$, thereby updating the results of~\cite{Notari:2025kqq}, obtained with lower resolution ($N=3060$ and $L=80$). We have also simulated the population bias case with exactly the same lattice parameters and initial conditions, and for a population bias $\mathcal{B}_s$ specifically chosen to match the annihilation time of the potential-biased simulations. A comparison of the evolution of the FV fraction (left) and area parameter (right) is shown in Fig.~\ref{fig:FVApPotPob}, which shows that annihilation is more abrupt and faster in the potential bias scenario than in the population bias case. 

\begin{figure}[t]
    \begin{center}
    \includegraphics[width=0.48\textwidth]{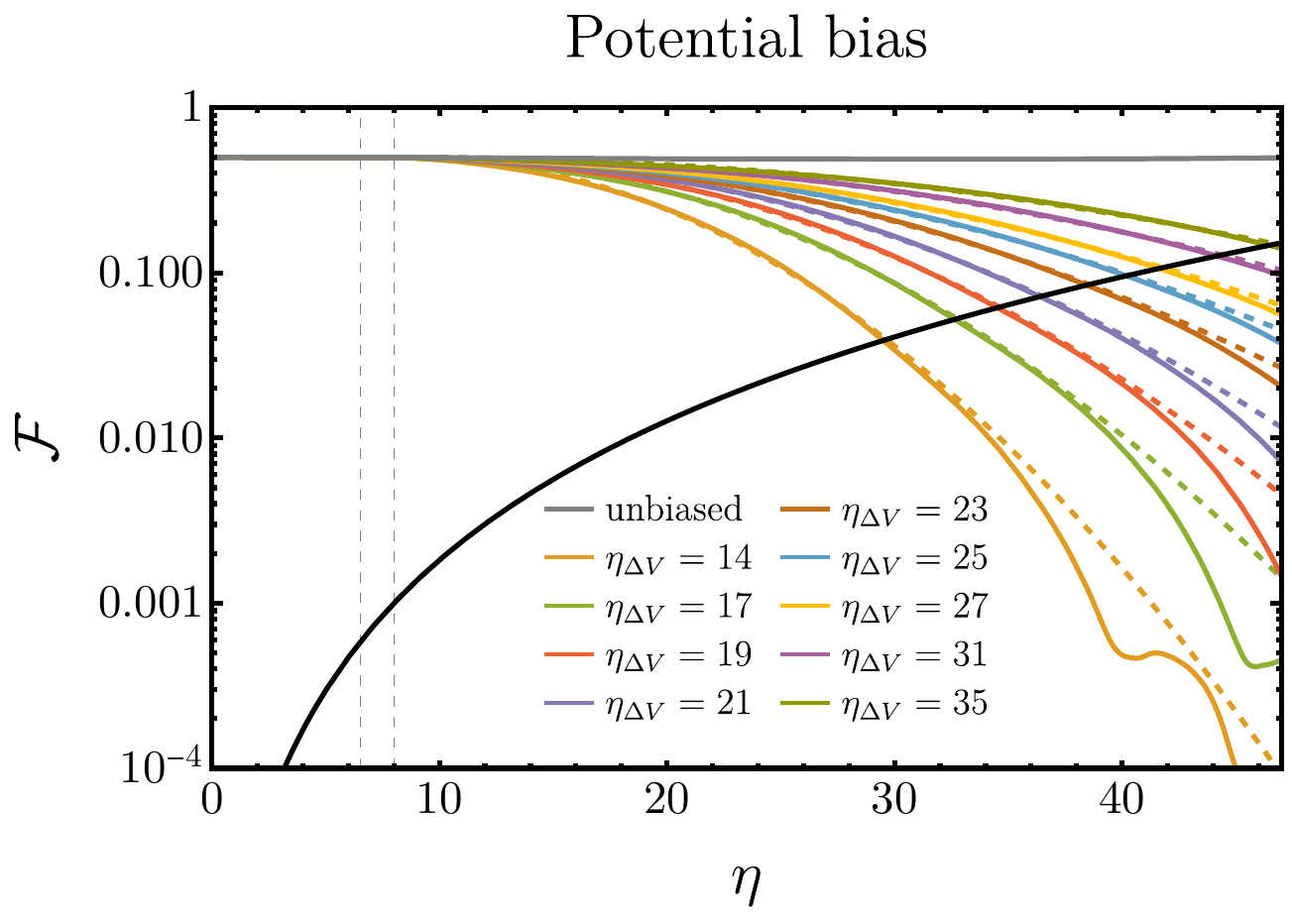} \,\includegraphics[width=0.5\textwidth]{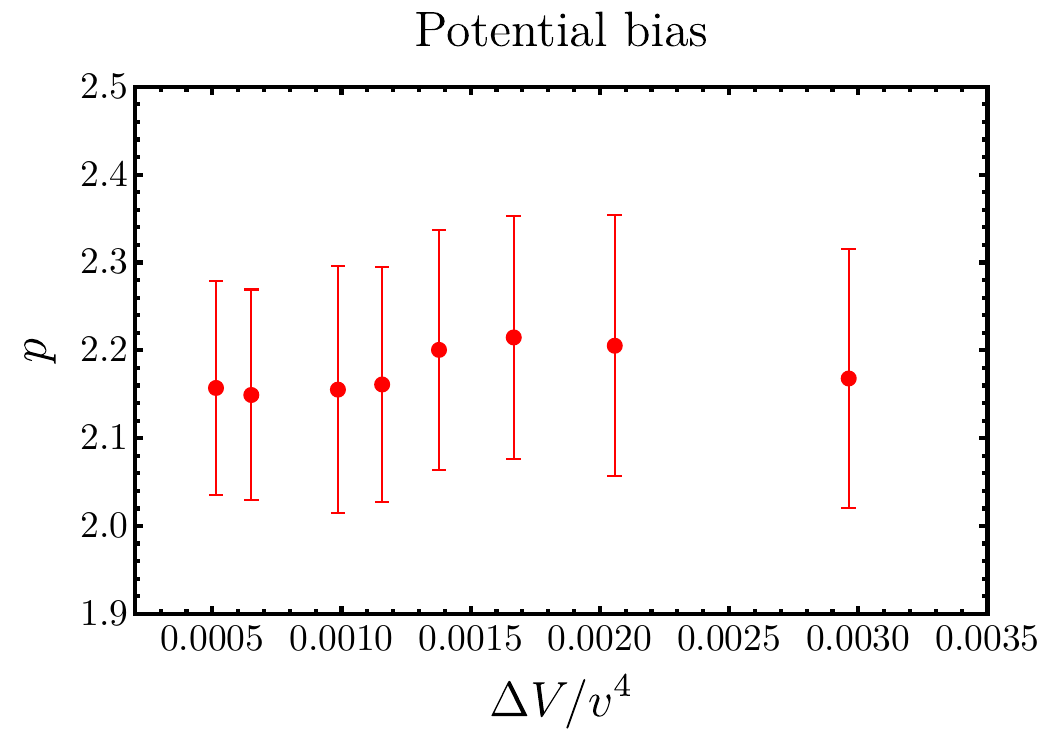}
    \end{center}
    \caption{\em{Left:}~Evolution of the FV fraction for different sizes of the potential bias, obtained with non-fattening lattice simulations with $N=5000$ and $L=90$. The dashed lines show the best fits to the lattice results using the template \eqref{eq:FfvFunc2} with $\alpha = 0$, and fitting only to the range $\eta>\eta_s$ and $\mathcal{F} > \mathcal{F}^{(H)}$, see Eq.~\eqref{eq:FfvHub}. \em{Right:} Values of $p$ extracted from fits of~\eqref{eq:FfvFunc2} to our simulation data.}
    \label{fig:p-potbias}
\end{figure}

To make this more quantitative, we have studied annihilation in the potential biased case for several choices of $q$, corresponding to different values of $\eta_{\Delta V}$, and extracted the fitting parameter $p$ from the template~\eqref{eq:FfvFunc2}. Results are shown in Fig.~\ref{fig:p-potbias}, based on physical simulations with $N=5000$ and $L=90$. Averaging over all three realizations, we find 
\begin{equation}
\label{eq:ppot}
\text{Potential bias:}\quad p=2.16 \pm 0.06,
\end{equation}
which is almost one unit larger than for population bias. In this case, we also find no evidence for a running $\alpha$ of the exponent, which we can thus set to $\alpha=0$ without affecting the inferred value of $p$. We notice that this result is appreciably smaller than what was previously reported in~\cite{Ferreira:2024eru}, i.e.~$p\approx 3$. We discuss possible causes and caveats in Sec.~\ref{sec:comparison}, together with additional comments on previous literature on potential bias. 

Finally, in scenarios where both population and potential biases are present (e.g. when the potential bias itself also causes a displacement of the maximum of the potential), we find that population bias is more effective than potential bias in inducing network annihilation unless $\BB_s^{1.6}\ll \Delta V / (\sigma\,H_s)$ (with $H_s$ the Hubble rate at the onset of scaling).

\section{Gravitational waves} \label{sec:GWs}

Biased domain wall networks act as efficient sources of GWs. In the case of potential bias, it has recently been understood that the amplitude and spectral shape of the GW signal are set by the collapse phase~\cite{Kitajima:2023cek, Ferreira:2024eru, Notari:2025kqq, Cyr:2025nzf, Babichev:2025stm} rather than by the scaling regime as previously thought. In this section, we provide the first results for GWs from the collapse of population-biased networks.

In our scenario, GWs $h_{ij} = h_{ij}  (\vec{x},t)$ obey the equation of motion in the transverse-traceless (TT) gauge
\be h''_{ij} - \nabla^2 h_{ij} + 2 \mathcal{H} h'_{ij} = \frac{2}{M_p^2} (\partial_i \phi \partial_j \phi )^{\rm TT} \ ,  \label{eq:eomGW} 
\ee
where the right-hand side of the equation contains the TT part of the anisotropic stress tensor of the scalar field. The energy density of the produced stochastic gravitational wave background is then
\be \rho_{\rm gw} \equiv \frac{M_p^2}{4 a^2} \langle h'_{ij} h'_{ij} \rangle_V \ , \ee
where $\langle \dots \rangle_V$ represents a volume average over a region large enough to capture the relevant wavelengths. In momentum space, this is
\begin{equation}  
\rho_{\rm gw} = \int \frac{d \rho_{\rm gw}}{d \log k} \, d \log k \ ,  \quad \text{with}\quad \frac{d \rho_{\rm gw}}{d \log k}  \equiv  \frac{M_p^2 k^3}{8 \pi^2 a^2 V} \int \frac{d \Omega_k}{4 \pi} h'_{ij} ({\bf k}, \eta) h'^*_{ij}  ({\bf k}, \eta) \ , 
\label{eq:RhoGWlogk}
\end{equation}
where $d \rho_{\rm gw} / d \log k$ is the energy density per logarithmic momentum interval, expressed as an angular integral in momentum space, with $d\Omega_k$ the corresponding solid angle differential. The energy density in GWs is also usually reported by normalizing to the critical energy density at a given time, $\rho_c \equiv 3 M_p^2 H^2$, with $M_p=1/\sqrt{8\pi G}\simeq 2.4\cdot 10^{18}~\text{GeV}$, i.e. $\Omega_{\rm gw} \equiv \rho_{\rm gw} / \rho_c $, and $d\Omega_{\rm gw}(f)/d\log(f)$ is the corresponding spectrum, defined as in~\eqref{eq:RhoGWlogk}. A simple dimensional analysis estimate based on the quadrupole approximation for scaling networks (i.e. with only one length/time scale of $\mathcal{O}(H^{-1})$) gives~\cite{ZambujalFerreira:2021cte}:
\be 
\label{eq:quadrupole}
\Omega_{\rm gw}^{\rm (scal)} (k_{\rm peak}, \eta ) = \epsilon~\Omega_\text{gw}^\text{quad}=\frac{3}{32 \pi} \epsilon \,  \alpha (\eta)^2 , \ee
where $\alpha(\eta) \equiv \rho_{\rm source}/\rho_c$ is the fraction of energy density stored in the source and $\epsilon$ is an efficiency factor to be estimated with numerical simulations, which quantifies deviations from the simple quadrupole expectation. In the scaling regime, we can take $\rho_\text{source}=\rho_{\rm dw}$ with $\mathcal{A} \approx 1$, see Eq.~\eqref{eq:area-parameter}. We then obtain
\be
\Omega_{\rm gw}^{\rm quad, s} \equiv \frac{3}{32 \pi} \left(\frac{2 \sigma H (\eta)}{\rho_c (\eta) } \right)^2 \ . \label{eq:quadApr1} \ee
For the purposes of comparing the efficiency of GW production in different biasing scenarios, it is interesting to consider the quadrupole prediction based on the total energy density in the scalar field, including contributions from scalar waves (which are known to be active sources of GWs at late times, see~\cite{Notari:2025kqq}). We therefore also consider
\be \widetilde{\Omega}_{\rm gw}^{\rm quad} \equiv \frac{3}{32 \pi} \left( \frac{\rho_\phi (\eta) }{ \rho_c (\eta)}\right)^2 \ , \label{eq:quadApr2} \ee
where $\rho_\phi$ is the actual total energy density stored in the scalar field at the time $\eta$. 

We compute the evolution of GWs using only physical simulations with $N=4228$ and $L=110$, by means of the dedicated module in \CL (see Sec.~8 of~\cite{Baeza-Ballesteros:2025tme} for implementation details). In order to capture the entire collapse of the DW network within the range of validity, we focus on the case $\mathcal{B}_s \equiv \mathcal{B} (\eta_s) =0.081$, which corresponds to setting $b_{\rm i} = 1.9 \cdot 10^{-2}$, and obtain five different realizations of initial conditions.\footnote{Note that, according to the prescription of Eq.~\eqref{eq:tmax}, our choice of lattice leads to a loss of DW resolution at times $\eta_{\rm max}^{\rm res} \approx 33$, while here we will extend our simulations to times $\eta \approx 50$. Indeed, the constraint $\eta < \eta_{\rm max}^{\rm res}$ turns out to be over-conservative for our purposes, as most of the gravitational waves produced at late times are mainly sourced by scalar field waves rather than by domain walls. See App.~\ref{app:Technicalities} for a comparison of simulations with different $\eta_{\rm max}^{\rm res}$.} Additionally, as mentioned above, in this set of simulations we use discretized spatial derivatives accurate up to fourth order in the lattice spacing.

\begin{figure*}
    \centering
    \includegraphics[width=0.65\textwidth]{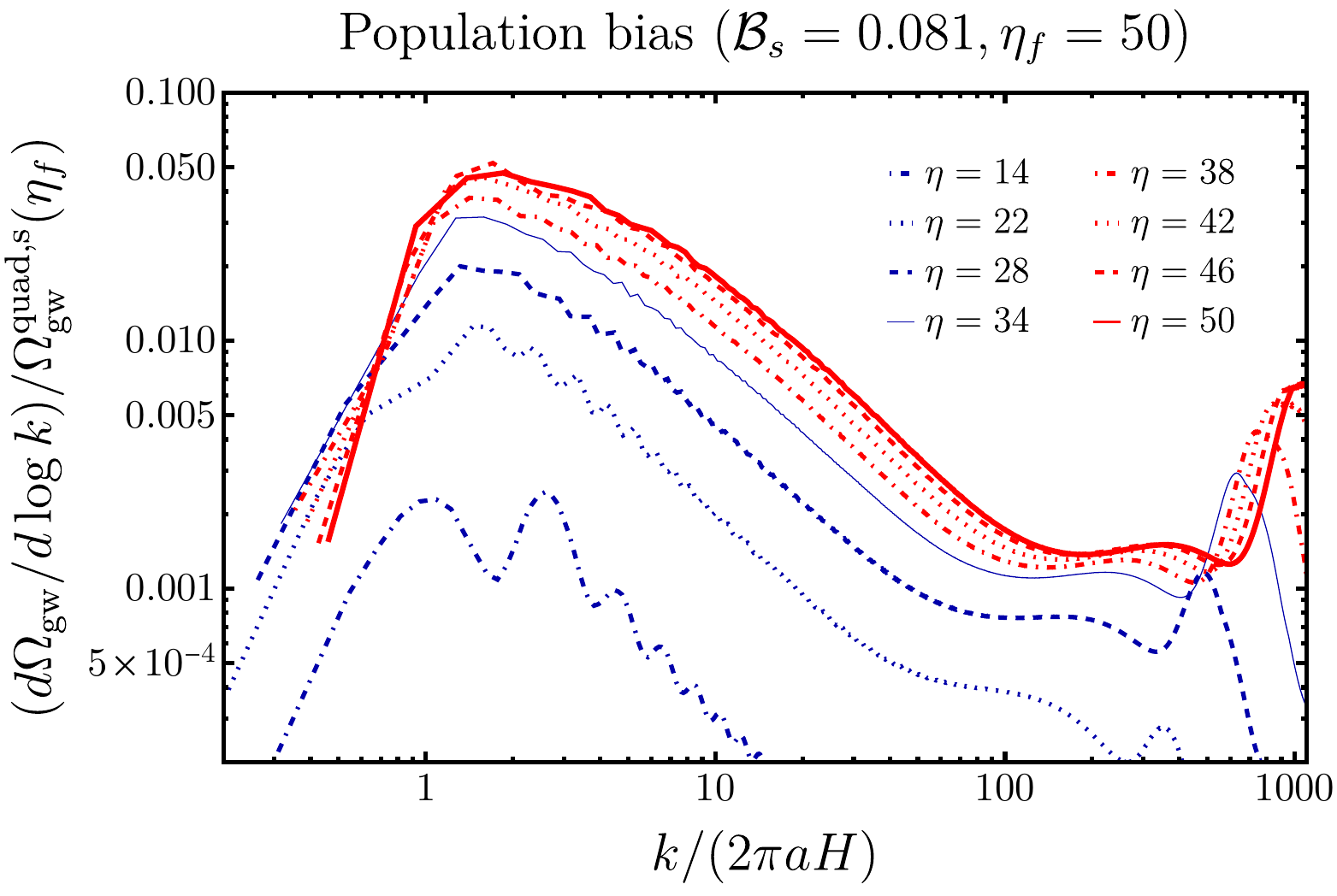} 
    \caption{Evolution of the GW spectrum for the initial bias $\mathcal{B}_s = 0.081$ obtained by averaging the results of five simulations with lattice parameters $N=4228$, $L=110$, and different random initial conditions. The spectrum is represented as a function of $x \equiv k /(2 \pi a H)$ and normalized by the quadrupole approximation \eqref{eq:quadApr1} at the time $\eta_f = 50$.} \label{fig:GWspectrum} 
\end{figure*}

The evolution of the GW spectrum, averaged over realizations, is shown in Fig.~\ref{fig:GWspectrum}, normalized to the quadrupole prediction~\eqref{eq:quadApr1} at the final time $\eta_f = 50$, where we observe approximate saturation of the spectrum. Let us discuss some of its qualitative features, before providing a detailed analysis.  As in the case of scaling networks, the spectrum features a peak at an approximately invariant position during the collapse, corresponding to a physical wavenumber that is not far from the Hubble scale at all times. The peak region is then followed by an extended UV tail, up to $x\equiv k/(2\pi a H) \approx 100$. This is followed by a plateau and a second peak appearing at wavenumbers $x \approx 10^3$. This second region is affected by the lack of UV resolution, as shown in~\cite{Notari:2025kqq}, with the second peak in particular being a numerical artifact, while the plateau region is effectively irrelevant for observational purposes, since in realistic scenarios it is located in the far UV region of the spectrum, corresponding to the inverse width of the walls. These features very closely mirror those already observed in the case of potential-biased networks~\cite{Notari:2025kqq}.

\begin{figure*}
    \centering
    \includegraphics[width=0.47\textwidth]{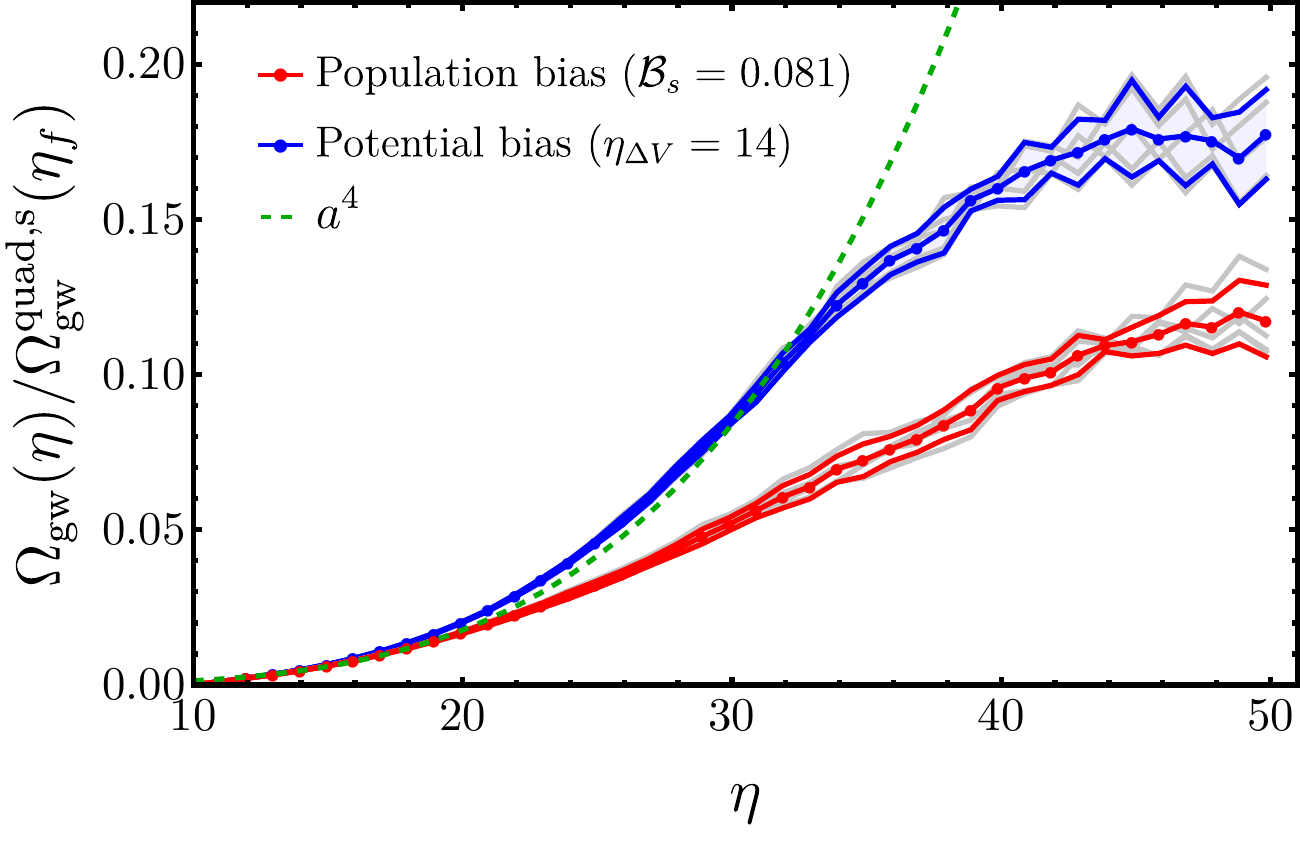}\ \hspace{0.3cm}
    \includegraphics[width=0.47\textwidth]{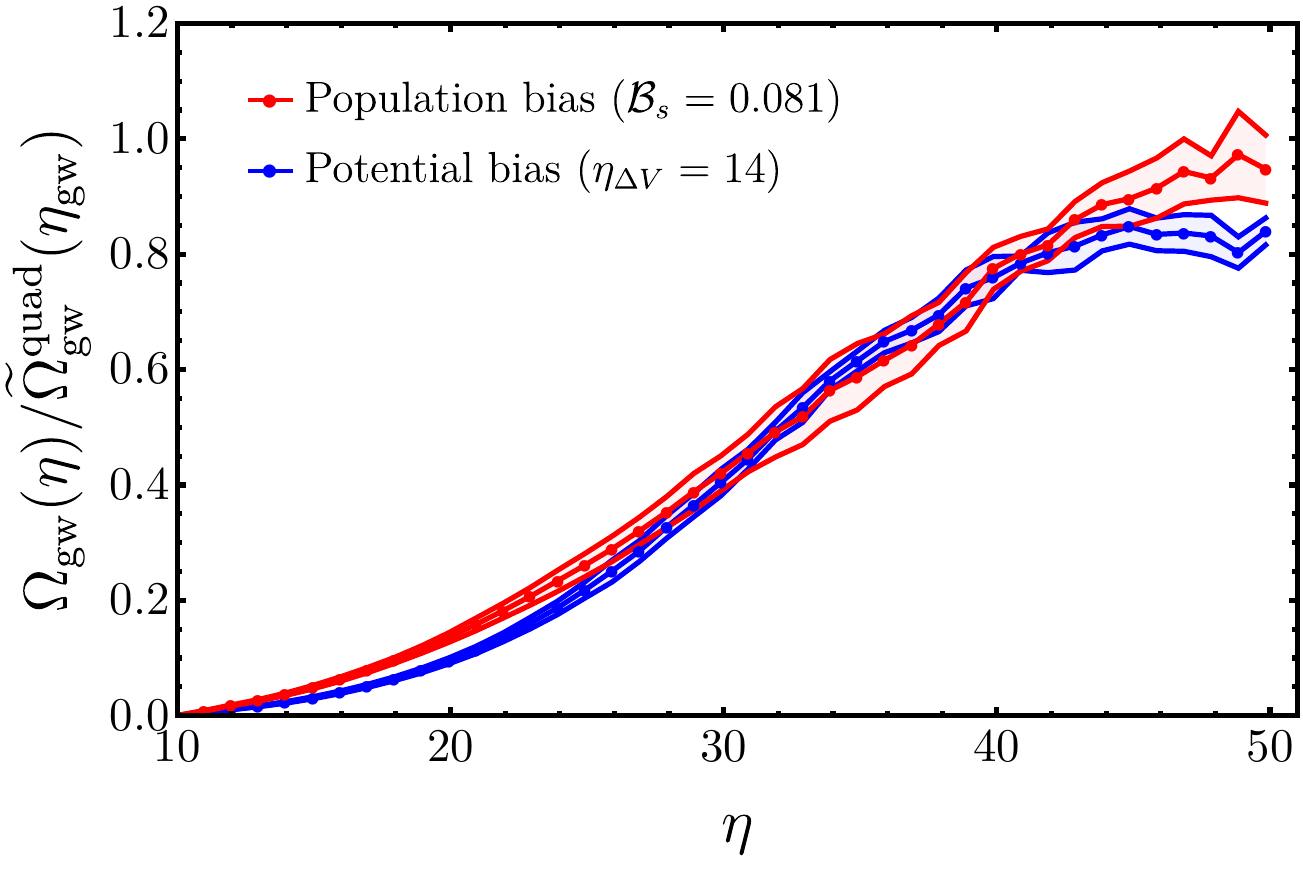}
    \caption{{\it Left}: Evolution of the GW energy density for the population and potential bias scenarios respectively (red and blue, respectively), normalized in both cases by the quadrupole approximation \eqref{eq:quadApr1} at $\eta_f = 50$. For each scenario, the shaded regions show the mean value and the one-standard-deviation uncertainty from averaging over five simulations with different random initial conditions. We also show each individual realization with gray lines. The dashed green line shows the expected scaling behaviour $\Omega_{\rm gw} \propto a^4$. {\it Right}: Same quantities, but normalized by the quadrupole approximation in terms of the total scalar field energy density \eqref{eq:quadApr2}.} \label{fig:OmegaGW} 
\end{figure*}

Let us now consider the total amplitude of the signal, i.e. the integrated energy density fraction in GWs, which we show in~Fig.~\ref{fig:OmegaGW} (left, red curve), normalized by \eqref{eq:quadApr1} at the final time $\eta_f = 50$. The deviation from the scaling expectation $\Omega_{\rm gw} \propto a^4$ (dashed curve) around $\eta\simeq 20$ is evident. When normalizing to the quadrupole estimate from the total energy density in the scalar field (right panel in Fig.~\ref{fig:OmegaGW}), we find the ``efficiency'' of GW emission to be close to unity.

We extract the time $\eta_{\rm gw}$ at which GW production stops, by determining the time at which the amplitude of the main spectral peak stops growing. Averaged over the five simulations, we find:
\begin{center}
Population Bias, $\mathcal{B}_s = 0.081$
\end{center}
\begin{align}
\nonumber    \Delta \eta_{\rm gw} \equiv \eta_{\rm gw}  - \eta_s =  37.3 \pm 2.9 \ , &\quad
\Delta \eta_{\rm ann}=  13.1 \pm 0.3 \ , \\
\Rightarrow\frac{\Delta \eta_{\rm gw}}{\Delta \eta_{\rm ann}} &= 2.85 \pm 0.23 \ ,
    \label{eq:delay}
    \end{align}
where $\eta_{\rm ann}$ is the average annihilation time, defined  by~\eqref{eq:FfvFunc2}. Note that, when comparing with $\Delta\eta_{\text{ann}}$, we subtract $\eta_s = 9$ from $\eta_{\text{gw}}$, i.e. $\Delta \eta_{\rm gw}\equiv \eta_{\text{gw}}-\eta_s$, since our determination of $\Delta\eta_{\text{ann}}$ is restricted to $\eta>\eta_s$. Eq.~\eqref{eq:delay} shows that GW production continues significantly after the start of annihilation, similarly to what happens for the potential bias scenario, see~\cite{Kitajima:2023cek,Ferreira:2024eru, Notari:2025kqq, Cyr:2025nzf, Babichev:2025stm}. 

\begin{figure*}
    \centering
    \includegraphics[width=0.8\textwidth]{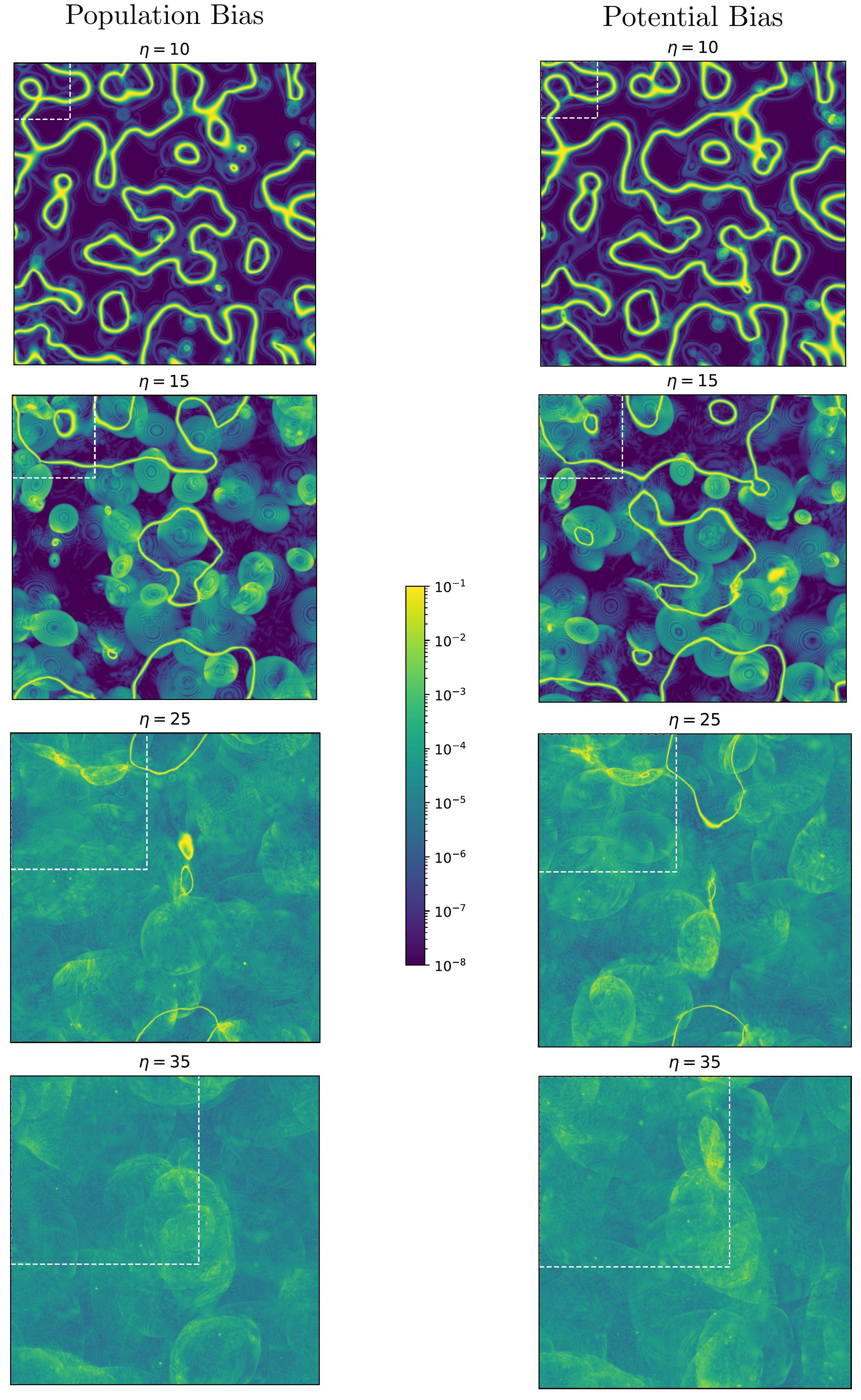}\
    \caption{Snapshots of the gradient energy density (in log scale) from our simulations of population-(left) and potential-(right) biased networks, at several different times. These are obtained as 2D slices of our 3D simulation box, with $N=4096$ and $L=59$ (the results for GWs presented in this section are obtained with higher resolution, see text). The simulations are performed with $\mathcal{B}= 0.081$ (left) and $\eta_{\Delta V}=14$ (right), and are thus characterized by a similar value of $\Delta \eta_{\text{ann}}\approx 13$. While we do not report them here, the corresponding snapshot for an unbiased network at time $\eta=10$ is almost indistinguishable from the one for potential bias. The squares in each panel have sides of length $1/H$, where $H$ is the Hubble rate.} \label{fig:gradients} 
\end{figure*}

In the same Fig.~\ref{fig:OmegaGW} (left and right panels, blue curves) we provide results for the GW energy density fraction from our new simulations of the potential bias scenario, with $\eta_{\Delta V}=14$. The growth saturates approximately at the same time in both cases (by our choice of $\mathcal{B}_s$, see Sec.~\ref{sec:dynamics}). We notice a larger saturation amplitude when we normalize by the domain wall contribution only, which is expected because in the potential bias case there is an additional contribution to the source of GWs with respect to population bias, namely the vacuum energy $\Delta V$. However, when normalizing to the quadrupole estimate from the total energy density available in the scalar field in each scenario separately (right panel), we observe very similar behaviours and saturation amplitudes, with the population bias scenario being only slightly stronger than the potential bias case. Following the same procedure as for the population bias scenario, we determine the average annihilation and saturation times:
\begin{center}
Potential Bias, $\eta_{\Delta V} = 14$
\end{center}
\begin{align}
\nonumber \Delta \eta_{\rm gw}  \equiv \eta_{\rm gw} - \eta_s &= 34.1 \pm 2.7 \ , \quad \Delta \eta_{\rm ann} \equiv \eta_{\rm ann} - \eta_s = 13.4 \pm 0.3 \ , \\
&\Rightarrow\frac{\Delta \eta_{\rm gw}}{\Delta \eta_{\rm ann}} \equiv \frac{ \eta_{\rm gw} - \eta_s}{\eta_{\rm ann} - \eta_s} = 2.55 \pm 0.21 \ . \hspace{0.5cm}
\label{eq:delaypot}
\end{align}
A comparison between~\eqref{eq:delay} and~\eqref{eq:delaypot} illustrates that the extension of the epoch of GW production is essentially independent of the physical mechanism that induces the annihilation of the network.\footnote{For potential bias, the timescale $\eta_{\Delta V}$ provides a simple prediction that is independent of the initial conditions and of the approach to scaling. It may therefore be more convenient to quantify the extension of GW production in terms of  $\eta_{\text{gw}}/\eta_{\Delta V}$, as in~\cite{Notari:2025kqq}, where  $\eta_{\text{gw}}/\eta_{\Delta V}\approx 3.2$ was reported. This estimate is confirmed by our new simulations. However, unlike \cite{Notari:2025kqq}, here we apply a shift to $\eta_{\text{gw}}$ in order to compare with $\Delta \eta_{\text{ann}}$, since $\eta_{\text{ann}}$ may depend on the application of a phase of friction, and is also affected by the initial formation epoch in the case of population bias. Clearly, when $\eta_{\text{gw}}, \eta_{\text{ann}}\gg \eta_s$, we expect $\Delta\eta_{\text{gw}}/\Delta\eta_{\text{ann}}\simeq \eta_{\text{gw}}/\eta_{\text{ann}}$, and the extension of GW production throughout the collapse phase to be independent of the bias size $\mathcal{B}_s$ and $\Delta V$. This might not be entirely the case in our simulations, given our current computational resources.} A graphical comparison between the evolution of the two types of biased networks is shown in Fig.~\ref{fig:gradients}, where we report 2D slices of the gradient energy density from simulations with slightly reduced resolution.

\begin{figure*}
    \centering
    \includegraphics[width=0.65\textwidth]{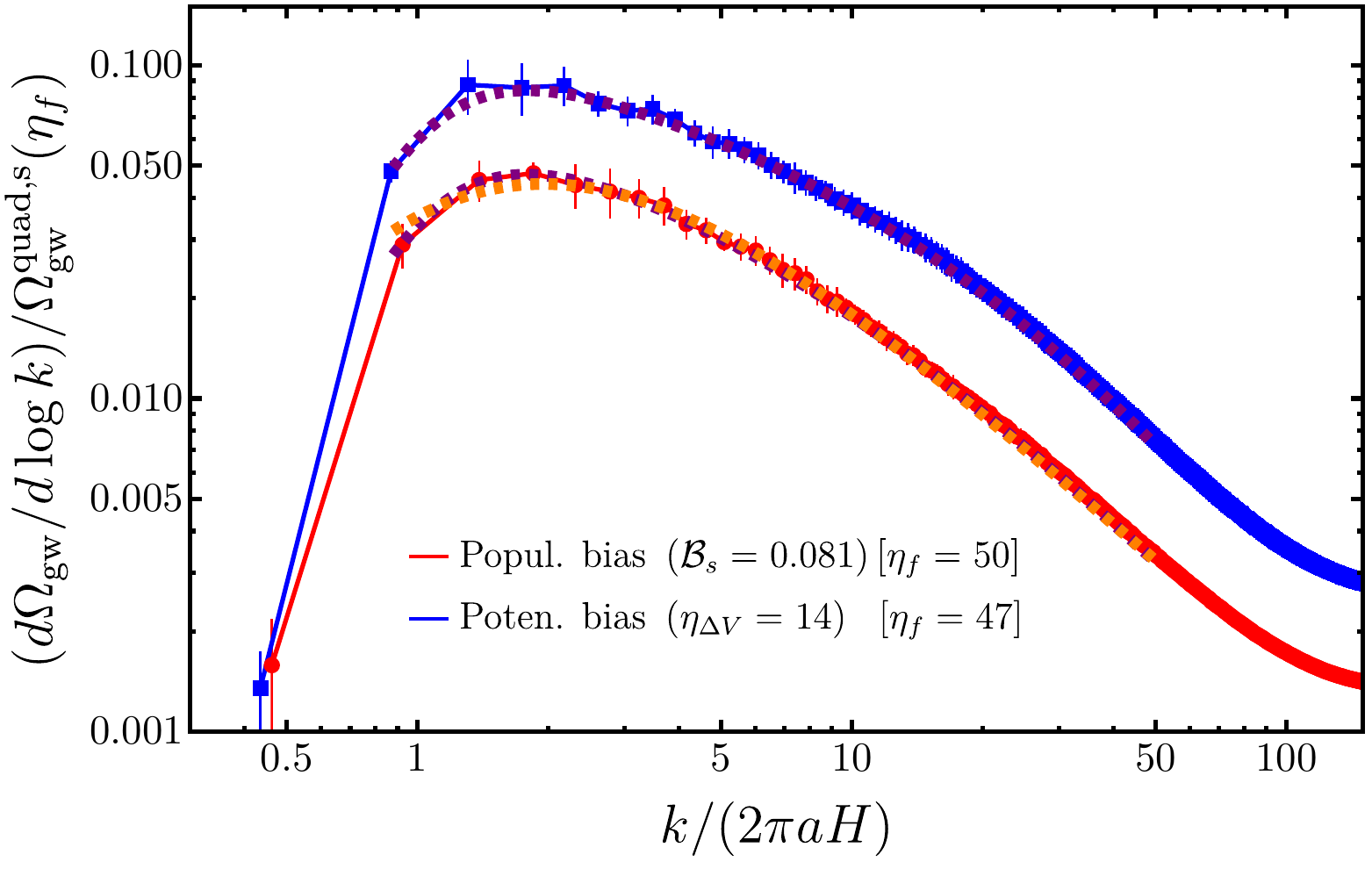}
    \caption{\textit{Red:} GW spectrum for the population bias scenario at time $\eta_f = 50$, normalized by the quadrupole approximation \eqref{eq:quadApr1} at the same time $\eta_f$, as a function of the wavenumber $x \equiv  k/(2\pi a H)$. \textit{Blue:} GW spectrum for the potential bias scenario at time $\eta_f = 47$, also normalized by the same quantity at $\eta_f=47$. In each case, the dots show the average of the different realizations in each spectral bin, with the error bars indicating one standard deviation. The purple dashed lines show the best fits of each data set to the doubly broken power law template \eqref{eq:templateDoubly}. For the population bias scenario, we also show the fit to the singly broken template \eqref{eq:templateSingly} in orange. } \label{fig:GWFit} 
\end{figure*}

Let us now provide a detailed analysis of the spectral shape of the GW signal.
In Fig.~\ref{fig:GWFit} we show the averaged GW spectrum at the final time of the simulations $\eta_f$, with the error bars on each point indicating the one-standard-deviation uncertainty of each spectral bin. Following~\cite{Notari:2025kqq}, we define:
\be \Omega_{\rm gw} (k, \eta_f) = \Omega_{\rm gw} (k_p,\eta_f) \times \mathcal{S} (x) \ , \hspace{0.5cm} x \equiv \frac{k}{2 \pi a H} \ , \label{eq:OGWetafS} \ee
where $k_p$ is the comoving momentum of the main spectral peak and $\mathcal{S}(x) $ is a given shape function such that $\mathcal{S} (x_p) = 1$ at the peak position $x_p \equiv k_p / (2 \pi a H)$. Here, we consider two possible templates for the shape function:
\begin{itemize}
\item {A single broken power law (following expectations on causality-limited sources, see e.g. ~\cite{Caprini:2019egz}):
\begin{equation}
    \label{eq:templateSingly}
    \mathcal {S} (x)= \frac{ (\alpha+\beta)^{\delta} }{\left( \beta\left(\frac{x}{x_p}\right)^{-\frac{\alpha}{\delta}}  +
  \alpha\left(\frac{x}{x_p}\right)^{\frac{\beta}{\delta}}\right)^{\delta} } \ , 
\end{equation}
where $x_{p}$ is the normalized wavenumber of the main spectral peak, $\alpha$ and $\beta$ represent the spectral slopes on the left (IR) and right (UV) tails of the peak, respectively, and $\delta$ is the width around the maximum.}
\item{A double broken power law, following previous findings of~\cite{Notari:2025kqq} for potential bias:
\begin{equation}
    \label{eq:templateDoubly}
    \mathcal {S} (x)=
    \frac{ \alpha+\beta + \left(\frac{x_p}{x_b}\right)^{\beta + \gamma}
    }{ \beta\left(\frac{x}{x_p}\right)^{-\alpha} + \alpha\left(\frac{x}{x_p}\right)^{\beta} + \left(\frac{x_b}{x_p}\right)^{-\beta} \left(\frac{x}{x_b}\right)^{\gamma} }
\end{equation}
which now includes an additional breaking point in the UV tail at the wavenumber $x_b\gtrsim x_p$. In this expression, $\beta$ and $\gamma$ approximately represent the powers of the UV tail in the wavenumber intervals $x_p \lesssim x \lesssim x_b$ and $x \gtrsim x_b$ respectively. We do not include a width parameter, since it is not favored by the data when using this template.}
\end{itemize}

For both templates, we fit only to data with $x \leq 50$ and discard the most IR bin due to the limited resolution in this region of the spectrum. We also fix $\alpha=3$ because of causality \cite{Caprini:2009fx}. The inferred values of the fitting parameters are reported in Table~\ref{tab:bias_fit_parameters}, while in Fig.~\ref{fig:GWFit} the curves corresponding to the central parameter values are reported (orange and purple dashed curves: single and double broken power laws, respectively).

Let us highlight the most relevant findings for population bias: first, we find the spectrum to be well described by a single broken power law, with no evidence for a second breaking point in our double broken power law fit. Second, we find the peak to be displaced with respect to the standard scaling expectation $x_p\simeq 1$ by approximately a factor of $2$. This finding mirrors the situation for potential bias~\cite{Notari:2025kqq}, and can be interpreted as arising from the larger abundance of sub-Hubble structures upon collapse. Third, we find a roughly linearly decreasing UV tail, only marginally shallower than in the scaling case (where $\beta\simeq 1.2$~\cite{Notari:2025kqq}). Finally, we find $\delta\approx 3$ for the width parameter, almost twice as large as in scaling~\cite{Notari:2025kqq}. Nonetheless, this result is affected by the lack of resolution in the IR, and should thus be taken with some caution. 

Let us compare with our new results for the potential bias case, which overall confirm the results of~\cite{Notari:2025kqq}. In particular, in this case we do find evidence for a breaking frequency at $x_b\simeq 2.5 x_p$, and thus we do not report results for the single broken fit (although the evidence is less significant than in~\cite{Notari:2025kqq}, and the breaking point is closer to the main peak).  We also confirm an initially shallow decrease of the spectrum $\sim k^{-0.4}$, followed by a steeper slope $\sim k^{-1.6}$, in agreement with~\cite{Notari:2025kqq}. The main peak position is similarly displaced from the scaling expectation.

\begin{table}
    \centering
    \begin{tabular}{|c|c|c|c|c|}
        \hline
        Parameter & \makecell{{\bf Popul. bias} \\ $\eta_f=50$  \\ \small{Double bpl}} & \makecell{{\bf Popul. bias}  \\ $\eta_f=50$ \\ \small{Single bpl}} & \makecell{{\bf Poten. bias}   \\ $\eta_f = 47$  \\ \small{Double bpl}  }  &  \makecell{{\bf Scaling}  \\ $\eta_f = 30$  \\ \small{Single bpl}  }  \\
        \hline
        $\epsilon_f\equiv\epsilon(\eta_f) $    & $0.045 \pm 0.006$ & $ 0.045 \pm 0.002 $ & $0.084 \pm 0.005$ & $0.249 \pm 0.017$  \\
       \hline $x_p$    & $2.33 \pm 0.68$   & $ 1.97 \pm 0.13 $ & $1.90  \pm 0.18 $ & $1.24 \pm 0.06$\\
        $x_b$    & $2.35 \pm 0.73$ & & $ 4.69 \pm 1.18 $ &  \\
        $\beta$  & $0.23 \pm 0.25$ & $ 1.04 \pm 0.09 $ & $ 0.37 \pm 0.15 $ & $1.19 \pm 0.02$\\
        $\gamma$ & $1.29 \pm 0.12$ & & $1.59 \pm 0.24$ & \\
        $\delta$ & & $2.83 \pm 0.79$ & & $1.69 \pm 0.28$ \\      
        \hline
    \end{tabular}
    \caption{Posterior values obtained for the population and potential bias scenarios, obtained by fitting the spectrum at time $\eta_f$ to the templates in~\eqref{eq:OGWetafS}. We also report results for the scaling regime from \cite{Notari:2025kqq}. }
    \label{tab:bias_fit_parameters}
\end{table}

We also estimate the efficiency of GW production: $\epsilon_f\equiv \epsilon(\eta_f)=\Omega_{\rm gw} (k_p,\eta_f) / \Omega_{\rm gw}^{\rm quad, s} (\eta_f) $, with respect to the quadrupole estimate from domain walls in scaling~\eqref{eq:quadApr1}.  The efficiency parameter $\epsilon_\text{gw}\equiv \epsilon(\eta_{\text{gw}})$ which is relevant to estimate the amplitude of the signal today, is obtained by noticing that the numerator $\Omega_{\rm gw} (k_p,\eta)$ is approximately constant between $\eta_{\text{gw}}$ and $\eta_f$, whereas the denominator $\Omega_{\rm gw}^{\rm quad, s} (\eta_f)$ grows as $\eta^{4}$. Therefore, we estimate
\begin{align}
\textbf{Population Bias}&:\quad \epsilon(\eta_{\text{gw}})=\left(\frac{a(\eta_f)}{a(\eta_\text{gw})}\right)^4\epsilon_f = 0.06\pm 0.02 \\
\textbf{Potential Bias}&:\quad \epsilon(\eta_{\text{gw}})=\left(\frac{a(\eta_f)}{a(\eta_\text{gw})}\right)^4\epsilon_f = 0.12\pm 0.03
\end{align}
As expected, we find a difference of roughly a factor $2$ between the two annihilation mechanisms, due to the additional energy density contribution in the potential bias case.

\begin{figure*}
    \centering
        \includegraphics[width=0.65\textwidth]{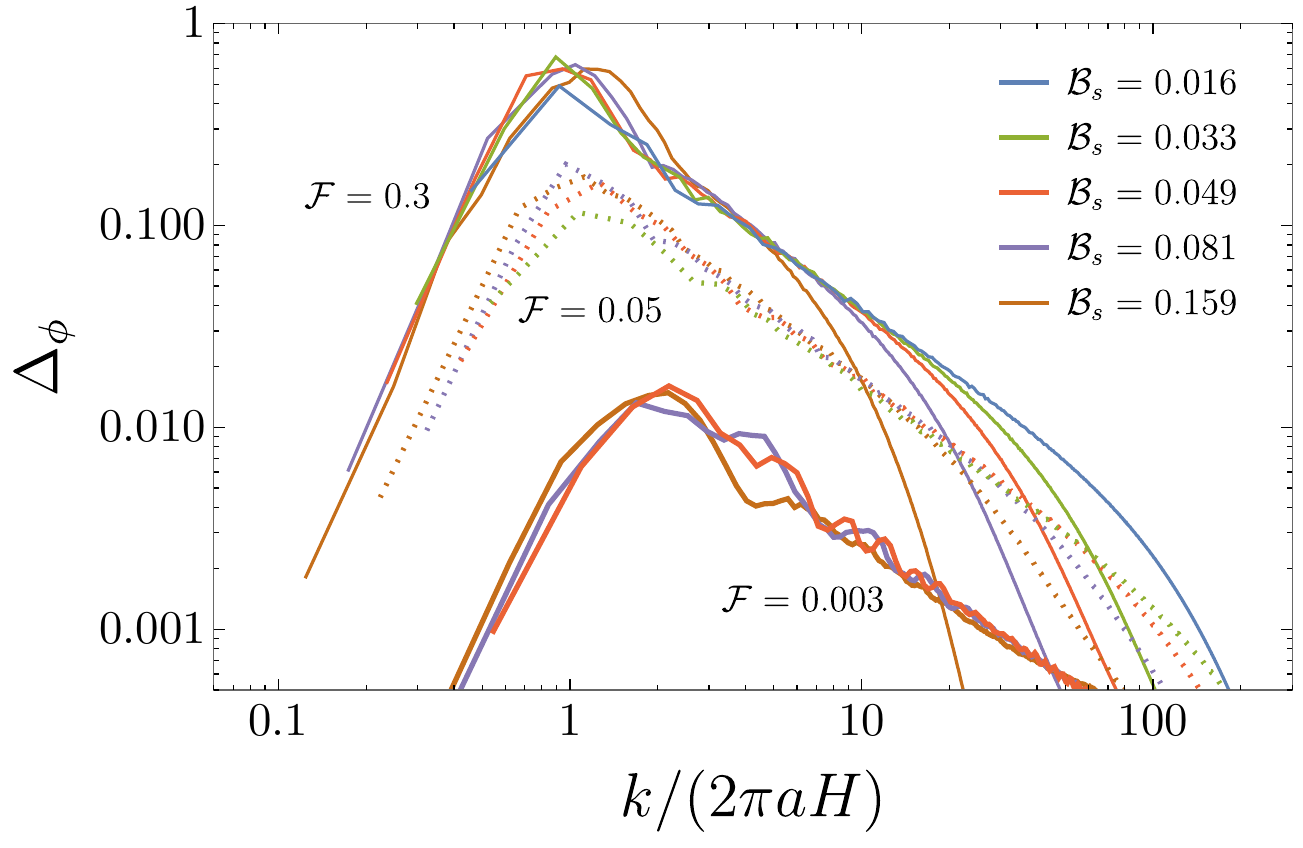} \hspace{0.2cm} 

    \caption{Evolution of the scalar field power spectrum $\Delta_{\phi} = k^3 |\phi_k|^2 / (2 \pi^2)$ for different initial population biases. In order to illustrate the independence of the spectral shape on the bias size, we depict the spectra when the false vacuum fraction is, in each case, $\mathcal{F}= 0.3$, $0.05$, and $0.003$ respectively (note that for the larger biases, we do not depict the last spectra due to the lack of domain wall resolution).}   \label{fig:BiasComp} 
\end{figure*}

Our GW results have been obtained for one specific value of the bias $\mathcal{B}_s$ because of computational limitations (see however App.~\ref{app:Technicalities}, where we present less reliable results for a smaller value of $\mathcal{B}_s$). In order to understand their validity for a generic bias size, we study the evolution of the scalar field power spectrum for different biases, which is computationally much less demanding. In Fig.~\ref{fig:BiasComp}  we compare the spectrum for different choices of initial bias at three different times, given by the values of the FV fraction $\mathcal{F} = 0.3$, $0.05$, and $0.003$ for each independent simulation. We clearly observe that, for identical false vacuum fractions, the spectral shape around the main peak is very similar for different initial bias choices. Since GWs are sourced by the scalar field, this analysis strongly suggests that our results for the GW spectrum extend to general values of the population bias at the onset of domain wall evolution.

\section{Comments on previous work on potential bias}
\label{sec:comparison}

Let us comment on the partial disagreement between the results on potential bias  obtained in~\cite{Notari:2025kqq} 
with other results presented in the literature~\cite{Cyr:2025nzf, Babichev:2025stm}. 
In this paper, we have presented new simulations of the potential bias case in Sec.~\ref{sec:PotBias} and Sec.~\ref{sec:GWs}, that confirm the results of~\cite{Notari:2025kqq}  while being characterized by even higher resolution ($N=4228, L=110$ vs $N=3060, L=80$ of~\cite{Notari:2025kqq}, which was already  higher than that of~\cite{Cyr:2025nzf, Babichev:2025stm}).
Further, both in~\cite{Ferreira:2024eru} and in the present paper, we have simulated smaller bias sizes than in~\cite{Cyr:2025nzf, Babichev:2025stm}, thereby more reliably achieving the scaling regime before annihilation.
In addition, in this paper, as well as in~\cite{Notari:2025kqq}, we use a fourth-order discretization algorithm to better resolve UV features. 

Let us then recap the main novel features of the GW spectrum from networks with a potential bias, as compared to scaling networks, and provide our explanation for the observed discrepancies with~\cite{Cyr:2025nzf, Babichev:2025stm}:

\begin{itemize}
\item\textbf{Amplitude:} An extended GW production during the annihilation phase leads to an enhancement of the amplitude of the GW signal with respect to the instaneous annihilation estimate, i.e.
\begin{equation}
\frac{\Omega_\text{gw}}{\Omega^{\text{inst}}_\text{gw}}\propto \left(\frac{T_\text{gw}}{T_{\Delta V}}\right)^{-4}=\left(\frac{\eta_\text{gw}}{\eta_{\Delta V}}\right)^{4}\approx 3^4.
\end{equation}
This result confirms the trend observed in~\cite{Kitajima:2023cek, Ferreira:2024eru}, while being somewhat larger, and also agrees  with~\cite{Cyr:2025nzf} (and with \cite{Babichev:2025stm} to a lesser extent).

\item\textbf{Peak frequency:} A shift of the peak frequency of the signal at emission
\begin{equation}
f_p\simeq H~(\text{scaling)} \rightarrow f_p\approx 2H~(\text{biased}) \, .
\end{equation}
This result also agrees with~\cite{Kitajima:2023cek} and is valid also for population bias. However, it disagrees with~\cite{Cyr:2025nzf, Babichev:2025stm}, where the shift is not observed. This discrepancy may be explained by the better IR and peak resolution of our simulations, that allows to determine the position of the peak more precisely. Nonetheless, an even higher resolution would help in clarifying this aspect.

\item\textbf{Spectral shape,} whose main new features are: the spectrum is distorted and an initial shallower decrease of the spectrum in the region $f\gtrsim f_p$ is observed. In particular, we confirm our previous results \cite{Notari:2025kqq}, i.e. 
\begin{equation}
\Omega_\text{gw}(f) ~(\text{scaling)} \propto f^{-(1.2-1.3)}\quad \rightarrow \quad \Omega_\text{gw}(f)\propto f^{-(0.4-0.5)}~(\text{potential biased}) 
\end{equation}
A flattening of the UV slope in the biased case with respect to the scaling network is observed in~\cite{Cyr:2025nzf, Babichev:2025stm} as well, although these works disagree quantitatively with ours, since they both find $\Omega_\text{gw}(f)\propto f^{-1}$ in the biased case. A secondary interesting finding of our work is the occurrence of a breaking point at  $f_b\simeq 5H$ (at emission), as already noticed in~\cite{Notari:2025kqq} (but still well before the numerically-affected far UV region), after which the spectrum decreases more steeply, roughly as $\Omega_\text{gw}(f\gtrsim f_b)\propto f^{-1.6}$. This feature is not observed in~\cite{Cyr:2025nzf, Babichev:2025stm}.

\begin{figure*}[t]
    \begin{center}
        \includegraphics[width=0.46\textwidth]{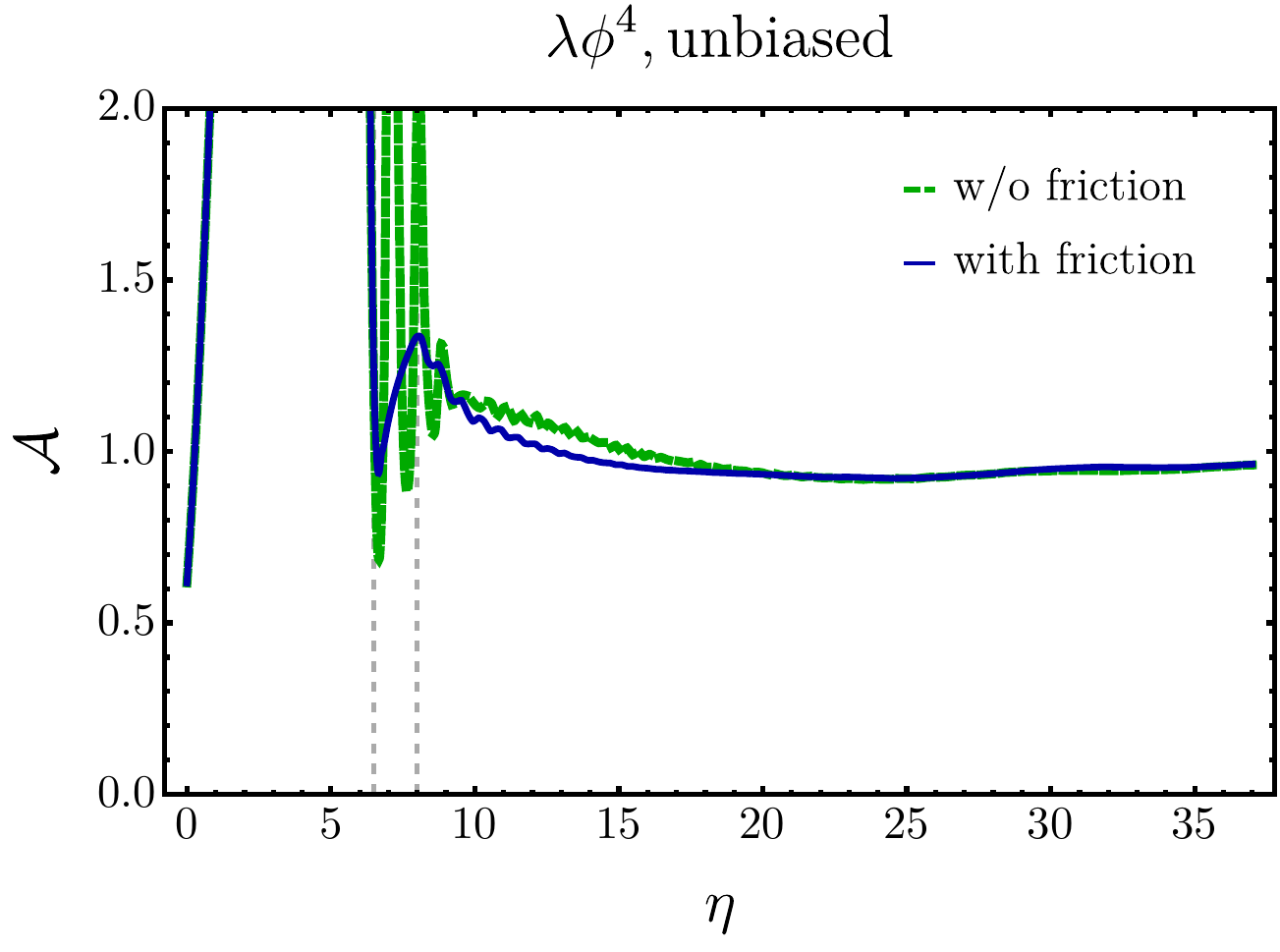} \,\,
        \includegraphics[width=0.46\textwidth]{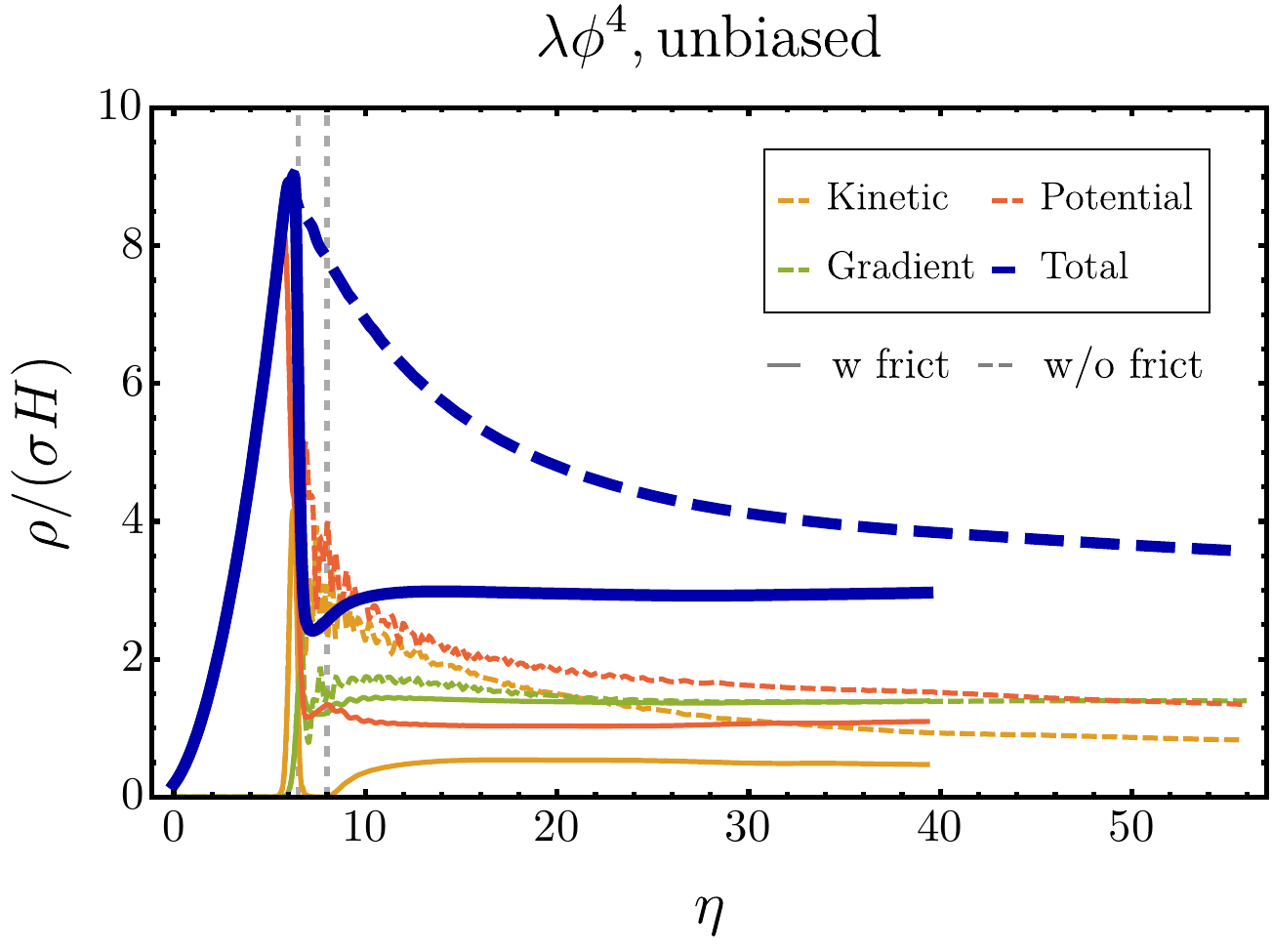}
    \end{center}
    \caption{\textit{Left:} Evolution of the area parameter for two simulations of the unbiased DW network (i.e.~without potential and population bias) attaining the scaling regime, one including the friction term (solid curve; obtained in a lattice of $N=3060$ and $L=70$) and the other without it (dashed curve; obtained in a lattice of $N=4960$ and $L=70$). \textit{Right:} Evolution of the components of the energy density for the same two simulations. The vertical dashed lines in both panels show the range when friction is active. Figure taken from~\cite{Notari:2025kqq}.}
    \label{fig:frict-comp}
\end{figure*}

\end{itemize}

The discrepancies in the spectral shape of the signal are explained, at least partially, by an important difference between our simulations (as well as those of the previous~\cite{Notari:2025kqq}), and those of~\cite{Cyr:2025nzf, Babichev:2025stm}: as already mentioned in Sec.~\ref{sec:model}, we introduce an additional friction term in the equations of motion (active only in the short time interval $6.5\leq \eta\leq 8$). The aim of this strategy is twofold: first, friction accelerates the approach of the network to the scaling regime, thereby leading to a more realistic initial condition for the annihilating network. This is clearly illustrated for an unbiased network in Fig.~\ref{fig:frict-comp}, taken from the Appendix of~\cite{Notari:2025kqq}, and reported here for convenience. As seen in the left panel, the area parameter reaches its asymptotic value significantly earlier when friction is included. Second, friction damps all the contributions to the energy density of the scalar field that are not due to DWs, as illustrated by the right panel of Fig.~\ref{fig:frict-comp}.  When friction is not used (as in the simulations of~\cite{Cyr:2025nzf, Babichev:2025stm}), the components of the energy density of the network do not achieve their scaling values within the simulation. This is to be attributed to the abundant presence of scalar waves produced in the early stages of domain wall formation and during the approach to scaling, which contribute in particular to the potential and kinetic energies of the system, as already noticed in~\cite{Ferreira:2024eru}. In a realistic network that has undergone a long epoch of scaling, those contributions are subdominant just before annihilation, since they are diluted roughly as non-relativistic matter.

\begin{figure*}[t]
    \begin{center}
        \includegraphics[width=0.46\textwidth]{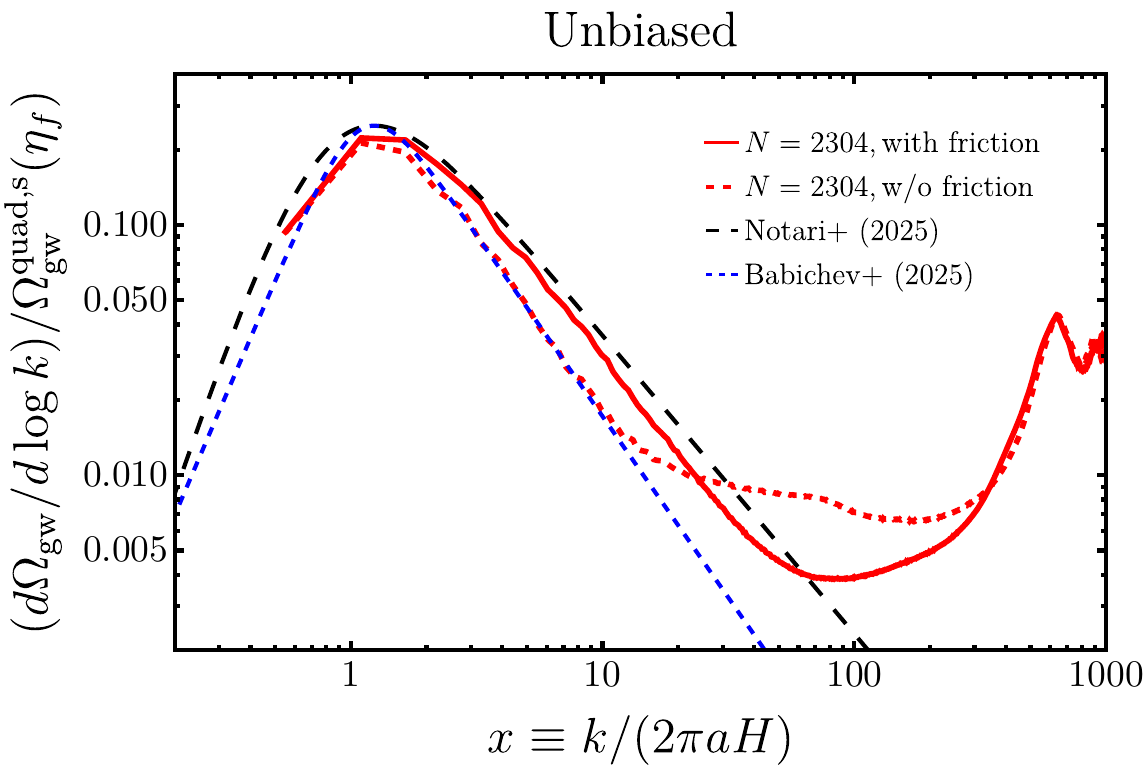} \,\,
        \includegraphics[width=0.46\textwidth]{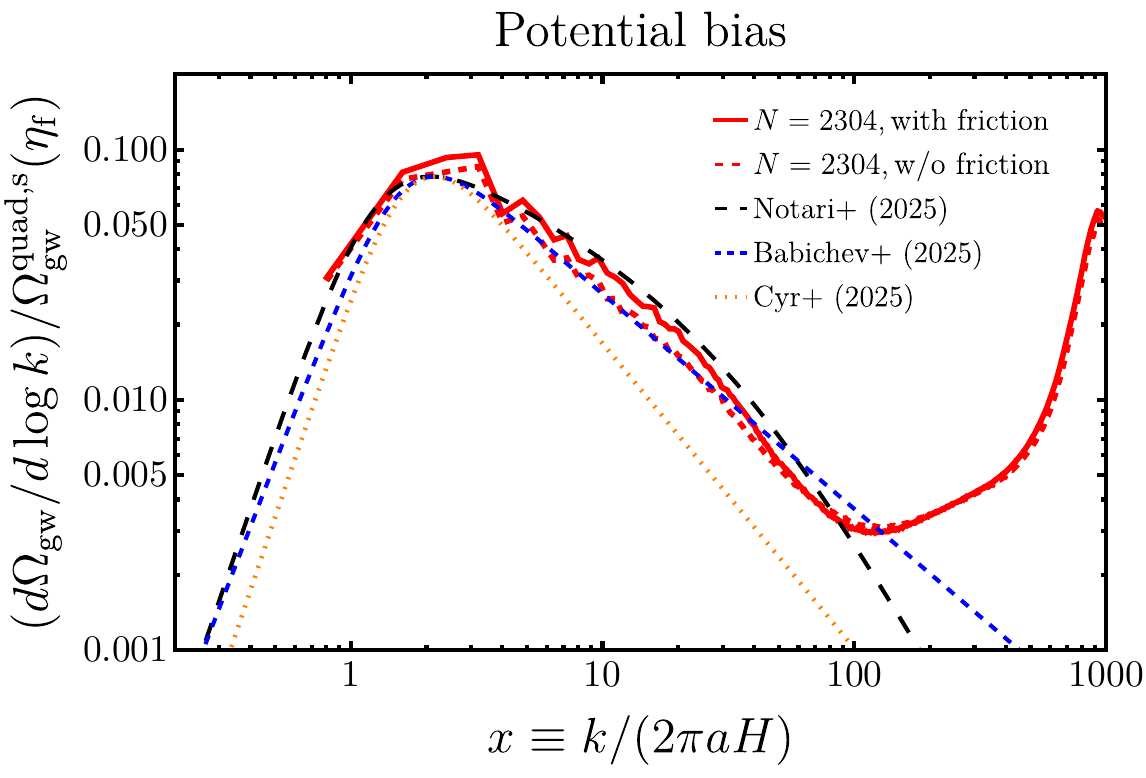}
    \end{center}
    \caption{\textit{Left:} GW spectrum for the unbiased DW network in scaling  at the time $\eta_f = 32$, extracted from simulations with and without an initial stage of friction, with $N=2304$ and $L=60$. We also plot the parametrizations obtained in \cite{Notari:2025kqq} and \cite{Babichev:2025stm} (see Table 2 of that work). For better comparison of the UV tails, we have shifted the fit from \cite{Babichev:2025stm} so that the peak's position and amplitude match those of \cite{Notari:2025kqq}. \textit{Right:} GW spectrum for the biased quartic potential with $\eta_{\Delta V} = 14$ at $\eta_f = 47$, extracted again from simulations with and without friction with the same resolution. We also plot the spectral parametrizations from \cite{Notari:2025kqq}, \cite{Babichev:2025stm} (see Table 2 of that work; case $\epsilon=0.05$), and also \cite{Cyr:2025nzf} (see Table III of that work, case $\epsilon / \lambda = 10^{-2.4}$). Again, we shift the fits from \cite{Babichev:2025stm} and \cite{Cyr:2025nzf} so that the peak's position and amplitude match the one of \cite{Notari:2025kqq}.}
    \label{fig:frict_comp_gw}
\end{figure*}

The impact of our strategy on GW spectra from annihilating walls is shown in Fig.~\ref{fig:frict_comp_gw}, for unbiased (leftmost) and biased (rightmost) networks. When friction is included (solid and dotted red curves: simulation results with and without friction respectively; dashed curves: fits), a flattening of the near-peak region of the spectrum is observed, in both unbiased and biased networks. Additionally, a doubly broken power-law feature emerges. In the case of an unbiased network, the leftmost plot shows that simulations without friction exhibit an intermediate fictitious ``plateau'', which subtracts power from the near-peak region, thereby leading to a steeper decrease. In the biased case, a similar loss of power in the near-peak region is observed. For comparison, we show the fits provided by~\cite{Babichev:2025stm} (with the amplitude fixed such as to match our simulations, to highlight differences in the spectral shape), which indeed correspond to the behaviour obtained in our simulations without friction.

Let us now comment on some additional claims and discrepancies of~\cite{Cyr:2025nzf} and~\cite{Babichev:2025stm}:

\begin{itemize}

\item{\textbf{Potential-induced population bias:} We notice that~\cite{Cyr:2025nzf} uses a biased potential that differs from ours, in that it also includes a linear term. Such a term displaces the maximum of the potential, and thus unavoidably introduces a population bias if the initial condition is kept at $\langle\phi\rangle=0$. The authors of~\cite{Cyr:2025nzf} argue that this population bias has a negligible impact if $\Delta V\lesssim 0.15$, based on the expectation that the network would still percolate according to the critical threshold predicted by percolation theory. However, our work shows that initial percolation is not the right criterion to understand the impact of population bias in 3D networks. In particular, we have shown that population bias is effective even when ``false vacuum'' regions initially percolate. 
This may also contribute to the discrepancy in the spectral shape, since indeed the results of~\cite{Cyr:2025nzf} are closer to those obtained in our population bias simulations.} 

\begin{figure*}[t]
\begin{center}
\includegraphics[width=0.6\textwidth]{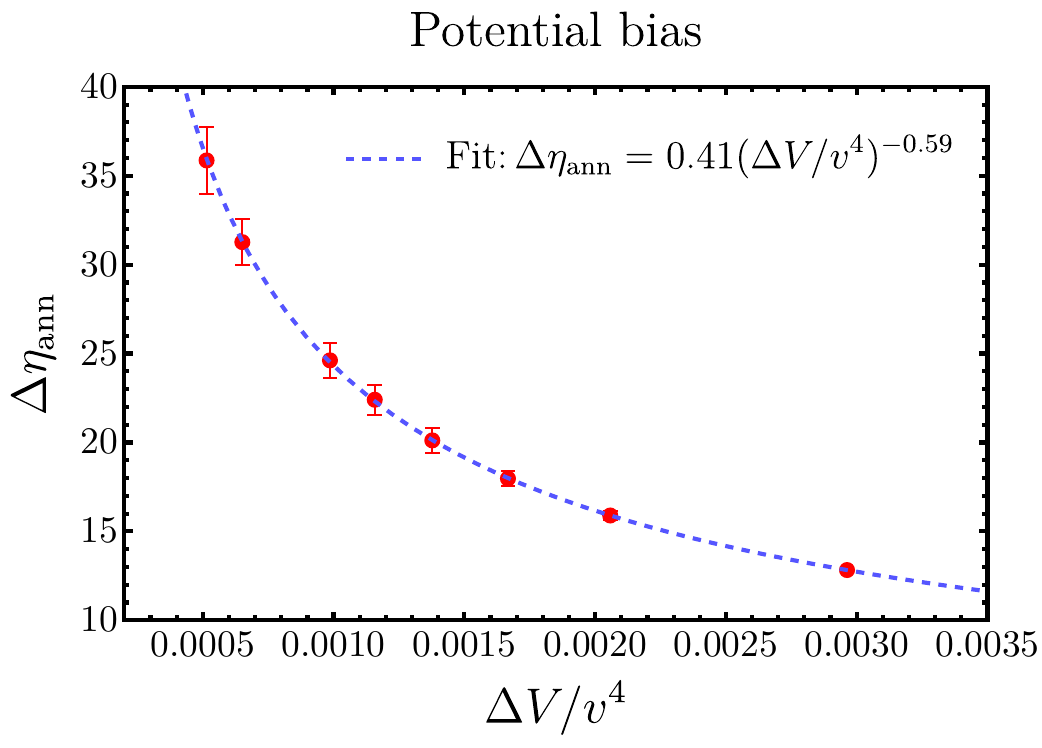}
\end{center}
    \caption{Values of $\Delta \eta_{\rm ann}$ extracted from fitting the template~\eqref{eq:FfvFunc2} to the decreasing false vacuum fraction in our potential bias simulations, see also Sec.~\ref{sec:PotBias}. For each bias size, we simulated three initial condition realizations. The dashed blue curve represents the result of a power law fit.}
    \label{fig:DeltaVetaann}
\end{figure*}

\item{\textbf{Dependence of $T_{\text{gw}}$ on $\Delta V$:} It is claimed in \cite{Babichev:2025stm} that biased domain wall networks annihilate at higher temperatures than those expected from the simple physical argument $\sigma H=\Delta V$. In particular, the claim is that $T_{\text{ann}}\propto \Delta V^{1/3}$ rather than $T_{\text{ann}}\propto \Delta V^{1/2}$. If true, this would significantly decrease the amplitude of the GW signal from DWs when $\Delta V$ is small, since $\Omega_{\text{gw}}\propto T_{\text{ann}}^{-4}$}. 
The simulations of~\cite{Babichev:2025stm} in fact exhibit a growing value of the exponent in the relation $T_{\text{ann}}\propto \Delta V^{\alpha}$, with smaller biases leading to larger values of $\alpha$, closer to $0.5$. Therefore, even the simulations and fits presented in~\cite{Babichev:2025stm} suggest that the claim is based on networks that do not fully achieve the scaling regime before annihilating. Nonetheless, we have investigated the issue further with new simulations with smaller potential bias sizes and with friction. The results are reported in Fig.~\ref{fig:DeltaVetaann}. The value of $\eta_{\text{ann}}\propto T_{\text{ann}}^{-1}$ is extracted by fitting the false vacuum fraction using the function~\eqref{eq:FfvFunc2}, see Sec.~\ref{sec:PotBias}. We find $T_{\text{ann}}\propto \Delta V^{0.59}$(with negligible statistical uncertainties), which we consider compatible with the physically expected behaviour, given the limited range of our simulations and systematic uncertainties.

In addition, we notice that the dependence suggested by~\cite{Babichev:2025stm} faces a significant physical challenge: in order to be correct, it would imply that the network annihilates when $\Delta V\ll \sigma H$. The authors argue that this can be explained by a matter-like dilution of biased domain walls (i.e. $\rho\propto a^{-3}$), though this is in clear contrast with the results from simulations, see e.g.~Fig.~2 of~\cite{Notari:2025kqq}. We conclude that the physically-motivated behavior $T_{\text{gw}}\propto \Delta V^{0.5}$  is supported by simulations and can be used to extrapolate the results of~\cite{Notari:2025kqq} and our work to general values of $\Delta V$.

\end{itemize}

Finally, let us comment on the discrepancy in the inferred value of the FV fraction exponent $p$ with respect to~\cite{Ferreira:2024eru}, where $p=3.0\pm 0.3$ was reported, which is significantly away from our new result $p=2.16\pm 0.06$, see~\eqref{eq:ppot}. The result of~\cite{Ferreira:2024eru} was obtained by including the early times of the simulations in the fit to the FV fraction, which was justified since a phase of friction was not employed in~\cite{Ferreira:2024eru}, and significantly smaller lattices were used, with lower resolution. These arguments provide partial support for our new result. However, a detailed assessment, including improved analytical modelling, requires a dedicated study, with smaller bias sizes and reaching smaller FV fractions in super-Hubble structures, which we leave for future work. Importantly, we notice that the parameter $\Delta \eta_{\text{ann}}$, which is key for the GW observables discussed in the next section, is rather insensitive to the inferred value of $p$.

\section{Conclusions and Observational Outlook} \label{sec:summary}

This work was devoted to the study of the evolution of biased domain wall networks, with a focus on scenarios where one vacuum is initially slightly more populated than the other. Similarly to scenarios with a potential bias, these networks annihilate after a possibly long epoch of standard scaling evolution, and thus evade the so-called domain wall problem. We have started by providing a qualitative physical understanding of why a population-biased network annihilates, despite the exactness of the underlying discrete symmetry, as a consequence of a non-vanishing average value of the extrinsic curvature term over the network.

We have then investigated the evolution of and signals from the population-biased network quantitatively, via numerical simulations (in this section we remove the shift $\eta_s$ in the definition of time scales related to the annihilation of network, since in a relevant cosmological scenario those are much larger than $\eta_s$). Our key results concern:
\begin{itemize}
\item{The decay of the ``false vacuum'' (the initially less populated one) as a function of time.
Specifically, we have found the FV to decay as
\begin{equation}
\FF\simeq \frac12 \,\exp\left[ - \left( \frac{\eta}{\eta_{\text{ann}}}\right)^{p(\eta)} \right]
\end{equation}
with a `running' exponent $p(\eta)$ that evolves from $p\simeq 1.4$ to $p\simeq 1.7$ from the start of the (quasi-)scaling regime to the end of our simulations. We interpret this time dependence  in $p$ as linked to the fact that a biased network interpolates between two very distinct regimes. Initially it is percolated and a small vacuum asymmetry $\BB = 1/2-\FF$ is expected to grow as a power law. However once percolation is lost the network evolves very differently, as a collection of closed walls. An exponential decay with some exponent is then naturally  expected.}
\item{The dependence of the annihilation temperature on the initial population bias. In particular, our results allow to relate the annihilation time $\eta_{\text{ann}}$ to the initial bias as
\begin{equation}
\eta_{\text{ann}}\simeq 0.2\eta_s\mathcal{B}_s^{-0.8},
\end{equation}
where $\eta_s$ here represents the time of the onset of the scaling regime, and we have extrapolated~\eqref{eq:EtaAnn} to $\eta_{\text{ann}}\gg \eta_s$.}
\item{The emission and spectrum of gravitational waves from the annihilating network. We have found the emission of GWs to continue throughout the annihilation of the network, with a maximal energy density fraction $\Omega_\text{gw}$ that is achieved at the time
\begin{equation}
\eta_{\text{gw}}\simeq 2.6 \eta_{\text{ann}} \simeq 0.5\eta_s \mathcal{B}_s^{-0.8}.
\end{equation}
Correspondingly, the spectrum is well characterized by a single broken power law (see below).}
\end{itemize}
Let us now relate these results to GW observations. In the radiation dominated Universe $\eta\sim T^{-1}g_*^{-1/6}$, where $g_*(T)$ is the number of relativistic degrees of freedom at the temperature $T$ (and we take $g_{*,s}(T)=g_*(T$)). Therefore we have:
\begin{equation}
\label{eq:tgw}
\textbf{Population Bias:}\quad T_{\text{gw}}\simeq 0.4~T_{\text{ann}}\simeq 126~\text{MeV}\left(\frac{g_*(T_s)}{g_*(T_\text{gw})}\right)^{\frac{1}{6}}\left(\frac{T_s}{10^6~\text{GeV}}\right) \left(\frac{\mathcal{B}_s}{10^{-9}}\right)^{0.8} 
\, . \end{equation}
It is in principle possible to express $T_{\text{ann}}$ in terms of the temperature of the symmetry breaking phase transition $T_c\sim v$ and the initial bias $b_i$ at $T_c$~\eqref{eq:biaspar}. However, the actual separation between $T_s$ and $T_c$ is model-dependent: in particular it depends on whether the field originally interacts with the thermal bath, in which case $T_s\ll T_c$, and on the self-coupling $\lambda$, since strictly speaking the network forms at the Ginzburg temperature $T_G$, such that $T_c - T_G\sim \lambda T_c$ (see e.g.~\cite{Vilenkin:2000jqa}). Additionally, if the transition is non-thermal and $H_i\simeq m$ when domain walls form (as in our simulations), the approach to the scaling regime may be faster than in the standard thermal case with $H_i\ll m$.  Furthermore, as discussed in Sec.~\ref{sec:dynamics}, the bias grows between $T_c$ and $T_s$ differently from the growth during scaling. For these reasons, we prefer to stick to the more model-independent characterization \eqref{eq:tgw}. 

For the potential bias scenario, we have overall confirmed the results previously obtained in~\cite{Notari:2025kqq} (within the uncertainties), finding:
\begin{equation}
\label{eq:tgwpot}
\textbf{Potential Bias:}\quad T_{\text{gw}}\simeq 0.4~T_{\Delta V}\simeq 180~\text{MeV}\left(\frac{g_*(T_\text{gw})}{17.25}\right)^{-\frac{1}{4}}\left(\frac{\Delta V^{1/4}}{100~\text{MeV}}\right)^{2}\left(\frac{10^5~\text{GeV}}{\sigma^{1/3}}\right)^{\frac{3}{2}},
\end{equation}
In both scenarios, the peak frequency of the GW signal is related to $T_\text{gw}$ by the usual relation~\cite{Notari:2025kqq}:
\begin{equation}
\label{eq:pfreqtoday}
\nonumber f_p \simeq 24~x_p~\text{nHz}\left(\frac{g_*(T_\text{gw})}{100}\right)^{1/6}\left(\frac{T_\text{gw}}{150~\text{MeV}}\right) \ , 
\end{equation}
with $x_p\simeq 2$ for both population and potential bias scenarios, 

The GW spectrum from biased domain walls today can then be expressed via the usual expression $\Omega_\text{gw}(f)h^2 = \Omega_\text{gw}(f_p)h^2\times \mathcal{S}(f/f_p)$, where the amplitude at the peak frequency $f_p$ is given by~\cite{Ferreira:2024eru, Notari:2025kqq}:
\begin{align}
\label{eq:GWtoday}
\nonumber \Omega_\text{gw}(f_p)h^2\simeq 10^{-10}&\epsilon_{\text{gw}}\left(\frac{g_*(T_\text{gw})}{10.75}\right)^{-1/3}\left(\frac{\alpha_\text{gw}}{0.01}\right)^{2} \, , \\
\textbf{Population Bias:}\quad \epsilon_{\text{gw}}\simeq 0.06, &
\quad \textbf{Potential Bias:}\quad \epsilon_{\text{gw}}\simeq 0.1,
\end{align}
where $\alpha_\text{gw}\equiv 2\sigma H_\text{gw}/(3H_\text{gw}^2M_p^2)$ and $H_\text{gw}\equiv H(T_\text{gw})$.
Concerning the spectral shape, we have found the signal from population bias to be well described by a single broken power law:
\begin{equation}
\textbf{Population Bias:}\quad    \mathcal {S} (f)= \frac{ (\alpha+\beta)^{\delta} }{\left( \beta\left(\frac{f}{f_p}\right)^{-\frac{\alpha}{\delta}}  +
  \alpha\left(\frac{f}{f_p}\right)^{\frac{\beta}{\delta}}\right)^{\delta} }, \quad \alpha=3,~\beta\approx 1,~\delta\approx 2.8,
\end{equation}
where we report only the central values of the parameters, see Table~\ref{tab:bias_fit_parameters} for full results. While the near-peak slope is close to the usual result in the scaling regime, the width of the peak is significantly broader (although the inference of this parameter is subject to statistical uncertainties on the IR tail of the signal), and the peak position is also roughly twice as large.
For potential bias, we confirm the evidence for a double broken spectrum, with parameters that roughly agree with~\cite{Notari:2025kqq}:
\begin{align}
\textbf{Potential Bias:}\quad 
    \mathcal {S} (f) &=
    \frac{ \alpha+\beta + \left(\frac{f_p}{f_b}\right)^{\beta + \gamma}
    }{ \beta\left(\frac{f}{f_p}\right)^{-\alpha} + \alpha\left(\frac{f}{f_p}\right)^{\beta} + \left(\frac{f_b}{f_p}\right)^{-\beta} \left(\frac{f}{f_b}\right)^{\gamma} },\alpha=3,~\beta\approx 0.4,~\gamma\approx 1.6,
\end{align}
and a breaking frequency $f_b\approx 2.5 f_p$.
The difference in spectral shapes of population- and potential-biased networks may potentially help in distinguishing between the two annihilation mechanisms, although it will likely be challenging to extract from the data. 

\begin{figure*}[t]
    \begin{center}
        
        \includegraphics[width=0.55\textwidth]{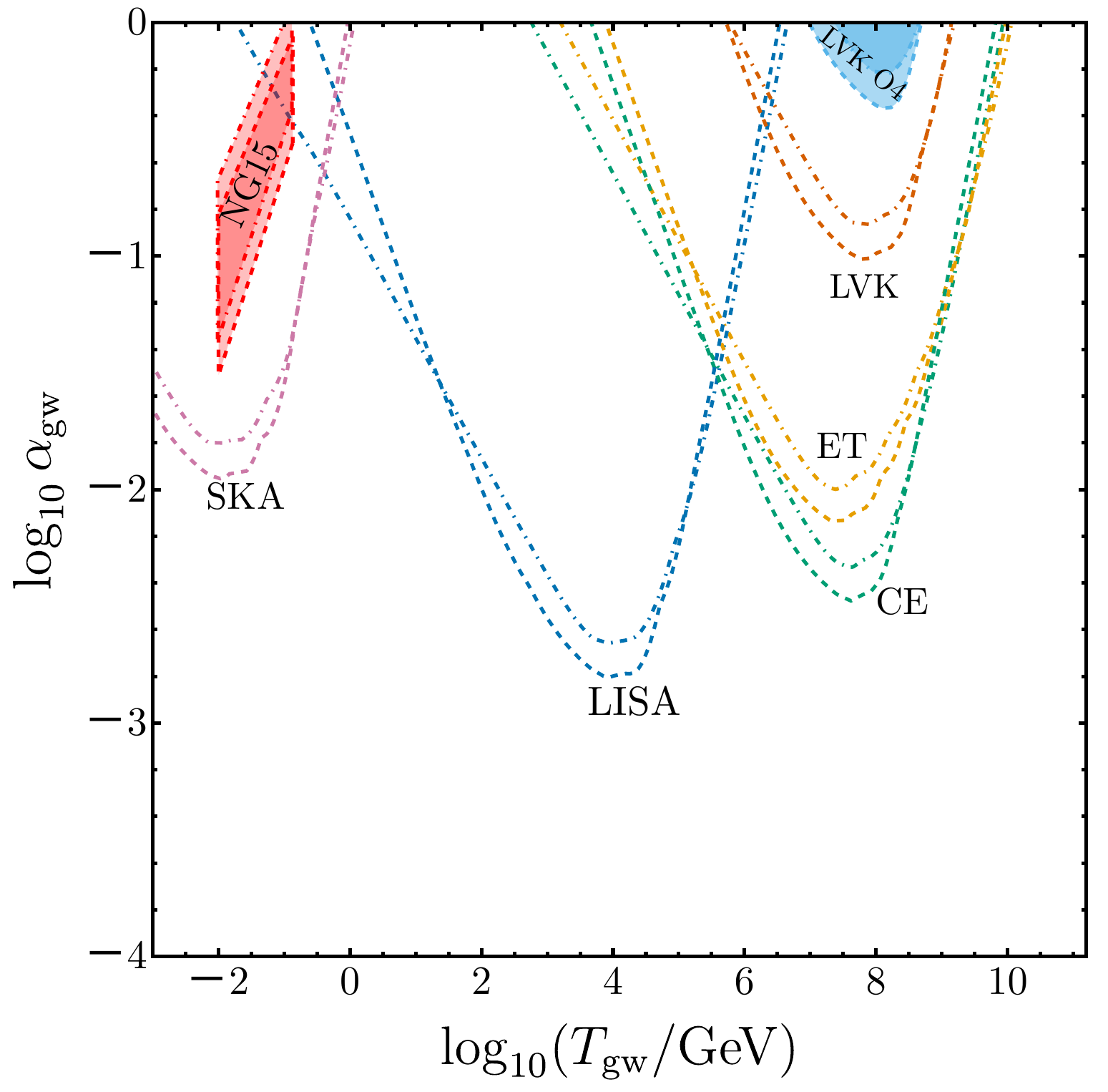}
        
    \end{center}
    \caption{Constraints from and sensitivity regions of GW observatories for:  population-(regions within dot-dashed curves) and potential-(within dashed curves)biased networks. The power-law integrated sensitivity curves of SKA, LISA, CE and ET are taken from~\cite{Caprini:2024ofd}, and partially account for astrophysical foregrounds. For LIGO-Virgo-KAGRA (LVK),  the O4 results (blue-shaded) are taken from~\cite{LIGOScientific:2025bgj}, while the design sensitivity is taken from~\cite{Schmitz:2020syl}. The pink-shaded region roughly corresponds to the evidence region from NANOgrav 15 yrs~\cite{NANOGrav:2023gor}.}
    \label{fig:alphatgw}
\end{figure*}

\begin{figure*}[t]
    \begin{center}
        
        \includegraphics[width=0.5\textwidth]{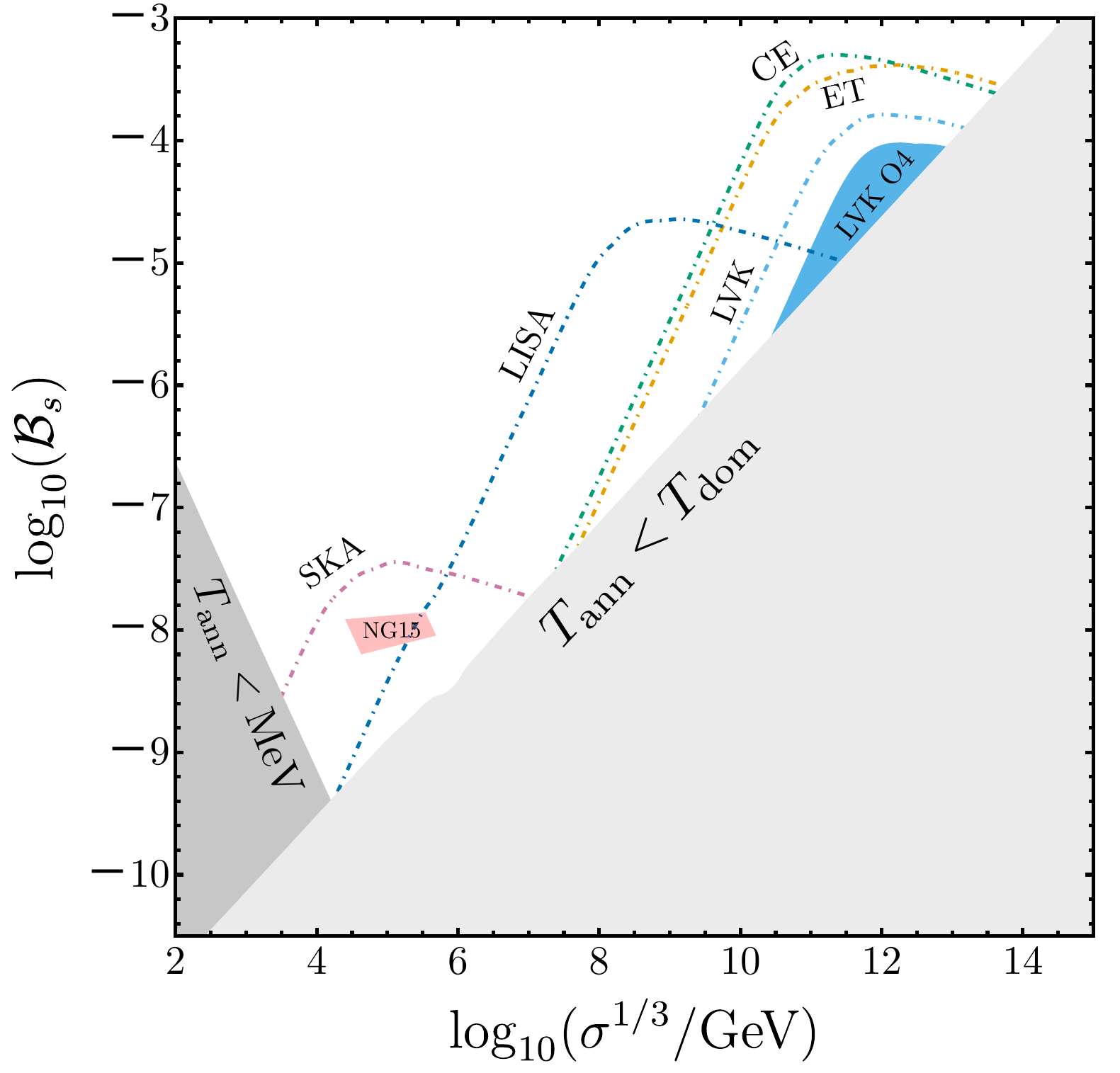}
        \hspace{0.1em}\includegraphics[width=0.48\textwidth]{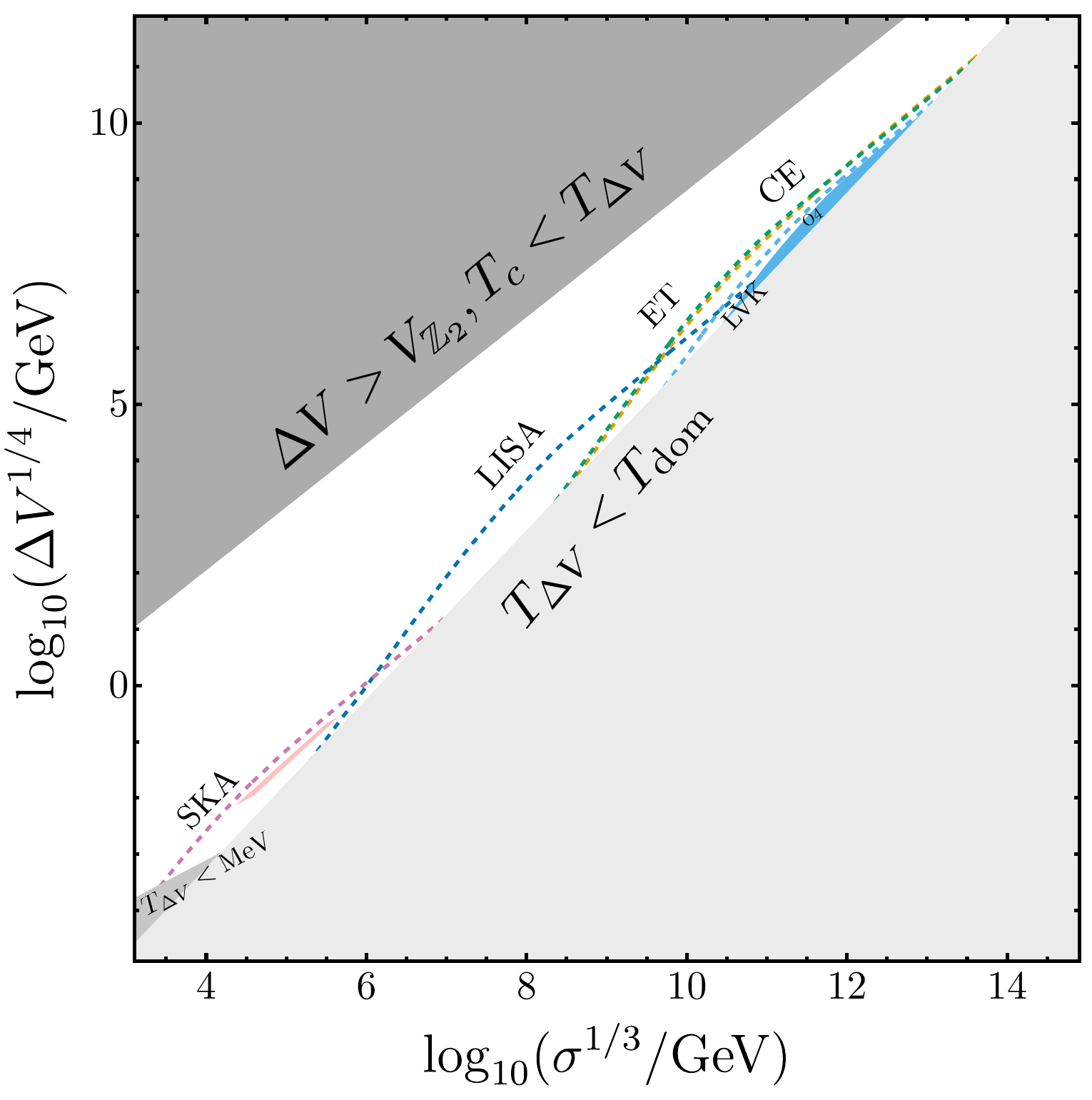} \,
        
    \end{center}
    \caption{Constraints from (shaded blue region) and sensitivity regions (bounded by dotted and dot-dashed curves) of GW observatories for:  population- (\textit{Left}) and potential- (\textit{Right}) biased networks. The pink-shaded region corresponds roughly to the evidence region from NANOgrav 15 yrs~\cite{NANOGrav:2023gor} (both in the left and right panels). The gray coloured regions represent constraints from: domain wall domination before annihilation ($T_{\rm dom} > T_{\Delta V, \text{ann}}$); annihilation during/after BBN ($T_{\Delta V, \text{ann}} < \rm MeV$) (this applies only to scenarios where DWs decay to Standard Model particles); potential bias larger than symmetric potential. For population-biased networks, we have fixed $\lambda = 1/2$.}
    \label{fig:sigma_vs_bias}
\end{figure*}

The regions of $\alpha_\text{gw}$ and $T_{\text{gw}}$ for biased networks, that are constrained or will be probed by (future) GW observatories are shown in Fig.~\ref{fig:alphatgw} (dot-dashed and dashed curves: population and potential bias respectively). 
These are mapped to the parameter spaces of the two annihilation mechanisms in Fig.~\ref{fig:sigma_vs_bias} (left and right panels: population and potential bias respectively).  In particular, we notice that both scenarios are able to provide an interpretation of the PTA background (in the figures, the red-shaded region is a rough indication of the values of parameters for which DWs can provide a good interpretation of the PTA GW background, as  reported by the NANOGrav collaboration~\cite{NANOGrav:2023gor}, other values of parameters close to this region may also provide a good fit) suggesting in both cases a (UV) particle physics scale around $10-1000~\text{TeV}$. Detailed searches in the datasets, especially those from PTAs, (as well as the latest data from LVK~\cite{LIGOScientific:2025kry}) can be performed using our results.

Our work also clarifies on some recent claims in the literature~\cite{Cyr:2025nzf, Babichev:2025stm} concerning the GW spectrum from potential-biased networks. In particular, we have highlighted that their discrepancies with respect to~\cite{Notari:2025kqq} (and the present work) likely stem from the more limited range of bias values analysed in~\cite{Cyr:2025nzf, Babichev:2025stm}, as well as from contamination from non-DW sources of GWs in~\cite{Cyr:2025nzf, Babichev:2025stm}, which~\cite{Notari:2025kqq} and this current work avoid via an initial friction epoch that accelerates the onset of a proper DW scaling regime.

Additional numerical resources, as well as improved numerical strategies, can help in addressing some remaining caveats in our work. First, our results for GW observation from population bias in this section are obtained  by extrapolating the outcome of our simulations to arbitrary small values of the bias $\mathcal{B}_s$. We expect this extrapolation to be roughly valid, based on the heuristic arguments presented in the first section, but it would be useful to extend the analysis to even smaller biases. 
Secondly, an improved resolution of the IR tail of the GW spectrum would allow to clarify the width of the peak in the population bias case. The fattening algorithm, which we have not used when computing GWs in this work, may be useful in this respect, since it allows for a smaller IR cutoff, although it may also affect the emission of GWs from the smaller scales in the network. A dedicated study might be interesting to pursue.

Finally, we expect the results of our work to have an important impact on the possibility to form Primordial Black Holes (PBHs) from the collapse of the network, as described in~\cite{Ferrer:2018uiu, Ferreira:2024eru} for the potential bias case (see also~\cite{Gouttenoire:2023gbn}). We stress that, among other necessary steps, a dedicated study of the behaviour of the FV fraction in super-Hubble structures at late times is required to reliably estimate the PBH fraction. We leave such an interesting task for future work.

\acknowledgments

We thank R.~Z.~Ferreira for sharing a script that Fig.~\ref{fig:alphatgw} is partially based on and H. Quelquejay Leclere for support with the sensitivity curves used in this work. The work of D.B. is supported by the predoctoral grant PREP2023-001785. The work of F.R.~is supported by the grant RYC2021-031105-I from the Ministerio de Ciencia e Innovación (Spain) (MICINN), through the Spanish State Research Agency (AEI). F.T.~is supported by the \textit{Atracción de Talento César Nombela} fellowship No 2025-T1/COM-36104 funded by Comunidad de Madrid (Spain). We acknowledge use of the following computational clusters: the Port d'Informació Científica (PIC), the UC3M C3 Cluster, the use of the latter co-financed through action EQC2021-007184-P and the NyX cluster at ICCUB.
D.B., F.R. and O.P. acknowledge support from the Spanish Ministry of Science and Innovation (MICINN) through the Spanish State Research Agency under the R\&D\&i project PID2023-146686NB-C31 funded by MICIU/AEI/10.13039/501100011033/ and by ERDF/EU, and under Severo Ochoa Centres of Excellence Programme 2025-2029 (CEX2024001442-S). IFAE is partially funded by the CERCA program of the Generalitat de Catalunya.
MICIIN supported this study with funding from the European Union NextGenerationEU (PRTR-C17.I1) and by the Generalitat de Catalunya. 

\newpage

\appendix

\section{Consistency checks} \label{app:Technicalities}

In this appendix, we provide various consistency checks to assess the accuracy of the simulations, related to the lattice resolution, initial fluctuations, and the bias size.

\begin{figure*}
    \centering
    \includegraphics[width=0.48\textwidth]{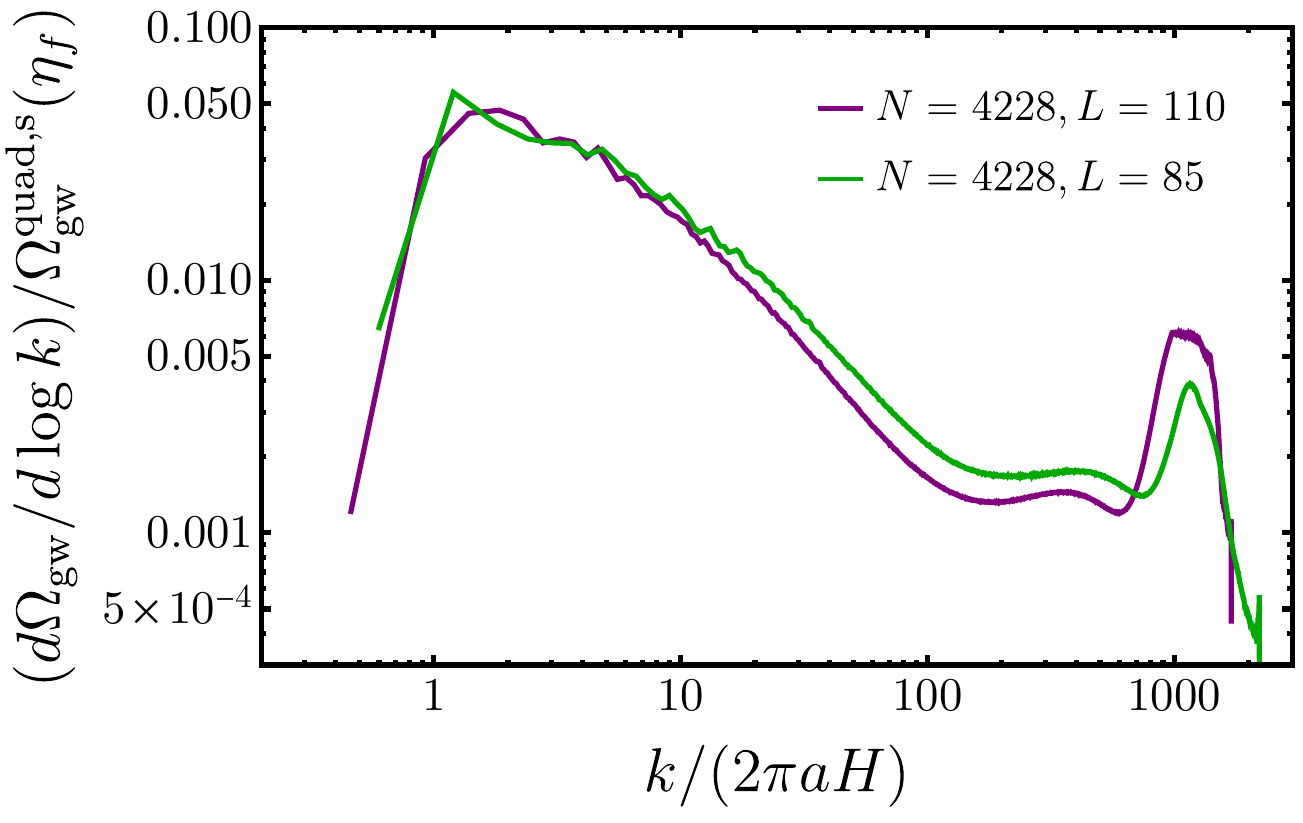} \,
    \includegraphics[width=0.46\textwidth]{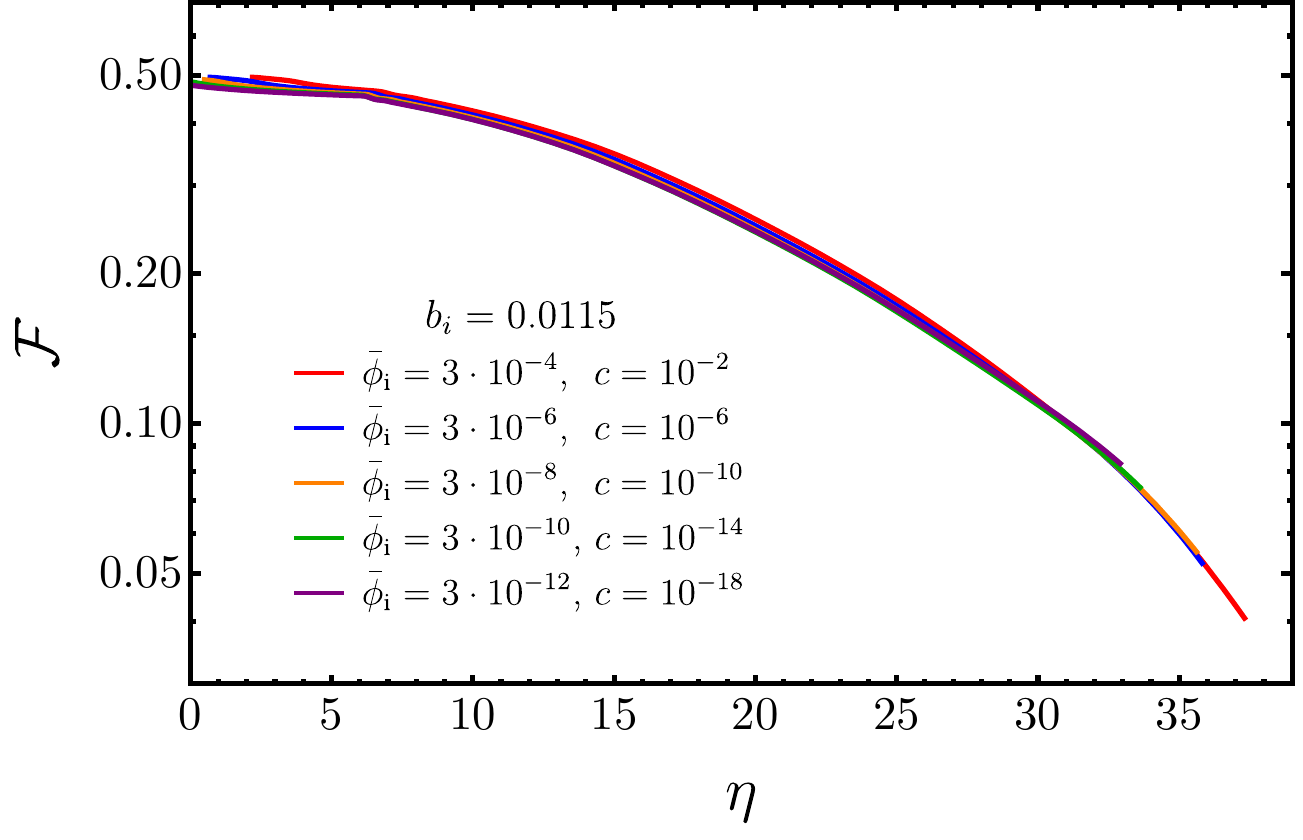} \\ \vspace{0.3cm}
    \includegraphics[width=0.48\textwidth]{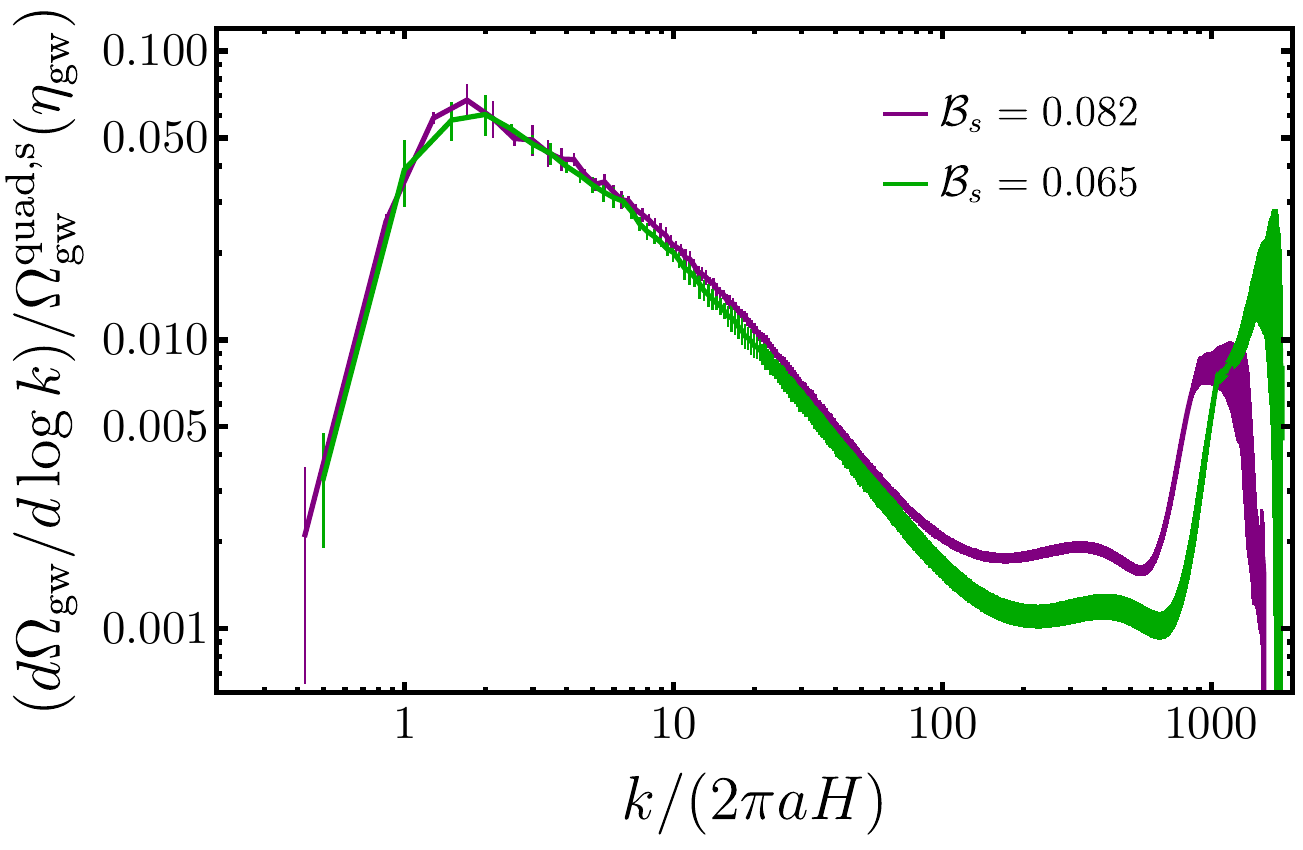}
    \caption{Consistency checks for our lattice simulations. The panels display the effects of different lattice resolutions (top-left), initial fluctuations (top-right), and the dependence of the GW spectral shape on the bias size (bottom). A detailed description of each panel is provided in Appendix \ref{app:Technicalities}.} \label{fig:NLtest} 
\end{figure*}

\begin{itemize}
    \item {\bf Lattice resolution:} Our parametrization for the GW spectrum has been obtained from simulations on lattices with $N = 4228$ and $L = 110$. Following the definition of \eqref{eq:tmax}, this leads to a maximum simulation time $\eta_{\rm max}^{\rm res} \approx 33$. By this time, the area parameter for our benchmark choice $\mathcal{B}_s \approx 0.081$ has decreased from $\mathcal{A} \approx 1$ to $\mathcal{A} \approx 0.15$, meaning that most of the annihilation has already taken place. Nevertheless, as discussed in Sec.~\ref{sec:dynamics}, we extend the simulation to later times and extract the spectrum at $\eta_f \approx 50$, since the late-time enhancement is primarily sourced by scalar waves. To assess the robustness of this assumption, in the top-left panel of Fig.~\ref{fig:NLtest} we compare the spectra obtained for $N=4228$ and $L=85$, which corresponds to $\eta_{\rm max}^{\rm res} \approx 42$. The spectra agree quite well for wavenumbers $x \lesssim 20$ but slightly diverge for larger wavenumbers due to the different UV resolutions.

    \item {\bf Initial fluctuations:} The bias parameter \eqref{eq:biaspar} depends on both the initial displacement of the field’s homogeneous mode $\bar{\phi}_i$ and the amplitude of the initial fluctuations (set by the dimensionless constant $c$, fixed to $c=10^{-10}$ in the simulations presented in the bulk text) through the ratio $b_i \propto \bar{\phi}_i / \sqrt{c}$. To verify that the DW evolution depends solely on $b_i$, in the top-right panel of Fig.~\ref{fig:NLtest} we compare the evolution of the FV fraction for different combinations of $\bar{\phi}_i$ and $c$ yielding the same bias parameter $b_i \propto \bar{\phi}_i / \sqrt{c} = 0.0115$. The simulations have been carried out in lattices of $N=2560$, $L=60$ and without friction, and the plotting times for each line have been slightly shifted to account for the slightly different domain wall formation times. We observe that the evolution of the FV is basically identical in all cases as expected.

    \item {\bf Dependence of the GW spectral shape on the bias:} In the bulk text we have parametrized the GW spectrum for the initial bias $\mathcal{B}_s = 0.081$. We have argued that, as the scalar field's spectral shape is very similar for different population biases for identical values of the FV fraction, see Fig.~\ref{fig:BiasComp}, we expect a similar result for the GW spectrum. To further confirm this, in the bottom panel of Fig.~\ref{fig:NLtest} we compare the GW spectrum for two choices of initial population bias, $\mathcal{B}_s =0.081, 0.065$, evaluated when $\eta - \eta_s = 2.9 \Delta \eta_{\rm ann}$, which corresponds approximately to the time when the GW spectrum has saturated, c.f.~\eqref{eq:delay}. Note that this condition holds at a later time for $\mathcal{B}_s =0.065$ than for $\mathcal{B}_s =0.081$, so the lattice resolution is significantly worse for the smaller initial bias. In each case we have simulated three initial condition realizations, with the error bars show the standard deviation for each spectral bin.  In any case, we observe that both spectra coincide at the same time for $x \lesssim 100$, suggesting that the parametrization presented in the main text for  $\mathcal{B}_s =0.081$ is also valid for smaller initial biases.
\end{itemize}

\section{Fattening} \label{app:fattening}

Here we discuss details of the fattening strategy employed in parts of this work.

The PRS algorithm \cite{Press:1989yh} modifies the EOM of the scalar field in a precise way, such that the dynamics of domain walls, in the thin-wall limit, match the evolution predicted by the physical EOM \cite{Sousa:2010zza}. For 3D simulations, this is achieved by introducing the constraint $\alpha + \beta/2 = 3$. The PRS algorithm does not guarantee that physical results are reproduced when domain walls are not present, either due to the complete annihilation of the domain wall network, or because they have not formed yet. While for simulations of the scaling regime this is of little practical importance, it instead limits the applicability of fattening to collapsing domain wall networks.

\begin{figure*}
    \centering
    \includegraphics[width=0.47\textwidth]{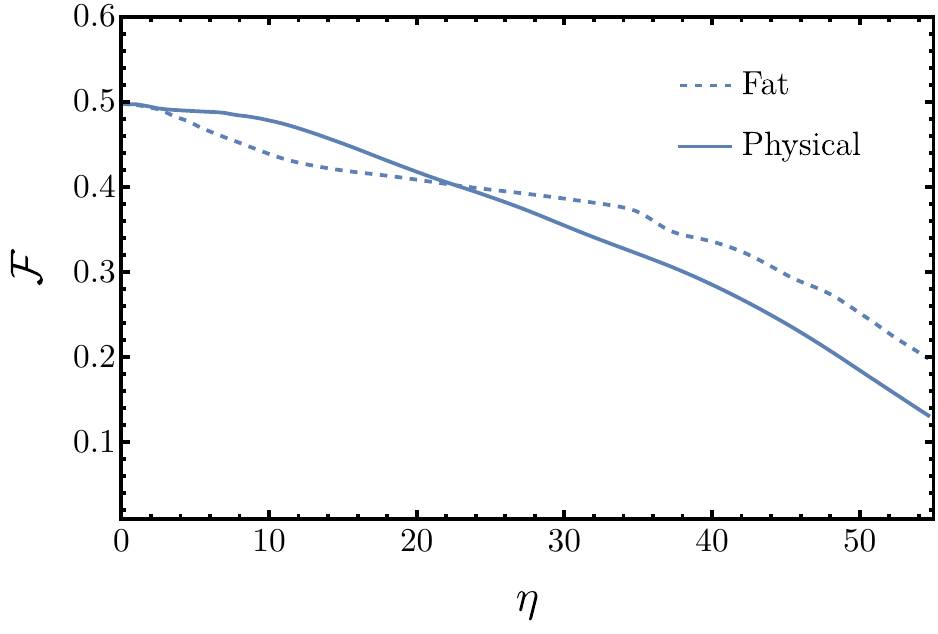}\
    \caption{Evolution of the FV fraction for two simulations with identical initial conditions with $N=3840$ and $L=80$, one (dashed) evolved accordingly to Eq.~\eqref{eq:EOMfattening} (setting $\alpha=3,\,\beta=0$) and the other (solid) with Eq.~\eqref{eq:EOMsc}. Here fattening is active since the initial time, and deviations in the evolution of the system accumulate, leading to significantly different results at later times.} \label{fig:fat_nofat_initial} 
\end{figure*}

In the early stages of the evolution, simulations with fattening exhibit delayed formation of domain walls,
and an initial evolution of the bias that differs from physical simulations, as can be observed in Fig.~\ref{fig:fat_nofat_initial}.

\begin{figure*}
    \centering
    \includegraphics[width=0.41\textwidth]{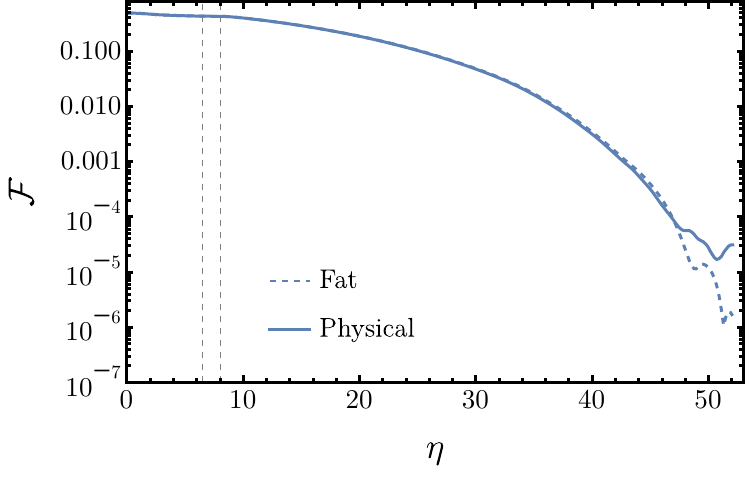} \hspace{0.3cm}
    \includegraphics[width=0.42\textwidth]{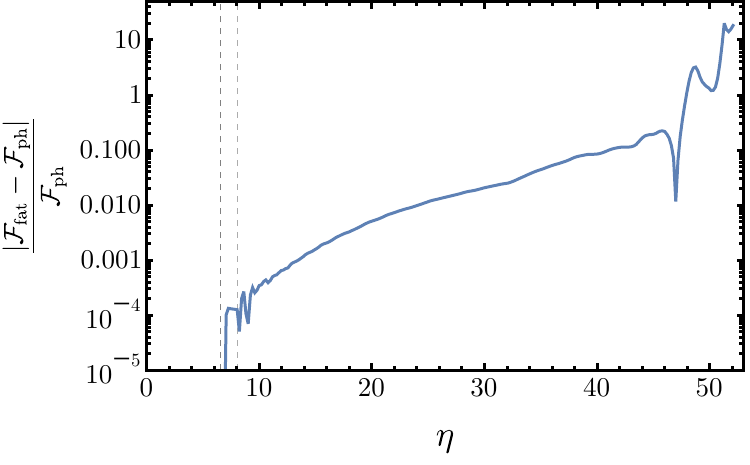}
    \includegraphics[width=0.39\textwidth]{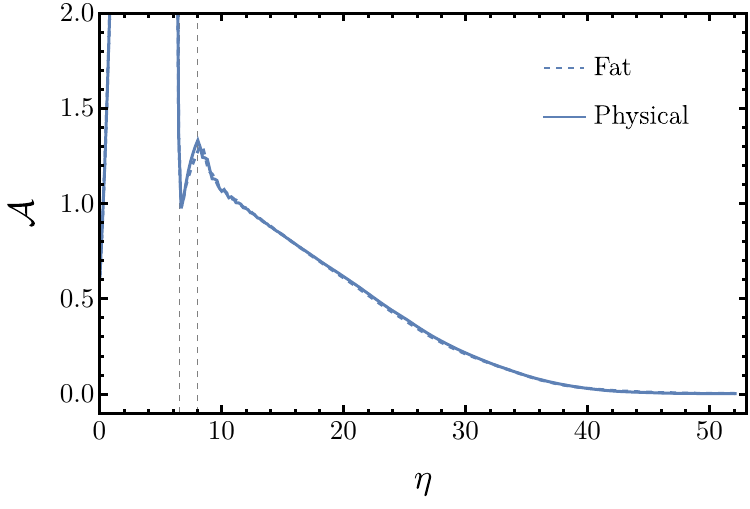} \hspace{0.3cm}
    \includegraphics[width=0.42\textwidth]{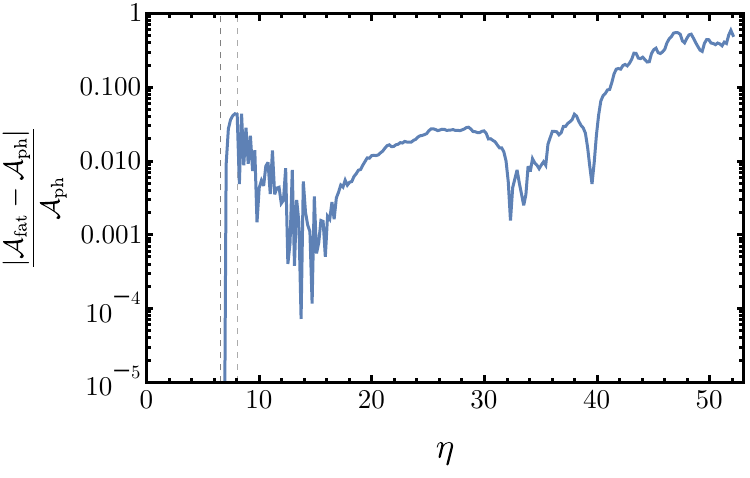}
    \includegraphics[width=0.41\textwidth]{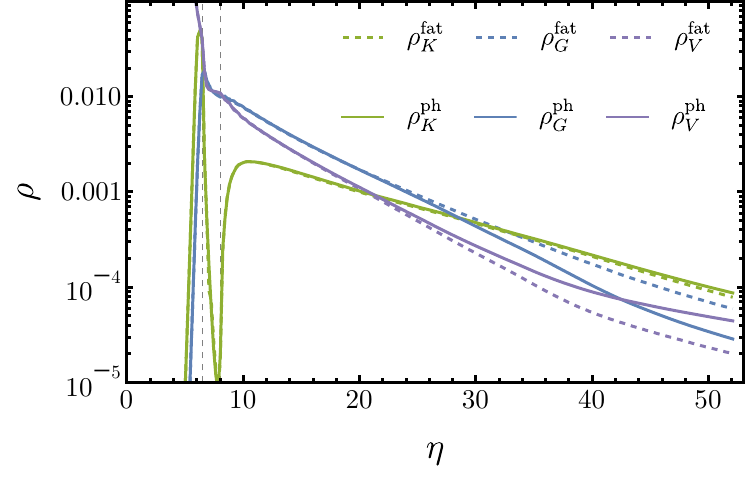} \hspace{0.3cm}
    \includegraphics[width=0.42\textwidth]{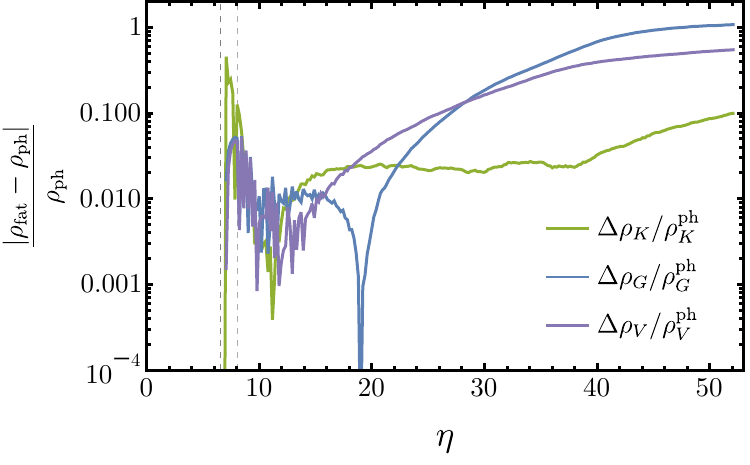}
    \caption{Comparison between two simulations with $N=4248$, $L=110$, $\mathcal{B}_s = 0.082$, and the same initial conditions, one evolved with fattening activated at $\eta=7$, and the other using physical equations of motion. Comparing with Fig. \ref{fig:fat_nofat_initial} we see that matching has vastly improved. At late times we start observing significant deviations in all quantities. As later explained, we attribute this to the collapse of the domain wall network.} \label{fig:nofat_fatact} 
\end{figure*}

Our objective is to leverage fattening to effectively extend the results obtained with physical simulations. Because of this, provided the same initial conditions and bias at $\eta = 0$, we wish to match the value of the bias $\mathcal{B}_s$ at the onset of scaling. To achieve this, we initially evolve the field with the physical equations of motion, thereby preserving the same dynamics during the formation of domain walls, and only activate fattening once the network has formed, as explained in Sec.~\ref{sec:dynamics}. This enables us to closely match the evolution of simulations with and without fattening, as can be appreciated in Fig.~\ref{fig:nofat_fatact}, where we compare the evolution of the energy density, false vacuum fraction, and area parameters.

In the late stages of collapse, the equivalence between the physical and fattened simulations breaks down, as domain walls become a subleading component of the system, and scalar waves dominate. This limitation imposes a minimum value of $\mathcal{A}$ and $\mathcal{F}$ that can be studied with fattening, which we estimate to be around $\mathcal{A} \sim 0.1$ and $\mathcal{F} \sim 0.01$. Notice that this is of no relevance for the fattening simulations presented in the main text, as such small values of $\mathcal{F}$ are never reached before intersecting the line $\mathcal{F}^{(H)}$.

\begin{figure*}
    \centering
    \includegraphics[width=0.41\textwidth]{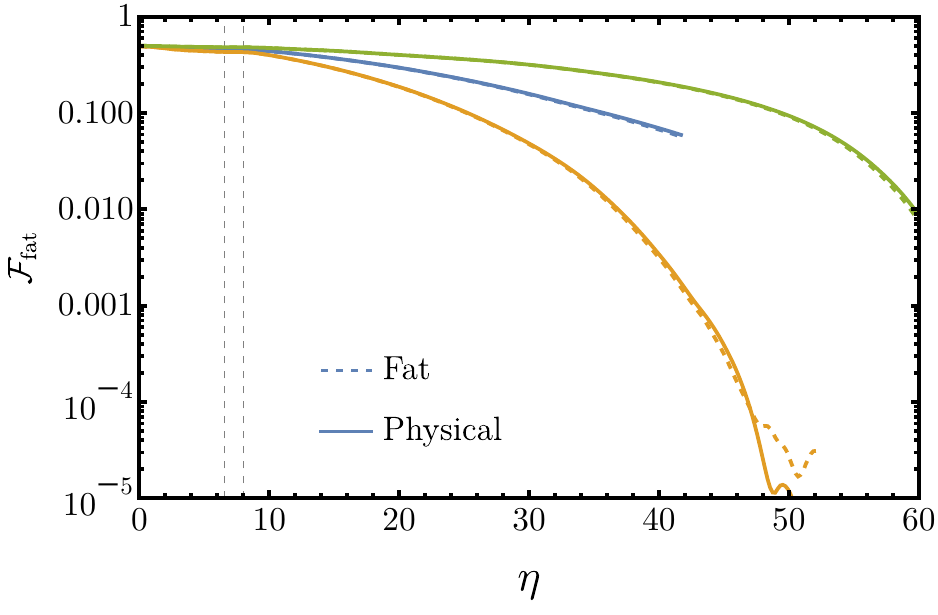} \hspace{0.3cm}
    \includegraphics[width=0.42\textwidth]{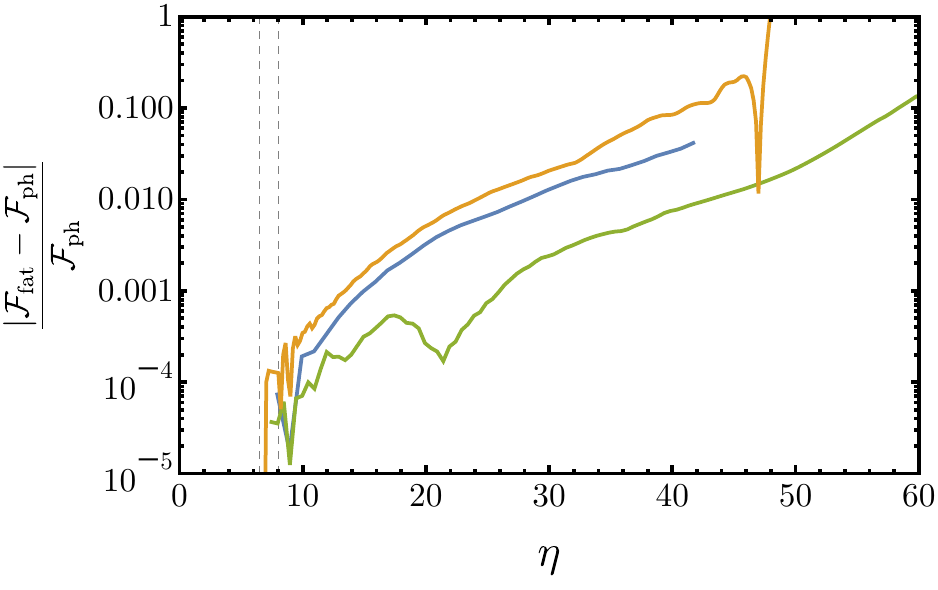}
    \caption{Comparison between different simulations with and without fattening for different values of the initial bias. The drastic growth of the error for the smallest value of the bias (the orange line) is due to the complete annihilation of the domain wall network. The green line corresponds to a simulation with $N=4096$, $L=59$ and $\mathcal{B}_s=0.022$, the blue line corresponds to a simulation with $N=3840$, $L=80$ and $\mathcal{B}_s=0.049$, the orange line corresponds to a simulation with $N=4248$, $L=110$ and $\mathcal{B}_s=0.082$.} \label{fig:fat_nofat_biases} 
\end{figure*}

We also notice that the relative errors between fattening and physical simulations grow with time. To better understand this behaviour, we have performed the same comparison between fattening and non fattening simulations for different values of the initial bias, which we report in Fig.~\ref{fig:fat_nofat_biases}. These simulations show that the growth does not only depend on time, but also on the false vacuum fraction. We understand that fattening simulations with larger annihilation times match physical simulations for longer epochs, justifying the use of fattening to study long lived networks.

A drawback of activating fattening after domain wall formation is the need to take into account the initial redshift of the comoving width of domain walls. In particular, the values of $N$ and $L$ must be chosen in such a way that domain walls are still resolved at the time of fattening activation $\eta_{F}$. Once $N$ is fixed, this introduces an upper bound on $L$, more relaxed than that of fully physical simulations. To take better advantage of fattening, we keep $L$ as large as possible, while still maintaining a good match with physical simulations. We have found that good results are reproduced for $L = N/16$ and an activation time of $\eta_F = 7$. This yields a ratio of $k_{max} / k_{w} \approx 2$ at $\eta_F$, which remains constant thereafter.

\bibliography{References.bib,extra.bib}
\bibliographystyle{utphys}

\end{document}